\documentclass[a4paper,11pt]{article}

\usepackage{amsmath,amsthm,amsbsy,amssymb,amsfonts,graphicx,mathrsfs,bm,cite,color,arydshln}
\usepackage{hyperref}
\usepackage[scr=boondoxo]{mathalfa}

\makeatletter
\catcode`\@=11
\@addtoreset{equation}{section}

\makeatother

\makeatletter

\@addtoreset{table}{section}
\makeatother

\usepackage[utf8]{inputenc}

\renewcommand{\thefootnote}{\arabic{footnote}}

\newcommand{\Exp}[1]{\operatorname{e}^{#1}}
\newcommand{\diag}{\operatorname{diag}}

\newcommand{\abs}[1]{\lvert {#1} \rvert}
\newcommand{\rmd}{{\mathrm{d}}}

\newcommand{\nn}{\nonumber}
\newcommand{\Lie}{\pounds}
\newcommand{\gLie}{\hat{\pounds}}

\newcommand{\cA}{\mathcal A}\newcommand{\cB}{\mathcal B}
\newcommand{\cC}{\mathcal C}\newcommand{\cD}{\mathcal D}

\newcommand{\cG}{\mathcal G}\newcommand{\cH}{\mathcal H}
\newcommand{\cI}{\mathcal I}\newcommand{\cJ}{\mathcal J}
\newcommand{\cK}{\mathcal K}\newcommand{\cL}{\mathcal L}
\newcommand{\cM}{\mathcal M}\newcommand{\cN}{\mathcal N}

\newcommand{\cR}{\mathcal R}

\newcommand{\cX}{\mathcal X}

\newcommand{\bba}{\alpha}
\newcommand{\bbb}{\beta}
\newcommand{\bbc}{\gamma}
\newcommand{\bbd}{\delta}

\newcommand{\dbba}{\dot\alpha}
\newcommand{\dbbb}{\dot\beta}
\newcommand{\dbbc}{\dot\gamma}
\newcommand{\dbbd}{\dot\delta}

\newcommand{\rr}{{\tt r}}

\newcommand{\hTheta}{\widehat{\Theta}}
\newcommand{\tTheta}{\widetilde{\Theta}}

\newcommand{\SL}{\text{SL}}
\newcommand{\SO}{\text{SO}}
\newcommand{\GL}{\text{GL}}
\newcommand{\OO}{\text{O}}

\newcommand{\hA}{\hat{A}}
\newcommand{\hB}{\hat{B}}
\newcommand{\hC}{\hat{C}}
\newcommand{\hD}{\hat{D}}
\newcommand{\hE}{\hat{E}}
\newcommand{\hF}{\hat{F}}

\newcommand{\hM}{\hat{M}}
\newcommand{\hN}{\hat{N}}
\newcommand{\hP}{\hat{P}}
\newcommand{\hQ}{\hat{Q}}
\newcommand{\hR}{\hat{R}}

\newcommand{\sfa}{\mathsf{a}}
\newcommand{\sfb}{\mathsf{b}}
\newcommand{\sfc}{\mathsf{c}}
\newcommand{\sfd}{\mathsf{d}}
\newcommand{\sfe}{\mathsf{e}}

\allowdisplaybreaks[3]

\begin{document}

\begin{titlepage}
\renewcommand{\thefootnote}{\fnsymbol{footnote}}

\vspace*{1cm}

\centerline{\Large\textbf{Half-maximal Extended Drinfel'd Algebras}}%

\vspace{1.5cm}

\centerline{\large Yuho Sakatani}

\vspace{0.2cm}

\begin{center}
{\it Department of Physics, Kyoto Prefectural University of Medicine,}\\
{\it 1-5 Shimogamohangi-cho, Sakyo-ku, Kyoto, Japan}\\
{\small\texttt{yuho@koto.kpu-m.ac.jp}}
\end{center}

\vspace*{2mm}

\begin{abstract}
Extended Drinfel'd algebra (ExDA) is the underlying symmetry of non-Abelian duality in the low-energy effective theory of string theory. Non-Abelian $U$-dualities in maximal supergravities have been studied well, but there has been no study on non-Abelian dualities in half-maximal supergravities. In this paper, we construct the ExDA for half-maximal supergravities in $d\geq 4$. We also find an extension of the homogeneous classical Yang--Baxter equation in these theories.
\end{abstract}

\thispagestyle{empty}
\end{titlepage}

\setcounter{footnote}{0}

\newpage

\setcounter{tocdepth}{2}
\tableofcontents

\newpage

\section{Introduction}

Recently, the Poisson--Lie (PL) $T$-duality \cite{hep-th:9502122} has been clarified and extended by using the duality-covariant formulations of supergravities, such as double field theory (DFT) \cite{hep-th:9302036,hep-th:9305073,hep-th:9308133,0904.4664,1006.4823} and exceptional field theory (EFT) \cite{1008.1763,1111.0459,1206.7045,1208.5884,1308.1673,1312.0614,1312.4542,1406.3348}. 
The initial progress was made in \cite{1707.08624} and further clarifications of the PL $T$-duality were made in \cite{1810.11446,1903.12175}. 
More recently, the PL $T$-duality in the presence of higher-derivative corrections has been studied in \cite{2007.07897,2007.07902,2007.09494}. 

The PL $T$-duality is based on a Lie algebra, called the Drinfel'd double, which is closely related to the $\OO(D,D)$ $T$-duality group. 
An extension of the Drinfel'd double that is based on the $\SL(5)$ $U$-duality group was proposed in \cite{1911.06320,1911.07833}, and it is called the exceptional Drinfel'd algebra (EDA). 
In \cite{2007.08510}, the $\SL(5)$ EDA was extended to the case of the $E_{6(6)}$ $U$-duality group, and it was further extended up to $E_{8(8)}$ $U$-duality group in \cite{2009.04454}. 
In \cite{2009.04454}, the EDA was formulated in terms of both M-theory and type IIB theory. 
Using these algebras, various concrete examples of non-Abelian $U$-dualities among solutions in 11D supergravity and type IIB supergravity were provided in \cite{2012.13263}. 
The non-Abelian $U$-duality in membrane sigma model was also studied in \cite{2001.09983}. 
However, at this time, non-Abelian $U$-dualities have been studied only in maximal supergravities. 

If we consider heterotic or type I supergravities compactified on a $D$-torus $T^D$ ($D\equiv 10-d$)\,, we can realize $d$-dimensional half-maximal supergravities. 
The purpose of this paper is to provide the algebraic basis for non-Abelian dualities in half-maximal supergravities. 
In $d\geq 5$, the duality group has been known to be $\cG=\mathbb{R}^+\times \OO(D,D+n)$\,, and it is enhanced to $\cG=\SL(2)\times \OO(6,6+n)$ in $d=4$. 
Extended field theories (ExFT) associated with these duality groups are known as the heterotic DFT \cite{hep-th:9302036,hep-th:9305073,hep-th:9308133,1103.2136,1109.4280,1201.2924} or the $\SL(2)$ DFT \cite{1612.05230}. 
Here, using these ExFTs and the general construction \cite{2009.04454} of extended Drinfel'd algebras (ExDA) for a wide class of duality group $\cG$\,, we construct the half-maximal ExDA
\begin{align}
 T_{\hA}\circ T_{\hB} = X_{\hA\hB}{}^{\hC}\,T_{\hC}\,.
\end{align}
By the definition, any ExDA has a maximally isotropic subalgebra $\mathfrak{g}$ generated by $T_a$ (which is a Lie algebra).
Using a group element $g=\Exp{x^a\,T_a}\in G\equiv \exp\mathfrak{g}$\,, we can systematically construct generalized frame fields $E_{\hA}{}^{\hM}\in \cG\times\mathbb{R}^+$ that satisfy the algebra
\begin{align}
 [E_{\hA},\,E_{\hB}]_{\text{D}} = -X_{\hA\hB}{}^{\hC}\,E_{\hC}\,,
\end{align}
where $[\cdot ,\,\cdot]_{\text{D}}$ denotes the generalized Lie derivative (or the D-bracket) in ExFT. 
For each duality group $\cG$\,, we identify the parameterization of the generalized frame fields, and find that they consist of several generalized Poisson--Lie structures, such as $\pi^{mn}$ and $\pi^m_{\cI}$\,. 

In \cite{1104.3587}, the embedding tensors of half-maximal gauged supergravity (with $n=0$) were obtained by acting a $\mathbb{Z}_2$ truncation to the embedding tensors of maximal gauged supergravity. 
Using the same $\mathbb{Z}_2$ truncation, the $\SL(2)$ DFT (with $n=0$) can be derived from the $E_{7(7)}$ EFT \cite{1612.05230}. 
Similarly, we can obtain various half-maximal ExFTs from the $E_{D+1(D+1)}$ EFT through a $\mathbb{Z}_2$ truncation (see \cite{1707.00714,1805.04524} for related works).
Then, as one may naturally expect, we can obtain the half-maximal ExDA from an $E_{D+1(D+1)}$ EDA through the $\mathbb{Z}_2$ truncation. 
However, the converse is not true. 
The Leibniz identities (or the quadratic constraints) in the maximal theory are stronger than in the half-maximal theory and not all of the embedding tensors in the half-maximal supergravity have an uplift to the maximal supergravity. 
The uplift condition has been discussed, for example, in \cite{1104.3587,1109.0290,1203.6562}. 
In this paper, we study the condition that a half-maximal ExDA can be uplifted to a maximal EDA (for simple cases $D\leq 3$). 
We then find some concrete examples where the uplift condition is violated. 

In the case of the Drinfel'd double, it is known that some of the Leibniz identities can be regarded as the cocycle condition. 
By considering the coboundary ansatz which automatically satisfies the cocycle condition, the dual structure constants $f_a{}^{bc}$ can be expressed by using the structure constants $f_{ab}{}^c$ and a skew-symmetric tensor $\rr^{ab}$\,. 
In that case, the other Leibniz identities are equivalent to the (modified) classical Yang--Baxter equations (CYBE) for $\rr^{ab}$\,. 
Similarly, in any ExDA, we can express some of the Leibniz identities as the cocycle condition. 
We identify the coboundary ansatz for the half-maximal ExDAs and obtain the generalized CYBE as a sufficient condition for the Leibniz identities to be satisfied. 

This paper is organized as follows. 
In section \ref{sec:ExFT}, we fix our convention on the half-maximal ExFTs in $d\geq 4$ by defining the generalized Lie derivative. 
In section \ref{sec:Hm-ExDA}, we construct the half-maximal ExDA in each dimension. 
We then identify the whole set of Leibniz identities. 
After that, the cocycle condition, coboundary ansatz, and the generalized CYBE are identified for each ExDA. 
We also discuss the relation between a half-maximal ExDA and an $E_{D+1(D+1)}$ EDA. 
In section \ref{sec:frame}, we show the explicit parameterization of the generalized frame fields $E_{\hA}{}^{\hM}$ by introducing generalized Poisson--Lie structures. 
We then show that the generalized frame fields satisfy the algebra $[E_{\hA},\,E_{\hB}]_{\text{D}} = -X_{\hA\hB}{}^{\hC}\,E_{\hC}$\,. 
In section \ref{sec:uplift}, we discuss a reduction of the half-maximal ExDA to a Leibniz algebra called DD$^+$ \cite{2104.00007}, and study the condition that the DD$^+$ can be uplifted to the half-maximal ExDA. 
We also study the conditions for a half-maximal ExDA to be uplifted to an EDA. 
In section \ref{sec:examples}, we find various non-trivial examples of the half-maximal ExDA. 
Some are uplifted EDAs, some are uplifted to embedding tensors which do not have the form of EDA. 
Section \ref{sec:conclusion} is devoted to conclusion and discussion. 

\medskip

A Mathematica notebook {\tt EDA.nb} can be found as an Ancillary file on arXiv \cite{EDA.nb}. 
This computes $X_{\hA\hB}{}^{\hC}$ for given structure constants (such as $f_{ab}{}^c$ and $f_a{}^{bc}$) and the generalized frame fields $E_{\hA}{}^{\hM}$ for a given parameterization of the group element, such as $g=\Exp{x^a\,T_a}$. 

\section{Generalized Lie derivative in half-maximal ExFT}
\label{sec:ExFT}

In this section, we consider the half-maximal ExFT in $d\geq 4$ where the duality group is\footnote{In $d=6$, it is also possible to consider the duality group $\cG=\OO(D+1,D+1+n)$ (see section \ref{sec:section-cond}).} 
\begin{align}
 \cG = \begin{cases}
 \mathbb{R}_d^+\times \OO(D,D+n) & (d\geq 5)
\\
 \SL(2)\times \OO(6,6+n)\quad & (d=4) 
\end{cases},
\end{align}
where $D\equiv 10-d$ and we have added the subscript ${}_d$ to $\mathbb{R}^+$ to indicate that this scale symmetry is related to the dilaton in heterotic supergravity.
In these ExFTs, we parameterize the generalized coordinates as
\begin{align}
 x^{\hM} =
\begin{cases}
 x^{M} =\bigl(x^{m},\,x^{\cI},\,x_m\bigr) & (d\geq 6)
\\
 \bigl(x^{M},\,x^*\bigr) =\bigl(x^{m},\,x^{\cI},\,x_m,\,x^*\bigr)\quad & (d=5) 
\\
 x^{\dbba M} =\bigl(x^{\dbba m},\,x^{\dbba \cI},\,x^{\dbba}{}_m\bigr) & (d=4)
\end{cases},
\end{align}
where $M=1,\dotsc,2D+n$ is the vector index for $\OO(D,D+n)$\,, $m=1,\dotsc,D$\,, $\cI=\dot{1},\cdots,\dot{n}$, and $\dbba=+,-$ is the index for an $\SL(2)$ doublet. 
On the extended space, infinitesimal diffeomorphisms are generated by the generalized Lie derivative \cite{hep-th:9302036,hep-th:9305073,hep-th:9308133,1103.2136,1109.4280,1201.2924,1612.05230,1707.00714}\footnote{We can also consider deformations of the generalized Lie derivative similar to \cite{1103.2136}, but here we consider the undeformed (or ungauged) theories.}
\begin{align}
 \gLie_{V} W^{\hM} = V^{\hN}\,\partial_{\hN} W^{\hM} - W^{\hN}\,\partial_{\hN} V^{\hM} + Y^{\hM\hP}_{\hQ\hN}\,\partial_{\hP} V^{\hQ}\,W^{\hN}\,.
\end{align}
Here the $Y$-tensor $Y^{\hM\hP}_{\hQ\hN}$ is defined as
\begin{align}
 Y^{\hM\hN}_{\hP\hQ} =
\begin{cases}
 \eta^{MN}\,\eta_{PQ} & (d\geq 6) 
\\
 \hat{\eta}_*^{\hM\hN}\,\hat{\eta}^*_{\hP\hQ} + \hat{\eta}^{\hM\hN}_R\,\hat{\eta}_{\hP\hQ}^R & (d=5) 
\\
 \delta^{\dbba}_{\dbbd}\,\delta^{\dbbb}_{\dbbc}\,\eta^{MN}\,\eta_{PQ} + 2\,\epsilon^{\dbba\dbbb}\,\epsilon_{\dbbc\dbbd}\, \delta^{MN}_{PQ}\quad & (d=4)
\end{cases},
\label{eq:Y-def}
\end{align}
where $\epsilon_{+-}=\epsilon^{+-}=1$\,, and
\begin{align}
\begin{alignedat}{3}
 \hat{\eta}^*_{\hM\hN} &\equiv \begin{pmatrix} \eta_{MN} & 0 \\
 0 & 0
\end{pmatrix},&\qquad
 \hat{\eta}_{\hM\hN}^P &\equiv \begin{pmatrix} 0 & \delta^P_M \\
 \delta_N^P & 0 
\end{pmatrix},&\qquad
 \eta_{MN} &\equiv \begin{pmatrix} 0 & 0 & \delta_m^n \\
 0 & \delta_{\cI\cJ} & 0 \\
 \delta^m_n & 0 & 0 
\end{pmatrix},
\\
 \hat{\eta}_*^{\hM\hN} &\equiv \begin{pmatrix} \eta^{MN} & 0 \\
 0 & 0 
\end{pmatrix},&\qquad
 \hat{\eta}^{\hM\hN}_P &\equiv \begin{pmatrix} 0 & \delta_P^M \\
 \delta^N_P & 0 
\end{pmatrix},&\qquad
 \eta^{MN} &\equiv \begin{pmatrix} 0 & 0 & \delta^m_n \\
 0 & \delta^{\cI\cJ} & 0 \\
 \delta_m^n & 0 & 0 
\end{pmatrix} .
\end{alignedat}
\end{align}
In this paper, we raise or lower the index $\cI$ using the Kronecker delta $\delta_{\cI\cJ}$\,. 

We denote the generators of the duality group $\cG$ collectively as $t_{\bm{\dot a}}$ $(\bm{\dot a}=1,\dotsc,\dim\cG)$\,. 
By using the matrix representations of $t_{\bm{\dot a}}$ and their duals $t^{\bm{\dot a}}$ (whose definition is given in Appendix \ref{app:conv}), the $Y$-tensor can be also expressed as
\begin{align}
 Y^{\hM\hN}_{\hP\hQ} = \delta^{\hM}_{\hP}\,\delta^{\hN}_{\hQ} + (t^{\bm{\dot a}})_{\hP}{}^{\hN} \, (t_{\bm{\dot a}})_{\hQ}{}^{\hM} + \beta_d\,\delta^{\hM}_{\hQ}\,\delta^{\hN}_{\hP} \qquad \bigl(\beta_d\equiv\tfrac{1}{d-2}\bigr)\,.
\label{eq:Y-tensor}
\end{align}
This shows that the generalized Lie derivative generates an infinitesimal (coordinate-dependent) duality rotation and a scale symmetry $\mathbb{R}^+$ with weight $\beta_d$
\begin{align}
 \gLie_{V} W^{\hM} = V^{\hN}\,\partial_{\hN} W^{\hM} + (\partial V)^{\bm{\dot a}}\, (t_{\bm{\dot a}})_{\hN}{}^{\hM}\,W^{\hN}
 - \beta_d\, (\partial_{\hP} V^{\hP})\,(t_0)_{\hN}{}^{\hM}\,W^{\hN}\,,
\label{eq:gLie}
\end{align}
where $(\partial V)^{\bm{\dot a}}\equiv \partial_{\hP} V^{\hQ}\,(t^{\bm{\dot a}})_{\hQ}{}^{\hP}$ and $(t_0)_{\hM}{}^{\hN}\equiv -\delta_{\hM}^{\hN}$ is the generator of the scale symmetry $\mathbb{R}^+$\,.

\subsection{Section condition}
\label{sec:section-cond}

For the consistency of ExFT, we impose the section condition,
\begin{align}
 Y^{\hM\hN}_{\hP\hQ}\,\partial_{\hM}\otimes \partial_{\hN} = 0\,.
\end{align}
In $d\geq 6$, this is equivalent to $\eta^{MN}\,\partial_{M}\otimes \partial_{N} = 0$\,, and as is well-known in DFT, there is only one solution of the section condition up to an $\OO(D,D+n)$ rotation, i.e.,
\begin{align}
 \partial_{m}\neq 0\,.
\end{align}
In $d=5$, the section condition is decomposed into two conditions
\begin{align}
 \eta^{MN}\,\partial_{M}\otimes\partial_{N}=0\,,\qquad \partial_{M}\otimes\partial_{*} = 0\,,
\end{align}
and there are two inequivalent solutions \cite{1707.00714}
\begin{align}
 (i)\quad \partial_{m}\neq 0\,,\qquad\quad 
 (ii)\quad \partial_*\neq 0 \,.
\end{align} 
The former gives a five-dimensional section while the latter gives a one-dimensional section. 
In $d=4$, the section condition is decomposed as \cite{1612.05230}
\begin{align}
 \eta^{MN}\,\partial_{\dbba M}\otimes\partial_{\dbbb N}=0\,,\qquad \epsilon^{\dbba\dbbb}\,\partial_{\dbba[M|}\otimes\partial_{\dbbb |N]} = 0\,,
\end{align}
and again there are two inequivalent solutions \cite{1612.05230}
\begin{align}
 (i)\quad \partial_{+m}\neq 0\,,\qquad\quad 
 (ii)\quad \partial_{\pm 1}\neq 0 \,.
\end{align} 
The former is a six-dimensional solution while the latter is a two-dimensional solution. 

In $d=5$ (and $d=4$), the first solution $(i)$ is suitable for describing heterotic/type I theory compactified on $T^D$, where $x^{m}$ (and $x^{+m}$) play the role of coordinates on $T^D$. 
It is the same for the solution in $d\geq 6$.
On the other hand, the second solution $(ii)$ in $d=4,5$ describes a $T^{D-4}$ compactification of six-dimensional (2,0) supergravity \cite{1612.05230,1707.00714}. 
This series of solutions reduces to the 0-dimensional solution in $d=6$\,, where the duality group becomes $\cG=\OO(5,5+n)$ and the ExFT describes the six-dimensional $(2,0)$ supergravity. 
In this paper, we restrict ourselves to the former solution. 
For simplicity, in the following, when we consider $d=4$, we may use a short-hand notation, such as $x^m\equiv x^{+m}$ and $\partial_m\equiv\partial_{+m}$\,. 

\subsection{Generalized Lie derivative}

In $d\geq 6$\,, under the section $\partial_m\neq 0$\,, the generalized Lie derivative reduces to
\begin{align}
 \gLie_{V} W^{\hM} = \begin{pmatrix}
 [v,w]^m \\
 v\cdot w^{\cI} - w\cdot v^{\cI} \\
 v\cdot w_m + \partial_m v^n\,w_n - w^{n}\,(\rmd v)_{nm} + w_{\cI}\,\partial_m v^{\cI} 
 \end{pmatrix},
\end{align}
where we have parameterized the generalized vector fields, for example, as
\begin{align}
 V^{\hM} = \bigl(v^{m},\,v^{\cI},\,v_m\bigr)\,,
\end{align}
and have denoted $v{}\cdot{} \equiv v^m\,\partial_m$ and $(\rmd v)_{mn}\equiv 2\,\partial_{[m} v_{n]}$\,.
Here, two scale transformations, $\mathbb{R}^+_d$ (contained in the second term of Eq.~\eqref{eq:gLie}) and $\mathbb{R}^+$ (the last term of Eq.~\eqref{eq:gLie}), are cancelled out and the generalized Lie derivative generates an infinitesimal $\OO(D,D+n)$ transformation similar to DFT.

In $d=5$\,, under the section $(i)$, we have
\begin{align}
 \gLie_{V} W^{\hM} = \begin{pmatrix}
 [v,w]^m \\
 v\cdot w^{\cI} - w\cdot v^{\cI} \\
 v\cdot w_m + \partial_m v^n\,w_n - w^{n}\,(\rmd v)_{nm} + w_{\cI}\,\partial_m v^{\cI} 
\\
 v\cdot w^* + \partial_m v^m\,w^*
 \end{pmatrix},
\end{align}
where we parameterized the generalized vector fields as
\begin{align}
 V^{\hM} = \bigl(v^{m},\,v^{\cI},\,v_{m},\,v^*\bigr)\,.
\label{eq:V-5d}
\end{align}
In this case, a combination of $\mathbb{R}^+_d$ and $\mathbb{R}^+$ generates the scale transformation $\partial_m v^m\,w^*$ in the last line.
Due to this scale transformation, the last component $w^*$ behaves as a scalar density. 

In $d=4$\,, under the section $(i)$, we find
\begin{align}
 \gLie_{V} W^{\hM} = {\small\begin{pmatrix}
 [v^+, w^{+}]^m \\
 [v^{+}, w^{-}]^m + \epsilon_{\dbba\dbbb}\,\partial_{n} v^{\dbba n}\,w^{\dbbb m} \\
 v^{+}\cdot w^{+\cI} - w^{+}\cdot v^{+\cI} \\
 v^{+}\cdot w^{-\cI} - w^{-}\cdot v^{+\cI} + \epsilon_{\dbba\dbbb}\,\partial_n v^{\dbba n} \, w^{\dbbb \cI} \\
 v^{+}\cdot w^{+}{}_m + \partial_m v^{+n}\,w^{+}{}_n - w^{+n}\,(\rmd v^{+})_{nm} + w^{+}{}_{\cI}\,\partial_m v^{+\cI} 
\\
 v^{+}\cdot w^{-}{}_m + \partial_m v^{+n}\,w^{-}{}_n - w^{-n}\,(\rmd v^{+})_{nm} + w^{-}{}_{\cI}\,\partial_m v^{+\cI} + \epsilon_{\dbba\dbbb}\,\partial_n v^{\dbba n}\, w^{\dbbb}{}_m 
 \end{pmatrix}},
\end{align}
where we have used the parameterization
\begin{align}
 V^{\hM} = \bigl(v^{+m},\,v^{-m},\,v^{+\cI},\,v^{-\cI},\,v^{+}{}_m,\,v^{-}{}_m\bigr)\,.
\end{align}
Due to a combination of a part of $\SL(2)$ transformation and the scale symmetry $\mathbb{R}^+$\,, the minus components $v^{-M}$ behave as tensors densities. 

In the following, we denote the generalized Lie derivative as
\begin{align}
 [V,\,W]_{\text{D}} \equiv \gLie_{V} W \,,
\end{align}
which is called the D-bracket and is not skew symmetric: $[V,\,W]_{\text{D}}\neq -[W,\,V]_{\text{D}}$. 
The antisymmetric part is known as the C-bracket,
\begin{align}
 [V,\,W]_{\text{C}} \equiv \tfrac{1}{2}\,\bigl(\gLie_{V}W - \gLie_{W}V\bigr) = - [W,\,V]_{\text{C}}\,,
\end{align}
although we do not use this bracket in this paper. 

\section{Half-maximal ExDA}
\label{sec:Hm-ExDA}

In this section, we construct the half-maximal ExDA by using the generalized Lie derivative introduced in the previous section. 
We then study the Leibniz identities of the ExDA in section \ref{sec:Leibniz}. 
In section \ref{sec:coboundary-ExDA}, some of the Leibniz identities are interpreted as the cocycle condition.
By considering the coboundary-type ExDA, we find the generalized CYBE in section \ref{sec:g-CYBE}. 
The relation between the half-maximal ExDA and the $E_{D+1(D+1)}$ EDA is detailed in section \ref{sec:ExDA-EDA}.

\subsection{Algebra}
\label{sec:ExDA}

An ExDA is a Leibniz algebra
\begin{align}
 T_{\hA}\circ T_{\hB} = X_{\hA\hB}{}^{\hC}\,T_{\hC}\,,
\label{eq:ExDA-T}
\end{align}
with generators $T_{\hA}$ transforming in the vector representation of the duality group $\cG$\,. 
Similar to the curved index $\hM$, we decompose the ``flat'' index $\hA$ as
\begin{align}
 T_{\hA} =
\begin{cases}
 T_A =\bigl(T_a,\,T_I,\,T^a\bigr) & (d\geq 6)
\\
 \bigl(T_A,\,T_*\bigr) =\bigl(T_a,\,T_I,\,T^a,\,T_*\bigr)\quad & (d=5) 
\\
 T_{\bba A} =\bigl(T_{\bba a},\,T_{\bba I},\,T_{\bba}{}^a\bigr) & (d=4)
\end{cases},
\end{align}
where $A=1,\dotsc,2D+n$, $I=\dot{1},\dotsc,\dot{n}$, $a=1,\dotsc,D$, and $\bba=+,-$\,. 
We raise or lower the index $I$ or $A$ using $\delta_{IJ}$ or $\eta_{AB}$, respectively.\footnote{$\eta_{AB}$ has the same matrix form as $\eta_{MN}$\,.} 
To simplify notation in $d=4$\,, we may denote the index $T_{+a}$ as $T_{a}$\,. 
The structure constants $X_{\hA\hB}{}^{\hC}$ are defined such that there exist certain generalized frame fields $E_{\hA}{}^{\hM}\in\cG\times \mathbb{R}^+$ satisfying the same algebra by means of the D-bracket,
\begin{align}
 [E_{\hA},\,E_{\hB}]_{\text{D}} = - X_{\hA\hB}{}^{\hC}\,E_{\hC}\,.
\label{eq:ExDA}
\end{align}
In general, the coefficients on the right-hand side are non-constant and are called the generalized fluxes $\bm{X}_{\hA\hB}{}^{\hC}$, but here we consider case where the generalized fluxes are constant: $\bm{X}_{\hA\hB}{}^{\hC}=X_{\hA\hB}{}^{\hC}$\,. 
In such a situation, Eq.~\eqref{eq:ExDA} can be regarded as the condition for the generalized parallelizability \cite{0807.4527,1401.3360} and the inverse $E_{\hM}{}^{\hA}$ of the generalized frame fields plays the role of the twist matrix for the generalized Scherk--Schwarz reduction. 
The construction of such generalized frame fields is discussed in section \ref{sec:frame}, and here we focus on finding the explicit form of the structure constants $X_{\hA\hB}{}^{\hC}$\,. 

By the definition of the D-bracket and Eq.~\eqref{eq:ExDA}, the structure constants should be expressed as\footnote{The matrices $(t_{\bm{a}})_{\hB}{}^{\hC}$, $(t^{\bm{a}})_{\hB}{}^{\hC}$, and $(t_0)_{\hB}{}^{\hC}$ have the same form as the curved ones, such as $(t_{\bm{\dot a}})_{\hM}{}^{\hN}$\,. Since $(t_{\bm{a}})_{\hB}{}^{\hC}$ and $(t^{\bm{a}})_{\hB}{}^{\hC}$ are invariant tensors and $E_{\hA}{}^{\hM}\in\cG\times\mathbb{R}^+$\,, we can convert the flat indices to the curved indices using $E_{\hA}{}^{\hM}$: e.g., $(t^{\bm{a}})_{\hD}{}^{\hE}\,(t_{\bm{a}})_{\hB}{}^{\hC}\,E_{\hN}{}^{\hB}\,E_{\hC}{}^{\hP}\,E_{\hQ}{}^{\hD}\,E_{\hE}{}^{\hR}=(t^{\bm{\dot a}})_{\hQ}{}^{\hR}\,(t_{\bm{\dot a}})_{\hN}{}^{\hP}$\,.} (see \cite{2009.04454} for general discussion)
\begin{align}
 X_{\hA\hB}{}^{\hC} 
 = \Omega_{\hA\hB}{}^{\hC} + (t^{\bm{a}})_{\hD}{}^{\hE}\,(t_{\bm{a}})_{\hB}{}^{\hC}\,\Omega_{\hE\hA}{}^{\hD} -\beta_d\,\Omega_{\hD\hA}{}^{\hD}\,(t_0)_{\hB}{}^{\hC}\,,
\label{eq:X-Omega}
\end{align}
by using some constants $\Omega_{\hA\hB}{}^{\hC}$ which can be understood as the Weitzenb\"ock connection
\begin{align}
 W_{\hA\hB}{}^{\hC}\equiv E_{\hA}{}^{\hM}\,E_{\hB}{}^{\hN}\,\partial_{\hM} E_{\hN}{}^{\hC}\,,
\label{eq:W-def}
\end{align}
evaluated at a certain point: $\Omega_{\hA\hB}{}^{\hC}=W_{\hA\hB}{}^{\hC}\rvert_{x=x_0}$. 
In fact, there is a special point $x_0$ where $E_{\hA}{}^{m}=\delta_{\hA}^{a}\, E_{a}{}^{m}$,\footnote{Here $a$ is to be understood as $+a$ in $d=4$\,.} and we choose the $x_0$ as such a point. 
Then, since we are choosing the section $\partial_m \neq 0$\,, the only non-vanishing components of $\Omega_{\hA\hB}{}^{\hC}$ are $\Omega_{a\hB}{}^{\hC}$\,. 
Moreover, because of $E_{\hA}{}^{\hM}\in \cG\times\mathbb{R}^+$ and Eq.~\eqref{eq:W-def}, the constants $\Omega_{a\hB}{}^{\hC}$ are generally expanded as 
\begin{align}
 \Omega_{a\hB}{}^{\hC} = \Omega_{a}{}^{\bm{a}}\,(t_{\bm{a}})_{\hB}{}^{\hC} + \Omega_{a}{}^0\,(t_0)_{\hB}{}^{\hC}\,.
\end{align}

Now, we require that the generators $T_{a}$ form a subalgebra $\mathfrak{g}$
\begin{align}
 T_{a} \circ T_{b} = f_{ab}{}^c\,T_{c}\,.
\label{eq:g-subalgebra}
\end{align}
This requirement gives a strong constraint on $\Omega_{a}{}^{\bm{a}}$\,, and in the following, we determine the explicit form of $\Omega_{a}{}^{\bm{a}}$\,. 
Using this $\Omega_{a}{}^{\bm{a}}$ and the relation \eqref{eq:X-Omega}, we can compute the structure constants of the ExDA $X_{\hA\hB}{}^{\hC}$\,. 

In the following, we decompose the $\OO(D,D+n)$ generators as (see Appendix \ref{app:conv} for more details)
\begin{align}
 \bigl\{ \tfrac{R_{a_1a_2}}{\sqrt{2!}},\,R_a^I,\,K^{a_1}{}_{a_2},\, R_{IJ},\,R^a_I,\, \tfrac{R^{a_1a_2}}{\sqrt{2!}}\bigr\}\,.
\end{align}
We also denote the $\mathbb{R}^+_d$ generator in $d\geq 5$ as $R_*$ and the $\SL(2)$ generators in $d=4$ as $R^{\bba}{}_{\bbb}$\,. 
Using these generators, we determine the explicit form of $\Omega_{a}{}^{\bm{a}}$ in each dimension. 

\subsubsection{ExDA in $d\geq 6$}

In $d\geq 6$, the requirement \eqref{eq:g-subalgebra} is satisfied if $\Omega_{a}{}^{\bm{a}}$ and $\Omega_{a}{}^{0}$ are expanded as
\begin{align}
 \Omega_{a}{}^{\bm{a}}\,t_{\bm{a}}
 = (k_{ab}{}^c - Z_a\,\delta_b^c)\,K^b{}_c 
 + \tfrac{1}{2!}\,f_{a}{}^{IJ}\,R_{IJ}
 + f_{a}{}^b{}_I\,R_b^I
 + \tfrac{1}{2!}\,f_{a}{}^{bc}\,R_{bc} \,, \qquad
 \Omega_{a}{}^0 = - Z_a\,,
\end{align}
without using generators $R^a_I$ and $R^{a_1a_2}$\,. 
We note that since $R_*$ is proportional to $t_0$\,, we have absorbed the structure constant associated with $R_*$ into $Z_a$\,. 
By substituting these into Eq.~\eqref{eq:X-Omega}, the matrices $(X_{\hA})_{\hB}{}^{\hC}\equiv X_{\hA\hB}{}^{\hC}$ are found as
\begin{align}
\begin{split}
 X_{a} &= 
 f_{ab}{}^c\,K^b{}_c
 + \tfrac{1}{2!}\, f_a{}^{bc}\,R_{bc} 
 + f_a{}^b{}_I\,R_b{}^I 
 + \tfrac{1}{2!}\, f_a{}^{IJ}\,R_{IJ} 
 - Z_a\,(K + t_0) \,,
\\
 X_{I} &= f_{a}{}^b{}_I\, K^a{}_b 
 - f_{aI}{}^{J}\, R^a{}_J 
 - Z_a\,R^a{}_I \,,
\\
 X^a &= f_b{}^{ca}\,K^b{}_c 
 - f_b{}^{aI} \,R^b{}_I 
 + \bigl(\tfrac{1}{2}\,f_{bc}{}^a - 2\,Z_{[b}\,\delta_{c]}^a \bigr)\,R^{bc} \,,
\end{split}
\end{align}
where $f_{ab}{}^c\equiv 2\,k_{[ab]}{}^c$ and $K\equiv K^a{}_a$\,. 
Then the algebra \eqref{eq:ExDA-T} becomes
\begin{align}
\begin{split}
 T_{a}\circ T_{b} &= f_{ab}{}^c\,T_{c}\,,
\\
 T_{a}\circ T_{J} &= -f_{a}{}^c{}_J \,T_{c} + f_{aJ}{}^{K}\,T_{K} + Z_a\,T_{J} \,,
\\
 T_{a}\circ T^b &= f_a{}^{bc}\,T_{c} + f_a{}^{bK}\,T_{K} - f_{ac}{}^b\,T^c + 2\,Z_a\,T^b\,,
\\
 T_{I} \circ T_{b} &= f_b{}^{c}{}_I\,T_{c} - f_b{}_I{}^K\,T_{K} -Z_b\,T_{I} \,,
\\
 T_{I}\circ T_{J} &= f_{cIJ}\,T^c + \delta_{IJ}\,Z_c\,T^c\,,
\\
 T_{I}\circ T^b &= -f_c{}^b{}_I\,T^c\,,
\\
 T^a\circ T_{b} &= -f_b{}^{ac}\,T_{c} - f_b{}^{aK}\,T_{K} + \bigl(f_{bc}{}^a+2\,\delta^a_b\,Z_c-2\,\delta^a_c\,Z_b\bigr)\,T^c \,,
\\
 T^a\circ T_{J} &= f_c{}^{a}{}_J\,T^c \,,
\\
 T^a\circ T^b &= f_c{}^{ab}\,T^c \,.
\end{split}
\label{eq:ExDA-Odd}
\end{align}
This is the Leibniz algebra of the half-maximal ExDA in $d\geq 6$\,. 
We note that the symmetric part $k_{(ab)}{}^c$ does not appear in the algebra. 

We can neatly express the structure constants $X_{\hA\hB}{}^{\hC}=X_{AB}{}^C$ as
\begin{align}
 X_{AB}{}^C &= F_{AB}{}^C + \eta_{AB}\,\xi^C + \xi_A\,\delta_B^C - \delta^C_{A} \,\xi_B \,, 
\end{align}
where components of the 3-form $F_{ABC}\equiv F_{AB}{}^D\,\eta_{DC} = F_{[ABC]}$ and $\xi_A$ are
\begin{align}
\begin{split}
 F_{ab}{}^c &= f_{ab}{}^c + \delta^c_a\,Z_b - \delta^c_b\,Z_a \,,\quad 
 F_{aJK} = f_{aJK} \,,\quad 
 F_{aJ}{}^c = -f_{a}{}^c{}_J \,,\quad 
 F_{a}{}^{bc} = f_{a}{}^{bc}\,,
\\
 F_{abc} &= F_{IJK} = F_{IJ}{}^c = F_I{}^{bc} = F^{abc} =0 \,,\qquad
 \xi_{A} \equiv \bigl(Z_a,\, 0,\, 0\bigr)\,.
\end{split}
\label{eq:F-comp}
\end{align}
If we set $n=0$\,, the half-maximal ExDA reduces to the Leibniz algebra DD$^+$ that plays a key role in the Jacobi--Lie $T$-plurality \cite{2104.00007}. 

\subsubsection{ExDA in $d=5$}

In $d=5$, the requirement \eqref{eq:g-subalgebra} is satisfied by
\begin{align}
\begin{split}
 \Omega_{a}{}^{\bm{a}}\,t_{\bm{a}}
 &= \bigl(f_a + k_{ba}{}^b\bigr)\,R_*
 + (k_{ab}{}^c - Z_a\,\delta_b^c)\,K^b{}_c 
 + \tfrac{1}{2!}\,f_{a}{}^{IJ}\,R_{IJ}
 + f_{a}{}^b{}_I\,R_b^I
 + \tfrac{1}{2!}\,f_{a}{}^{bc}\,R_{bc} \,, 
\\
 \Omega_{a}{}^0 &= \tfrac{1}{3}\, \bigl(f_a + k_{ba}{}^b\bigr) - Z_a \,.
\end{split}
\label{eq:Omega-a-4d}
\end{align}
The embedding tensors can be obtained as
\begin{align}
\begin{split}
 X_{a} &= 
 f_{ab}{}^c\,K^b{}_c
 + \tfrac{1}{2!}\, f_a{}^{bc}\,R_{bc} 
 + f_a{}^b{}_I\,R_b{}^I 
 + \tfrac{1}{2!}\, f_a{}^{IJ}\,R_{IJ} 
 - Z_a\,(K + t_0)
 + f_a \, (R_*+\tfrac{1}{3}\,t_0) \,,
\\
 X_{I} &= f_{a}{}^b{}_I\, K^a{}_b 
 - f_{aI}{}^{J}\, R^a{}_J 
 - Z_a\,R^a{}_I 
 + f_a{}^a{}_I\,\bigl(R_*+\tfrac{1}{3}\,t_0\bigr) \,,
\\
 X^a &= f_b{}^{ca}\,K^b{}_c 
 - f_b{}^{aI} \,R^b{}_I
 + \bigl(\tfrac{1}{2}\,f_{bc}{}^a - 2\,Z_{[b}\,\delta_{c]}^a \bigr)\,R^{bc} 
 + f_b{}^{ba}\,\bigl(R_*+\tfrac{1}{3}\,t_0\bigr)\,, \qquad
 X_* = 0\,.
\end{split}
\end{align}
Again, only the antisymmetric part $f_{ab}{}^c\equiv 2\,k_{[ab]}{}^c$ appears in the embedding tensor although $\Omega_{a}{}^{\bm{a}}$ and $\Omega_{a}{}^0$ contain the symmetric part $k_{(ab)}{}^c$ as well. 

We find that the generators $T_A\equiv (T_a,\,T_I,\,T^a)$ form the subalgebra given in Eq.~\eqref{eq:ExDA-Odd}.
The products including the additional generator $T_*$ can be found as
\begin{align}
\begin{alignedat}{3}
 T_a &\circ T_* = (Z_a - f_a)\,T_*\,,\qquad&
 T_I &\circ T_* = - f_c{}^c{}_I\,T_*\,,\qquad&
 T^a &\circ T_* = - f_c{}^{ca}\,T_*\,,
\\
 T_* &\circ T_b =0\,,\qquad&
 T_* &\circ T_J =0\,,\qquad&
 T_* &\circ T^b =0\,.
\end{alignedat}
\end{align}
We can neatly express the non-vanishing components of the structure constants $X_{\hA\hB}{}^{\hC}$ as
\begin{align}
 X_{AB}{}^C &= F_{AB}{}^C + \eta_{AB}\,\xi^C + \xi_A\,\delta_B^C - \delta^C_{A} \,\xi_B \,, \qquad 
 X_{A*}{}^* = - 2\,\xi_A - 3\,\vartheta_A \,,
\label{eq:5d-X}
\end{align}
where $F_{AB}{}^C$ and $\xi_{A}$ are the same as Eq.~\eqref{eq:F-comp} and
\begin{align}
 \vartheta_{A} \equiv \tfrac{1}{3}\,\bigl(f_a - 3\,Z_a,\, f_b{}^{b}{}_I,\, f_b{}^{ba}\bigr)\,.
\end{align} 

We can compare the expression \eqref{eq:5d-X} with Eq.~(3.6) of \cite{hep-th:0602024}. 
Our $F_{AB}{}^C$ and $\xi_A$ correspond to their $-f_{AB}{}^C$ and $-\frac{1}{2}\,\xi_A$\,. 
The embedding tensor $\xi_{AB}$ of \cite{hep-th:0602024} is not present in our ExDA. 
On the other hand, our $\vartheta_{A}$ is not present there because the trombone symmetry $\mathbb{R}^+$ has not been gauged in \cite{hep-th:0602024}. 

\subsubsection{ExDA in $d=4$}

In $d=4$, we find that
\begin{align}
\begin{split}
 \Omega_{+a}{}^{\bm{a}}\,t_{\bm{a}}
 &\equiv f_{a\bba}{}^{\bbb}\,R^{\bba}{}_{\bbb} 
 + (k_{ab}{}^c - Z_a\,\delta_b^c)\,K^b{}_c 
 + \tfrac{1}{2!}\,f_{a}{}^{IJ}\,R_{IJ}
 + f_{a}{}^b{}_I\,R_b^I
 + \tfrac{1}{2!}\,f_{a}{}^{bc}\,R_{bc} \,,
\\
 \Omega_{+a}{}^0 &\equiv f_{a+}{}^+ - Z_a\,,\qquad
 f_{a+}{}^+ = -f_{a-}{}^- = \tfrac{1}{2}\,(k_{ba}{}^b + f_a) \,,
\end{split}
\end{align}
is consistent with Eq.~\eqref{eq:g-subalgebra}.
Using these parameterizations, we find
\begin{align}
 X_{+a} &= 
 f_{ab}{}^c\,K^b{}_c
 + f_{a-}{}^{+}\,R^{-}{}_{+} 
 + \tfrac{1}{2!}\, f_a{}^{bc}\,R_{bc} 
 + f_a{}^b{}_I\,R_b{}^I 
 + \tfrac{1}{2!}\, f_a{}^{IJ}\,R_{IJ} 
\nonumber\\
 &\quad + f_a\,\bigl(R^{+}{}_{+}+\tfrac{1}{2}\,t_0\bigr)
 - Z_a\,(K + t_0) \,,
\label{eq:ExDA-X-1}
\\
 X_{-a} &= -f_{b-}{}^{+}\,K^b{}_a - f_{a-}{}^{+}\,\bigl(R^{+}{}_{+}+\tfrac{1}{2}\,t_0\bigr) + f_a\,R^{+}{}_{-} \,,
\\
 X_{+I} &= f_{a}{}^b{}_I\, K^a{}_b 
 - f_{aI}{}^{J}\, R^a{}_J 
 - Z_a\,R^a{}_I 
 + f_a{}^a{}_I\,\bigl(R^{+}{}_{+}+\tfrac{1}{2}\,t_0\bigr) \,,
\\
 X_{-I} &= f_a{}^a{}_I\,R^{+}{}_{-}
 - f_{a-}{}^{+}\,R^a{}_I\,,
\\
 X_{+}{}^a &= f_b{}^{ca}\,K^b{}_c 
 - f_b{}^{aI} \,R^b{}_I 
 + \bigl(\tfrac{1}{2}\,f_{bc}{}^a - 2\,Z_{[b}\,\delta_{c]}^a \bigr)\,R^{bc} 
 + f_b{}^{ba}\,\bigl(R^{+}{}_{+}+\tfrac{1}{2}\,t_0\bigr)\,,
\\
 X_{-}{}^a &= f_b{}^{ba}\,R^{+}{}_{-} + f_{b-}{}^{+}\,R^{ab}\,.
\label{eq:ExDA-X-6}
\end{align}
We note that the constants $\Omega_{+a\hB}{}^{\hC}$ of the form
\begin{align}
 \Omega_{+a\hB}{}^{\hC} = k_{ad}{}^e\,(K^d{}_e)_{\hB}{}^{\hC} + k_{da}{}^d\,(R^+{}_+ + \tfrac{1}{2}\,t_0)_{\hB}{}^{\hC}\,,
\label{eq:Omega-anti-symmetric}
\end{align}
contributes to $X_{\hA\hB}{}^{\hC}$ only through the anti-symmetric part $k_{[ab]}{}^c=\frac{1}{2}\,f_{ab}{}^c$\,, and again, the symmetric part $k_{(ab)}{}^c$ does not show up in $X_{\hA\hB}{}^{\hC}$\,. 
Accordingly, the subalgebra $\mathfrak{g}$ with the structure constants $f_{ab}{}^c$ is a Lie algebra. 
Moreover, $f_{a+}{}^-$ does not appear in $X_{\hA\hB}{}^{\hC}$ and we ignore $f_{a+}{}^-$ in the following discussion. 

The explicit form of the half-maximal ExDA is as follows:
\begin{align}
\begin{split}
 T_{\bba a}\circ T_{\bbb b} &= \delta_{\bba}^+\,f_{ab}{}^c\,T_{\bbb c} 
 - \epsilon_{\bba\bbb}\,f_a\,T_{-b}
 + \delta_{\bbb}^-\,f_{a-}{}^+\,T_{\bba b}
 - \delta_{\bba}^-\,f_{b-}{}^+\,T_{\bbb a} \,,
\\
 T_{\bba a}\circ T_{\bbb J}
 &= \delta_{\bba}^+\,\bigl(-f_{a}{}^c{}_J \,T_{\bbb c} + f_{aJ}{}^{K}\,T_{\bbb K} + Z_a\,T_{\bbb J}\bigr)
 + \delta_{\bbb}^-\, f_{a-}{}^+\,T_{\bba J} 
 - \epsilon_{\bba\bbb}\, f_a \,T_{-J} \,,
\\
 T_{\bba a}\circ T_{\bbb}{}^b &= \delta_{\bba}^+\,\bigl(f_a{}^{bc}\,T_{\bbb c} + f_a{}^{bK}\,T_{\bbb K} - f_{ac}{}^b\,T_{\bbb}{}^c + 2\,Z_a\,T_{\bbb}{}^b \bigr)
\\
 &\quad - \epsilon_{\bba\bbb}\,f_a\,T_-{}^b 
 + \delta_{\bbb}^-\, f_{a-}{}^+\,T_{\bba}{}^b
 + \delta_{\bba}^-\,\delta_a^b\,f_{c-}{}^+\,T_{\bbb}{}^c\,,
\\
 T_{\bba I} \circ T_{\bbb b} &= \delta_{\bba}^+\,\bigl(f_b{}^{c}{}_I\,T_{\bbb c} - f_b{}_I{}^K\,T_{\bbb K} -Z_b\,T_{\bbb I} \bigr)
 - \epsilon_{\bba\bbb}\,f_d{}^{d}{}_I\,T_{-b} 
 -\delta_{\bba}^-\,f_{b-}{}^+\,T_{\bbb I} \,,
\\
 T_{\bba I}\circ T_{\bbb J} &= \delta_{\bba}^+\,\bigl(f_{cIJ} + \delta_{IJ}\,Z_c\bigr)\,T_{\bbb}{}^c
 -\epsilon_{\bba\bbb}\,f_d{}^d{}_I\,T_{-J} 
 + \delta_{\bba}^-\,\delta_{IJ}\,f_{c-}{}^+\,T_{\bbb}{}^c\,,
\\
 T_{\bba I}\circ T_{\bbb}{}^b &= -\delta_{\bba}^+\,f_c{}^b{}_I\,T_{\bbb}{}^c - \epsilon_{\bba\bbb}\,f_d{}^{d}{}_I\,T_-{}^b\,,
\\
 T_{\bba}{}^a\circ T_{\bbb b} &= -\delta_{\bba}^+\,\big[f_b{}^{ac}\,T_{\bbb c} + f_b{}^{aJ}\,T_{\bbb J} - \bigl(f_{bc}{}^a+2\,\delta^a_b\,Z_c-2\,\delta^a_c\,Z_b\bigr)\,T_{\bbb}{}^c \bigr]
\\
 &\quad
 - \epsilon_{\bba\bbb}\,f_d{}^{da}\,T_{-b} 
 + \delta_{\bba}^-\,\bigl(\delta^a_b\,f_{c-}{}^+ - \delta_c^a\,f_{b-}{}^+\bigr)\,T_{\bbb}{}^c \,,
\\
 T_{\bba}{}^a\circ T_{\bbb J} &= \delta_{\bba}^+\,f_c{}^{a}{}_J\,T_{\bbb}{}^c - \epsilon_{\bba\bbb}\,f_d{}^{da}\,T_{-J} \,,
\\
 T_{\bba}{}^a\circ T_{\bbb}{}^b &= \delta_{\bba}^+\,f_c{}^{ab}\,T_{\bbb}{}^c -\epsilon_{\bba\bbb}\, f_d{}^{da}\,T_{-}{}^b \,.
\end{split}
\end{align}
A more explicit expression is given in Appendix \ref{app:4D-ExDA}. 
It is noted that the generators $T_{A}\equiv (T_{+a},\,T_{+I},\,T_+{}^a)$ form a subalgebra which has the same form as Eq.~\eqref{eq:ExDA-Odd}. 

Now we can compare the algebra with the embedding tensor known in $\cN=4$, $d=4$ gauged supergravity \cite{hep-th:0602024}.
Using the trombone gauging $\vartheta_{\bba A}$ \cite{1612.05230}, we can parameterize the structure constants $X_{\hA\hB}{}^{\hC}$ as\footnote{For our convenience, we have chosen the sign of $F_{\bba AB}{}^C$ and $\xi_{\bba A}$ to be the opposite of that in \cite{hep-th:0602024,1612.05230}.}
\begin{align}
\begin{split}
 X_{\hA\hB}{}^{\hC} &= \delta_{\bbb}^{\bbc}\,F_{\bba AB}{}^C
 -\tfrac{1}{2}\,\bigl(\delta_A^C\,\delta_{\bbb}^{\bbc}\,\xi_{\bba B}
 -\delta_B^C\,\delta_{\bba}^{\bbc}\,\xi_{\bbb A}
 -\delta_{\bbb}^{\bbc}\,\eta_{AB}\, \xi_{\bba}^C
 +\epsilon_{\bba\bbb}\,\delta_B^C\,\xi_{\bbd A}\,\epsilon^{\bbd\bbc}\bigr)
\\
 &\quad + \delta_A^C\,\delta_{\bbb}^{\bbc}\,\vartheta_{\bba B} 
 - \delta_{\bbb}^{\bbc}\,\eta_{AB}\, \vartheta_{\bba}^C
 - \delta_{\bbb}^{\bbc}\,\delta_B^C\,\vartheta_{\bba A} \,.
\end{split}
\end{align}
By comparing this with Eqs.~\eqref{eq:ExDA-X-1}--\eqref{eq:ExDA-X-6}, we find
\begin{align}
\begin{split}
\begin{alignedat}{2}
 F_{+AB}{}^C&= F_{AB}{}^C\,,&\qquad 
 F_{-AB}{}^C&=0\,,
\\
 \xi_{+A} &= \bigl(f_a,\, f_b{}^{b}{}_I,\, f_b{}^{ba} \bigr)\,,&\qquad
 \xi_{-A} &= \bigl(f_{a-}{}^+,\, 0,\, 0 \bigr)\,,
\\
 \vartheta_{+A} &= \tfrac{1}{2}\,\bigl(f_a -2\,Z_a,\, f_b{}^{b}{}_I,\, f_b{}^{ba} \bigr)\,,&\qquad
 \vartheta_{-A} &= -\tfrac{1}{2}\,\bigl(f_{a-}{}^+,\, 0,\, 0 \bigr)\,,
\end{alignedat}
\end{split}
\end{align}
where $F_{AB}{}^C$ is the same as \eqref{eq:F-comp}. 
In our case, it may be more convenient to redefine $\xi_{\hA}$ as $\xi_{\hA}\to \frac{1}{2}\,\xi_{\hA}-\vartheta_{\hA}$\,. 
This yields
\begin{align}
\begin{split}
 X_{\hA\hB}{}^{\hC} &= \delta_{\bbb}^{\bbc}\,F_{\bba AB}{}^C
 - \bigl(\delta_A^C\,\delta_{\bbb}^{\bbc}\,\xi_{\bba B}
 -\delta_B^C\,\delta_{\bba}^{\bbc}\,\xi_{\bbb A}
 -\delta_{\bbb}^{\bbc}\,\eta_{AB}\, \xi_{\bba}^C
 +\epsilon_{\bba\bbb}\,\delta_B^C\,\xi_{\bbd A}\,\epsilon^{\bbd\bbc}\bigr)
\\
 &\quad + 2\,\bigl(\delta_{\bba}^{\bbc}\,\vartheta_{\bbb A}
 - \delta_{\bbb}^{\bbc}\,\vartheta_{\bba A}\bigr)\,\delta_B^C \,,
\end{split}
\label{eq:X-4d}
\end{align}
where
\begin{align}
\begin{split}
\begin{alignedat}{2}
 F_{+AB}{}^C&= F_{AB}{}^C\,,&\qquad 
 F_{-AB}{}^C&=0\,,
\\
 \xi_{+A} &= \bigl(Z_a,\, 0,\, 0\bigr)\,,&\qquad
 \xi_{-A} &= \bigl(f_{a-}{}^+,\, 0,\, 0 \bigr)\,,
\\
 \vartheta_{+A} &= \tfrac{1}{2}\,\bigl(f_a -2\,Z_a,\, f_b{}^{b}{}_I,\, f_b{}^{ba} \bigr)\,,&\qquad
 \vartheta_{-A} &= -\tfrac{1}{2}\,\bigl(f_{a-}{}^+,\, 0,\, 0 \bigr)\,.
\end{alignedat}
\end{split}
\end{align}
This neatly summarizes the structure constants of the half-maximal ExDA in $d=4$. 

In the literature, antisymmetric tensors
\begin{align}
 \Omega_{\hA\hB} \equiv \epsilon_{\bba\bbb}\,\eta_{AB}\,,\qquad
 \Omega^{\hA\hB} \equiv -\epsilon_{\bba\bbb}\,\eta_{AB}\,,
\end{align}
are used to raise or lower the indices $\hA,\hB$\,. 
For example, if we consider $(t_{\bm{a}})_{\hA\hB} \equiv (t_{\bm{a}})_{\hA}{}^{\hC}\,\Omega_{\hC\hB}$ we can check that this satisfies $(t_{\bm{a}})_{\hA\hB}=(t_{\bm{a}})_{\hB\hA}$\,. 
Using this index convention, we find that the so-called intertwining tensor $Z_{\hC\hA\hB}\equiv X_{(\hA\hB)\hC}=X_{(\hA\hB)}{}^{\hD}\,\Omega_{\hD\hC}$ takes the form,
\begin{align}
 Z_{\hC\hA\hB} = - \tfrac{1}{2}\,(t_{\bm{a}})_{\hA\hB}\,\bigl[ \Theta_{\hC}{}^{\bm{a}} + 2\,(t^{\bm{a}})_{\hC}{}^{\hD}\,\vartheta_{\hD} \bigr] \,.
\label{eq:Z-4d}
\end{align}
This can be compared with the intertwining tensor in $d=4$ maximal supergravity \cite{0809.5180} because the half-maximal supergravity can be realized as a $\mathbb{Z}_2$ truncation of the maximal supergravity (see section \ref{sec:ExDA-EDA}). 
Our generators $t^{\bm{a}}\,t_{\bm{a}}$ corresponds to their $-12\,t^{\bm{a}}\,t_{\bm{a}}$ and our $\Theta_{\hA}{}^{\bm{a}}\,t_{\bm{a}}$ corresponds to their $\hTheta_{\hA}{}^{\bm{a}}\,t_{\bm{a}}\equiv \bigl[\Theta_{\hA}{}^{\bm{a}} + 8\,(t^{a})_{\hA}{}^{\hB}\,\vartheta_{\hB}\bigr]\,t_{\bm{a}}$\,. 
Then our intertwinig tensor \eqref{eq:Z-4d} can be expressed as
\begin{align}
 Z_{\hC\hA\hB} = - \tfrac{1}{2}\,(t_{\bm{a}})_{\hA\hB}\,\bigl[ \Theta_{\hC}{}^{\bm{a}} -16\,(t^{\bm{a}})_{\hC}{}^{\hD}\,\vartheta_{\hD} \bigr] \,,
\end{align}
in thir notation. 
This expression matches with Eq.~(3.34) of \cite{0809.5180} and gives a non-trivial consistency check of our computation. 

\subsection{Leibniz identities}
\label{sec:Leibniz}

Since the D-bracket satisfies the Leibniz identities, the ExDA should satisfy
\begin{align}
 L(X,\,Y,\,Z) \equiv X\circ(Y\circ Z) - (X\circ Y)\circ Z - Y\circ (X\circ Z) = 0\,.
\end{align}
for arbitrary generators $X,Y,Z$ of the half-maximal ExDA. 
We here find the full set of identities by substituting the generators $T_{\hA}$ into $X$, $Y$, and $Z$\,. 

\subsubsection{Leibniz identities in $d\geq 5$}

By brute force computation, we identify the whole set of the Leibniz identities in $d\geq 6$ as
\begin{align}
\begin{split}
 &f_{ab}{}^c\,Z_c = 0\,,\qquad
  f_{[ab}{}^e\,f_{c]e}{}^d = 0\,,\qquad
  f_{ab}{}^c\,f_c{}^{IJ} + 2\,f_a{}^{K[I}\, f_b{}^{J]}{}_K = 0 \,,\qquad
  f_a{}^b{}_I\,Z_b =0\,, 
\\
 &f_a{}^b{}_{[I|}\,f_{b|JK]} = 0\,,\qquad
  2\,f_{[a}{}^{cJ} \,f_{b]IJ}
 +2\,f_{[a|}{}^d{}_I\,f_{d|b]}{}^c
 -f_{ab}{}^d\,f_d{}^c{}_I
 +2\,f_{[a|}{}^c{}_I\,Z_{|b]}=0\,,
\\
 &f_a{}^{bc}\,Z_c=0\,, \qquad
  4\,f_{[a}{}^{e[c}\,f_{b]e}{}^{d]}
  -f_{ab}{}^e\,f_e{}^{cd}
  +2\,f_a{}^{[c}{}_I\,f_b{}^{d]I}
  +4\,f_{[a}{}^{cd}\,Z_{b]}=0\,,
\\
 &f_a{}^{bc}\, f_{cIJ} + 2\,f_a{}^c{}_{[I|}\, f_c{}^b{}_{|J]} =0\,,\qquad 
  f_a{}^d{}_I\,f_d{}^{bc}+2\,f_a{}^{d[b}\,f_d{}^{c]}{}_I=0\,, 
\qquad 
  f_e{}^{[ab}\,f_d{}^{c]e}=0\,.
\end{split}
\label{eq:LI-6d-ExDA}
\end{align}
In $d=5$, the Leibniz identities involving the generator $T_*$ additionally give\footnote{The third identity can be also expressed as $f_c\,f_a{}^{cb} = f_a{}^{bI}\,f_c{}^c{}_I + f_{ca}{}^b\,f_d{}^{dc} + 2\,Z_a\,f_{c}{}^{cb}$\,.}
\begin{align}
\begin{split}
 &f_{ab}{}^c\,f_c = 0\,,\qquad 
 f_a{}^{b}{}_I\,f_b = f_{aIJ}\,f_b{}^{bJ} + Z_a\,f_b{}^{b}{}_I\,,
\\
 &f_a{}^{cd}\,f_{cd}{}^b = f_a{}^{bc}\,\bigl(f_c - f_{cd}{}^d\bigr) + f_a{}^c{}_I\,f_c{}^{bI}\,,\qquad
 f_b{}^a{}_I\,f_c{}^{cb} = 0\,.
\end{split}
\label{eq:LI-5d}
\end{align}

\subsubsection{Leibniz identities in $d=4$}
\label{sec:Leibniz-4d}

In $d=4$\,, the identification of the whole set of Leibniz identities is complicated.
However, we find that a combination
\begin{align}
 L(T_{-a},\,T^{+b},\,T_{+b}) + L(T_{-a},\,T_{+b},\,T^{+b}) = 0\,,
\end{align}
is equivalent to
\begin{align}
 f_{a-}{}^+\,(f_b-Z_b) = 0\,,
\label{eq:LI0}
\end{align}
and this shows that there are two branches: $f_{a-}{}^+=0$ and $f_a=Z_a$\,. 

\medskip

\noindent
\underline{\textbf{$\bullet$ The case $f_{a-}{}^+=0$}}

\noindent
When $f_{a-}{}^+=0$, the whole set of Leibniz identities are exactly the same as those in $d=5$:
\begin{align}
\begin{split}
 &f_{ab}{}^c\,f_c = 0\,,
\qquad
  f_{ab}{}^c\,Z_c = 0\,,
\qquad
  f_{[ab}{}^e\,f_{c]e}{}^d = 0\,,
\qquad
  f_{ab}{}^c\,f_c{}^{IJ} + 2\,f_a{}^{K[I}\, f_b{}^{J]}{}_K = 0 \,,
\\
 &f_a{}^b{}_I\,Z_b =0\,, 
\qquad
  f_a{}^b{}_I\,f_b = f_{aIJ}\,f_b{}^{bJ} + Z_a\, f_b{}^b{}_I \,,
\qquad
  f_a{}^b{}_{[I|}\,f_{b|JK]} = 0\,,
\\
 &2\,f_{[a}{}^{cJ} \,f_{b]IJ}
 +2\,f_{[a|}{}^d{}_I\,f_{d|b]}{}^c
 -f_{ab}{}^d\,f_d{}^c{}_I
 +2\,f_{[a|}{}^c{}_I\,Z_{|b]}=0\,,
\\
 &f_a{}^{bc}\,Z_c=0\,,
\qquad
  4\,f_{[a}{}^{e[c}\,f_{b]e}{}^{d]}
  -f_{ab}{}^e\,f_e{}^{cd}
  +2\,f_a{}^{[c}{}_I\,f_b{}^{d]I}
  +4\,f_{[a}{}^{cd}\,Z_{b]}=0\,,
\\
 &f_a{}^{cd}\,f_{cd}{}^b = f_a{}^{bc}\,\bigl(f_c - f_{cd}{}^d\bigr) + f_a{}^c{}_I\,f_c{}^{bI}\,, 
\qquad
  f_a{}^{bc}\, f_{cIJ} + 2\,f_a{}^c{}_{[I|}\, f_c{}^b{}_{|J]} =0\,,
\\
 &f_a{}^d{}_I\,f_d{}^{bc}+2\,f_a{}^{d[b}\,f_d{}^{c]}{}_I=0\,, 
\qquad
  f_b{}^a{}_I\,f_c{}^{cb} = 0\,,
\qquad
  f_e{}^{[ab}\,f_d{}^{c]e}=0\,.
\end{split}
\label{eq:LI-4d}
\end{align}

\medskip

\noindent
\underline{\textbf{$\bullet$ The case $f_{a-}{}^+\neq 0$}}

\noindent
When $f_{a-}{}^+\neq 0$\,, the Leibniz identities require very strong constraints on the other structure constants in a non-trivial manner. 
Since it is not easy to identify the identities for general $f_{a-}{}^+$\,, let us perform a $\GL(D)$ redefinition of generators such that $f_{a-}{}^+$ becomes $f_{a-}{}^+ = \delta_a^1$\,. 
In this case, we find that the general solution of the Leibniz identities is
\begin{align}
 f_{ab}{}^c = 2\,Z_{[a}\,\delta_{b]}^c \,,\quad 
 f_a{}^{bc} = f_a{}^b{}_I = f_a{}^I{}_J =0\,,\quad
 f_a = Z_a \,.
\label{eq:f-+nonzero}
\end{align}
They are independent of the particular direction $a=1$\,, and as one may naturally expect, if Eq.~\eqref{eq:f-+nonzero} is satisfied, the Leibniz identities are always satisfied for any $f_{a-}{}^+$\,. 
We thus conclude that the most general half-maximal ExDA in the branch $f_{a-}{}^+\neq 0$ is given by Eq.~\eqref{eq:f-+nonzero} with $f_{a-}{}^+$ arbitrary. 

Using the general expression \eqref{eq:X-4d}, we can express the structure constants as
\begin{align}
 X_{\hA\hB}{}^{\hC} = 
 \delta_{\bba}^{\bbc}\,\delta_B^C\,\xi_{\bbb A}
 - \delta_{\bbb}^{\bbc}\,\delta_A^C\,\xi_{\bba B}
 + \eta_{AB}\,\delta_{\bbb}^{\bbc}\, \xi_{\bba}^C\ \,,
\end{align}
where $\xi_{\bba A} \,\bigl(= -2\,\vartheta_{\bba A}\bigr)$ takes the form
\begin{align}
 \xi_{+A} = \bigl(Z_a,\, 0,\, 0\bigr) \,, \qquad
 \xi_{-A} = \bigl(f_{a-}{}^+,\, 0,\, 0 \bigr) \,,
\end{align}
and $Z_a$ and $f_{a-}{}^+$ can take arbitrary values. 

This half-maximal ExDA contains a subalgebra generated by $\{T_a\equiv T_{+a},\,T^a\equiv T_+{}^a\}$ which is independent of the structure constants $f_{a-}{}^+$\,,
\begin{align}
 T_{a}\circ T_{b} = Z_a\,T_b - Z_b\, T_a \,,\quad
 T_{a}\circ T^b = Z_a\,T^b + Z_c\,\delta_a^b\,T^c \,,\quad
 T^a\circ T_{b} = T^a\circ T^b = 0 \,.
\end{align}
When $Z_a\neq 0$\,, we can choose a particular basis $Z_a=-\delta_a^1$\,. 
For example, for $D=3$\,, this Leibniz algebra DD$^+$ corresponds to the Jacobi--Lie bialgebra $(\{5,-2\,T^1\}|\{1,0\})$ of \cite{1407.4236}.

\subsection{Coboundary ExDA}
\label{sec:coboundary-ExDA}

Here we explain that some of the Leibniz identities can be understood as the cocycle condition. 
Then, following the general discussion of \cite{2009.04454}, we find the explicit form of the coboundary ansatz which automatically solves the cocycle condition. 

Let us decompose the embedding tensor as
\begin{align}
 X_{\hA}=\Theta_{\hA}{}^{\bm{a}}\,t_{\bm{a}}+\vartheta_{\hA}\,t_{0}\,.
\end{align}
In $d\geq 6$ (where $\hA=A$), the generator $R_*$ contained in $\{t_{\bm{a}}\}$ is proportional to $t_{0}$\,, and this expression is to be understood under the identification
\begin{align}
 \Theta_{A}{}^{*}= 0\,,\qquad \vartheta_{A} = -\xi_A \,.
\end{align}
We then introduce a grading, called the level, to each generator $t_{\bm{a}}$ of the duality group $\cG$ as
\begin{align}
\begin{tabular}{|c||c|c|c|c|c|}\hline
 level $l_{\bm{a}}$ & $-2$ & $-1$ & $0$ & $1$ & $2$
\\\hline
 $t_{\bm{a}}$ & $R_{a_1a_2}$ & $R_a^I$ & $R_*,\,R^{\bba}{}_{\bbb},\,K^a{}_b,\,R_{IJ}$ & $R^a_I$ & $R^{a_1a_2}$
\\\hline
\end{tabular}\quad.
\end{align}
The level can be also defined by $[K,\,t_{\bm{a}}]=l_{\bm{a}}\,t_{\bm{a}}$ and the commutator of a level-$p$ generator and a level-$q$ generator has level $p+q$\,. 
In particular, when $\abs{p+q}>2$\,, the level-$p$ and level-$q$ generators commute with each other. 
The level-0 generators form a subalgebra and all of the generators transform under some representations of the subalgebra. 
Using this level, we decompose the embedding tensor $\Theta_{a}$ (which means $\Theta_{+a}$ in $d=4$) into two parts
\begin{align}
 \Theta_{a} = \Theta^{(0)}_{a} + \tTheta_{a}\,.
\end{align}
Here $\Theta^{(0)}_{a}$ contains the level-0 generators while $\tTheta_{a}$ contains the negative-level generators. 
More explicitly, we have
\begin{align}
 \Theta^{(0)}_{a} &= \bigl(f_{ab}{}^c-Z_a\,\delta_b^c\bigr)\,K^b{}_c 
 + \tfrac{1}{2!}\, f_a{}^{IJ}\,R_{IJ} + \begin{cases}
  0 & (d\geq 6)
\\
 f_a\, R_* & (d=5)
\\
 f_a\, R^{+}{}_{+} + f_{a-}{}^{+}\,R^{-}{}_{+} & (d=4)
 \end{cases}\,,
\\
 \tTheta_{a} &= \tfrac{1}{2!}\, f_a{}^{bc}\,R_{bc} + f_a{}^b{}_I\,R_b{}^I \,.
\label{eq:tTheta}
\end{align}
By considering the level, the Leibniz identity $[X_{\hA},\,X_{\hB}]=-X_{\hA\hB}{}^{\hC}\,X_{\hC}$ for $\hA=a$ and $\hB=b$ can be decomposed into
\begin{align}
 0&=[\Theta^{(0)}_{a},\,\Theta^{(0)}_{b}] + f_{ab}{}^c\,\Theta^{(0)}_{c}\,,
\label{eq:Theta0-alg}
\\
 0&=f_{ab}{}^c\,\vartheta_c \,,
\label{eq:theta0-alg}
\\
 0&=[\Theta^{(0)}_{a},\,\tTheta_{b}]-[\Theta^{(0)}_{b},\,\tTheta_{a}] + f_{ab}{}^c\,\tTheta_{c} + [\tTheta_{a},\,\tTheta_{b}] \,.
\label{eq:Theta-hat-alg}
\end{align}
The first two relations \eqref{eq:Theta0-alg} and \eqref{eq:theta0-alg} are equivalent to
\begin{align}
 &f_{[ab}{}^e\,f_{c]d}{}^f = 0\,,\qquad
  f_{ab}{}^c\,f_c{}^{IJ} + 2\,f_a{}^{K[I}\, f_b{}^{J]}{}_K = 0\,,\qquad 
  f_{ab}{}^c\,Z_c = 0 \quad (d\geq 4)\,,
\\
 &f_{ab}{}^c\,f_c = 0\quad (d=4,5)\,,\qquad
  2\,f_{[a|-}{}^+\,f_{|b]} + f_{ab}{}^c\,f_{c-}{}^+ =0\quad (d=4)\,,
\end{align}
while the last relation \eqref{eq:Theta-hat-alg} correspond to
\begin{align}
\begin{split}
 &2\,f_{[a}{}^{cJ} \,f_{b]IJ}
 +2\,f_{[a|}{}^d{}_I\,f_{d|b]}{}^c
 -f_{ab}{}^d\,f_d{}^c{}_I
 +2\,f_{[a|}{}^c{}_I\,Z_{|b]}=0\,,
\\
 &4\,f_{[a}{}^{e[c}\,f_{b]e}{}^{d]}
  -f_{ab}{}^e\,f_e{}^{cd}
  +2\,f_a{}^{[c}{}_I\,f_b{}^{d]I}
  +4\,f_{[a}{}^{cd}\,Z_{b]}=0\,.
\end{split}
\label{eq:cocycle}
\end{align}

In order to clarify the structure of Eq.~\eqref{eq:Theta-hat-alg}, let us define an operation
\begin{align}
 x\cdot f \equiv \bigl[f,\,x^a\,\Theta^{(0)}_{a}\bigr]\,,
\end{align}
where $x\equiv x^a\,T_a \in \mathfrak{g}$ and $f$ is an element of the duality algebra $g$ ($\cG=\exp g$). 
Using the relation \eqref{eq:Theta0-alg}, we can show that this operation satisfies
\begin{align}
 x\cdot (y\cdot f) - y\cdot (x\cdot f) = [x,y]\cdot f \,.
\end{align}
Then, using the notation
\begin{align}
 f(x) \equiv x^a\,\tTheta_{a}\,,
\end{align}
we can express Eq.~\eqref{eq:Theta-hat-alg} as the cocycle condition
\begin{align}
 \delta_1 f(x,y) \equiv x \cdot f(y) - y \cdot f(x) - f([x,y]) - [f(x),\,f(y)] = 0\,.
\end{align}
The family of coboundary operators $\delta_n$ satisfying $\delta_{n+1}\,\delta_n=0$ can be systematically constructed (see \cite{2009.04454}), and in particular, the $\delta_0$ can be found as
\begin{align}
 \delta_0 \rr(x) \equiv x^a\,(\Exp{\text{ad}_\rr}-1)\,\Theta^{(0)}_{a}
 = [\rr,\,x^a\,\Theta^{(0)}_{a}] +\tfrac{1}{2!}\,[\rr,\,[\rr,\,x^a\,\Theta^{(0)}_{a}]] + \cdots \,,
\label{eq:delta-0}
\end{align}
for $\rr=\rr^{\bm{a}}\,t_{\bm{a}}\in g$\,. 

Now, to get a trivial solution to the cocycle condition, let us suppose that the 1-cocycle $f(x)$ is given by a coboundary ansatz
\begin{align}
 f(x) \equiv x^a\,\tTheta_{a} = \delta_0 \rr(x) \qquad \bigl(\Leftrightarrow\ \Theta_a = \Exp{\text{ad}_\rr} \Theta^{(0)}_{a} \bigr) \,.
\label{eq:cob-ansatz}
\end{align}
Since $\tTheta_{a}$ is a linear combination of the negative-level generators, this identification is possible only when $\rr$ has the form
\begin{align}
 \rr = \rr^a_I\,R_a^I + \tfrac{1}{2!}\,\rr^{ab}\,R_{ab}\,.
\end{align}
Using this $\rr$ and Eq.~\eqref{eq:tTheta}, we can identify the structure constants as
\begin{align}
 f_a{}^b{}_I &= f_{ac}{}^b\,\rr^c_I - f_{aI}{}^{J} \,\rr^b_J - Z_a\,\rr^b_I \,,
\label{eq:f-r-1}
\\
 f_a{}^{bc} &= 2\,\rr^{[b|d}\,f_{ad}{}^{|c]} 
 + \delta^{IJ} f_{ad}{}^{[b}\,\rr^{c]}_I\,\rr^d_J
 + f_a{}^{IJ}\, \rr^b_I\,\rr^c_J 
 - 2\,Z_a\,\rr^{bc} \,.
\label{eq:f-r-2}
\end{align}
When the structure constants $f_a{}^b{}_I$ and $f_a{}^{bc}$ have this form, we call the ExDA a coboundary ExDA. 
In the following, we denote the structure constants $X_{\hA\hB}{}^{\hC}$ as $\cX_{\hA\hB}{}^{\hC}$ when we stress that the ExDA is of coboundary type. 

\subsection{Classical Yang--Baxter equations}
\label{sec:g-CYBE}

The coboundary ansatz \eqref{eq:cob-ansatz} is sufficient for the cocycle condition \eqref{eq:cocycle}, but still, the whole set of the Leibniz identities are not ensured. 
In the case of the Drinfel'd double, the closure of the algebra further imply the homogeneous classical Yang--Baxter equations (CYBE) for $\rr^{ab}$\,.\footnote{To be more precise, we can relax the homogeneous CYBE to the modified CYBE for the closure, but here we do not consider such relaxations and consider only a sufficient condition for the closure.}

Let us denote the structure constants $X_{\hA\hB}{}^{\hC}$ with $f_a{}^b{}_I=f_a{}^{bc}=0$ as $\mathring{X}_{\hA\hB}{}^{\hC}$\,,
\begin{align}
 \mathring{X}_{\hA\hB}{}^{\hC} \equiv X_{\hA\hB}{}^{\hC}\rvert_{f_a{}^b{}_I=f_a{}^{bc}=0}\,,
\end{align}
which is supposed to satisfy the Leibniz identities
\begin{align}
 f_{ab}{}^c\,f_c = 0\,,
\quad
 f_{ab}{}^c\,Z_c = 0\,,
\quad
  f_{[ab}{}^e\,f_{c]e}{}^d = 0\,,
\quad
  f_{ab}{}^c\,f_c{}^{IJ} + 2\,f_a{}^{K[I}\, f_b{}^{J]}{}_K = 0 \,.
\end{align}
We then consider a constant duality rotation and define
\begin{align}
 \mathfrak{X}_{\hA\hB}{}^{\hC} &\equiv \cR_{\hA}{}^{\hD}\,\cR_{\hB}{}^{\hE}\,(\cR^{-1})_{\hF}{}^{\hC}\,\mathring{X}_{\hD\hE}{}^{\hF}\,,
\\
 \cR_{\hA}{}^{\hB} &\equiv \bigl(\Exp{\rr^a_I\, R^I_a}\Exp{\frac{1}{2}\,\rr^{ab}\,R_{ab}}\bigr)_{\hA}{}^{\hB} \ \in\ \cG\,.
\label{eq:cR-def}
\end{align}
This $\mathfrak{X}_{\hA\hB}{}^{\hC}$ obviously satisfies the Leibniz identities because $\mathring{X}_{\hA\hB}{}^{\hC}$ satisfies those. 
One can easily see that $\mathfrak{X}_{a\hB}{}^{\hC}$ coincides with $\cX_{a\hB}{}^{\hC}$ but the other components do not match and the algebra defined by $\mathfrak{X}_{\hA\hB}{}^{\hC}$ is not an ExDA. 
Now we require that the whole components coincide
\begin{align}
 \cX_{\hA\hB}{}^{\hC} = \mathfrak{X}_{\hA\hB}{}^{\hC}\,.
\label{eq:X-CYBE}
\end{align}
Then we get a coboundary ExDA $\cX_{\hA\hB}{}^{\hC}$ that automatically satisfies the Leibniz identities. 
By explicitly computing all of the components of \eqref{eq:X-CYBE} in $d\geq 6$\,, we find that \eqref{eq:X-CYBE} is equivalent to the following set of conditions for $\rr^a_I$ and $\rr^{ab}$\,:
\begin{align}
 &\rr^a_I\,Z_a = 0\,, \qquad
 \rr^{ab}\,Z_b = 0\,, \qquad
  f_{a[IJ}\,\rr^{a}_{K]} = 0\,,
\label{eq:YB-0}
\\
 &\rr^{b}_I\,\rr^c_J\,f_{bc}{}^a + \bigl(\rr^{ab} + \tfrac{1}{2}\,\delta^{KL}\,\rr^a_K\,\rr^b_L\bigr)\,f_{bIJ} =0\,,
\label{eq:YB-1}
\\
 &2\,\rr^{ac}\,f_{cd}{}^b\,\rr^d_I
 + \rr^{bc}\,f_{cI}{}^J\,\rr^a_J 
 - \tfrac{1}{4}\,f_{c}{}^{KL}\,\rr^c_I\,\rr^a_K\,\rr^b_L=0\,,
\label{eq:YB-2}
\\
 &3\,\rr^{[a|d|}\, \rr^{b|e|}\, f_{de}{}^{c]}
 -\tfrac{3}{4}\,\delta^{IJ}\,f_{de}{}^{[a} \,\rr^{b|d|}\,\rr^{c]}_I\,\rr^{e}_J = 0\,.
\label{eq:YB-3}
\end{align}
In $d=5$\,, there is an additional condition
\begin{align}
 \rr^a_I\,\bigl(f_{ba}{}^b + f_a\bigr) = f_{aI}{}^J\,\rr^a_J \,. 
\label{eq:YB-4}
\end{align}
In $d=4$\,, there are two branches, $f_{a-}{}^+= 0$ and $f_{a-}{}^+\neq 0$\,.
When $f_{a-}{}^+= 0$\,, we find
\begin{align}
 \bigl(\rr^{ab}+ \tfrac{1}{2}\,\delta^{IJ}\,\rr^a_I\,\rr^b_J\bigr)\,\bigl(f_{cb}{}^c + f_b\bigr) = \rr^{cd}\,f_{cd}{}^a \,,
\label{eq:YB-5}
\end{align}
in addition to Eqs.~\eqref{eq:YB-0}--\eqref{eq:YB-4}. 
When $f_{a-}{}^+\neq 0$\,, where the structure constants are given by Eq.~\eqref{eq:f-+nonzero}, the condition \eqref{eq:X-CYBE} is equivalent to
\begin{align}
 \rr^a_I\,Z_a = 0\,, \quad
 \rr^{ab}\,Z_b = 0\,, \quad
 \rr^a_I\,f_{a-}{}^+ = 0\,, \quad
 \rr^{ab}\,f_{b-}{}^+ = 0\,, 
\end{align}
which leads to $f_a{}^b{}_I = f_a{}^{bc} = 0$ as expected. 
The conditions \eqref{eq:YB-0}--\eqref{eq:YB-3} may be regarded as the homogeneous CYBE for $\rr^a_I$ and $\rr^{ab}$\,.
Indeed, they reduce to the standard homogeneous CYBE when $n=0$ and $Z_a=0$\,. 

\subsection{Relation to $E_{D+1(D+1)}$ EDA}
\label{sec:ExDA-EDA}

In a particular case $n=0$\,, the half-maximal supergravity can be reproduced from the maximal supergravity through a $\mathbb{Z}_2$ truncation \cite{1104.3587,1612.05230}. 
We here consider reproducing the half-maximal ExDA with $n=0$ through a truncation of the $E_{D+1(D+1)}$ EDA in the type IIB picture \cite{2009.04454}. 

The $E_{D+1(D+1)}$ EDA ($D\leq 6$) in the type IIB picture is generated by
\begin{align}
 T_{\cA} = \{T_{\sfa},\,T^{\sfa}_{\bm{\alpha}},\,\tfrac{T^{\sfa_1\sfa_2\sfa_3}}{\sqrt{3!}},\,\tfrac{T_{\bm{\alpha}}^{\sfa_1\cdots \sfa_5}}{\sqrt{5!}},\,T^{\sfa_1\cdots \sfa_6,\sfa}\}\,,
\end{align}
where $\sfa=1,\dotsc,D$ and $\bm{\alpha}=\bm{1},\bm{2}$\,. 
For convenience, let us show the explicit form of the EDA for $D\leq 4$ (the algebra for higher $D$ is given in section 6 of \cite{2009.04454}):
\begin{align}
\begin{split}
 T_{\sfa}\circ T_{\sfb} &=f_{\sfa\sfb}{}^{\sfc}\,T_{\sfc}\,,
\\
 T_{\sfa}\circ T^{\sfb}_\beta &= f_{\sfa}{}_{\beta}^{\sfc\sfb}\,T_{\sfc}
 + f_{\sfa\beta}{}^{\gamma}\,T_\gamma^{\sfb} - f_{\sfa\sfc}{}^{\sfb}\,T_\beta^{\sfc} 
 +2\,Z_{\sfa}\,T^{\sfb}_\beta \,,
\\
 T_{\sfa}\circ T^{\sfb_1\sfb_2\sfb_3} &=
 f_{\sfa}{}^{\sfc\sfb_1\sfb_2\sfb_3}\, T_{\sfc} + 3\,\epsilon^{\gamma\delta}\,f_{\sfa}{}_{\gamma}^{[\sfb_1\sfb_2}\, T_{\delta}^{\sfb_3]} - 3\,f_{\sfa\sfc}{}^{[\sfb_1}\, T^{\sfb_2\sfb_3]\sfc} 
 +4\,Z_{\sfa}\,T^{\sfb_1\sfb_2\sfb_3}\,,
\\
 T^{\sfa}_\alpha\circ T_{\sfb} &= 
 f_{\sfb}{}_{\alpha}^{\sfa\sfc} \, T_{\sfc} 
 + 2\,\delta^{\sfa}_{[\sfb}\,f_{\sfc]\alpha}{}^{\gamma}\, T_\gamma^{\sfc}
 + f_{\sfb\sfc}{}^{\sfa}\,T_\alpha^{\sfc} 
 +4\,Z_{\sfc}\,\delta^{[\sfa}_{\sfb}\,T^{\sfc]}_\alpha\,,
\\
 T^{\sfa}_\alpha\circ T^{\sfb}_\beta &= - f_{\sfc}{}_{\alpha}^{\sfa\sfb}\,T_\beta^{\sfc} 
 - f_{\sfc\alpha}{}^\gamma\,\epsilon_{\gamma\beta}\,T^{\sfc \sfa\sfb} 
 + \tfrac{1}{2}\,\epsilon_{\alpha\beta}\,f_{\sfc_1\sfc_2}{}^{\sfa}\,T^{\sfc_1\sfc_2\sfb}
 -2\,\epsilon_{\alpha\beta}\,Z_{\sfc}\,T^{\sfa\sfb\sfc} \,,
\\
 T^{\sfa}_\alpha\circ T^{\sfb_1\sfb_2\sfb_3}
 &= -3\,f_{\sfc}{}_{\alpha}^{\sfa[\sfb_1}\,T^{\sfb_2\sfb_3]\sfc} \,,
\\
 T^{\sfa_1\sfa_2\sfa_3} \circ T_{\sfb} 
 &= -f_{\sfb}{}^{\sfc \sfa_1\sfa_2\sfa_3}\,T_{\sfc} 
 - 6\,\epsilon^{\gamma\delta}\,f_{[\sfb|}{}_{\gamma}^{[\sfa_1\sfa_2}\,\delta_{|\sfc]}^{\sfa_3]}\,T_\delta^{\sfc}
\\
 &\quad + 3\,f_{\sfb\sfc}{}^{[\sfa_1}\, T^{\sfa_2\sfa_3]\sfc}
 + 3\, f_{\sfc_1\sfc_2}{}^{[\sfa_1}\,\delta_{\sfb}^{\sfa_2}\,T^{\sfa_3]\sfc_1\sfc_2}
 +16\,Z_{\sfc}\,\delta_{\sfb}^{[\sfa_1}\,T^{\sfa_2\sfa_3\sfc]} \,,
\\
 T^{\sfa_1\sfa_2\sfa_3} \circ T_\beta^{\sfb} 
 &= -f_{\sfc}{}^{\sfa_1\sfa_2\sfa_3 \sfb}\,T_\beta^{\sfc}
  + 3\, f_{\sfc}{}_{\beta}^{[\sfa_1\sfa_2}\,T^{\sfa_3] \sfb\sfc} \,,
\\
 T^{\sfa_1\sfa_2\sfa_3} \circ T^{\sfb_1\sfb_2\sfb_3} 
 &= -3\, f_{\sfc}{}^{\sfa_1\sfa_2\sfa_3 [\sfb_1}\, T^{\sfb_2\sfb_3] \sfc} \,.
\end{split}
\label{eq:IIB-EDA}
\end{align}
For a general case $D\leq 6$\,, we introduce a $\mathbb{Z}_2$ action
\begin{align}
\begin{split}
 \{T_{\sfa},\,T_{\bm{1}}^{\sfa},\,T_{\bm{2}}^{\sfa_1\cdots \sfa_5},\,T^{1\cdots 6,\sfa}\} 
 &\to +\{T_{\sfa},\,T_{\bm{1}}^{\sfa},\,T_{\bm{2}}^{\sfa_1\cdots \sfa_5},\,T^{\sfa_1\cdots \sfa_6,\sfa}\}\,,
\\
 \{T_{\bm{2}}^{\sfa},\,T^{\sfa_1\sfa_2\sfa_3},\,T_{\bm{1}}^{\sfa_1\cdots \sfa_5}\} 
 &\to -\{T_{\bm{2}}^{\sfa},\,T^{\sfa_1\sfa_2\sfa_3},\,T_{\bm{1}}^{\sfa_1\cdots \sfa_5}\}\,,
\end{split}
\label{eq:Z2-action}
\end{align}
which is an element of $E_{D+1(D+1)}$\,, and truncate the $\mathbb{Z}_2$-odd generators. 
Under this $\mathbb{Z}_2$ action, the $E_{D+1(D+1)}$ generators
\begin{align}
 \bigl\{R^{\bm{1}}_{\sfa_1\cdots \sfa_6},\,\tfrac{R_{\sfa_1\cdots \sfa_4}}{\sqrt{4!}},\, \tfrac{R^{\bm{2}}_{\sfa_1\sfa_2}}{\sqrt{2!}},\,R^{\bm{1}}{}_{\bm{2}},\,R^{\bm{2}}{}_{\bm{1}},\, \tfrac{R_{\bm{2}}^{\sfa_1\sfa_2}}{\sqrt{2!}},\,\tfrac{R^{\sfa_1\cdots \sfa_4}}{\sqrt{4!}},\,R_{\bm{1}}^{\sfa_1\cdots \sfa_6}\bigr\}\,,
\end{align}
have odd parity while the other $\mathbb{Z}_2$-even generators
\begin{align}
 \bigl\{R^{\bm{2}}_{\sfa_1\cdots \sfa_6},\, \tfrac{R^{\bm{1}}_{\sfa_1\sfa_2}}{\sqrt{2!}},\,R^{\bm{1}}{}_{\bm{1}},\,K^{\sfa}{}_{\sfb},\, \tfrac{R_{\bm{1}}^{\sfa_1\sfa_2}}{\sqrt{2!}},\,R_{\bm{2}}^{\sfa_1\cdots \sfa_6}\}\,,
\end{align}
form a subgroup that coincides with the duality group $\cG$ of the half-maximal theory with $n=0$.
In $d\geq 4$, the relation between the $\OO(D,D)$ generators $\{\frac{R_{ab}}{\sqrt{2!}},\,K^a{}_b,\,\frac{R^{ab}}{\sqrt{2!}}\}$ and the $\mathbb{Z}_2$-even generators can be identified as
\begin{align}
 K^a{}_b\equiv K^{\sfa}{}_{\sfb} - \delta^{\sfa}_{\sfb}\,\bigl(\tfrac{1}{8}\,K^{\sfc}{}_{\sfc} + \tfrac{1}{2}\,R^{1}{}_{1}\bigr) \,, \quad
 R^{ab} \equiv -R_{\bm{1}}^{\sfa\sfb}\,,\quad 
 R_{ab} \equiv -R^{\bm{1}}_{\sfa\sfb} \,.
\label{eq:cG-Enn-1}
\end{align}
In $d=4$\,, the $\SL(2)$ generators $R^{\bba}{}_{\bbb}$ can be identified as
\begin{align}
 R^{+}{}_{+} \equiv \tfrac{1}{8}\,K^{\sfa}{}_{\sfa} + \tfrac{1}{2}\,R^{1}{}_{1} \,,\quad 
 R^{+}{}_{-} \equiv - R_{\bm{2}}^{1\cdots 6}\,,\quad 
 R^{-}{}_{+} \equiv R^{\bm{2}}_{1\cdots 6}\,.
\label{eq:cG-Enn-2}
\end{align}
In $d\geq 5$\,, the generators $R^{\bm{2}}_{1\cdots 6}$ and $R_{\bm{2}}^{\sfa_1\cdots \sfa_6}$ vanish and we identify the $\mathbb{R}^+_d$ generator as
\begin{align}
 R_* \equiv \tfrac{1}{8}\,K^{\sfa}{}_{\sfa} + \tfrac{1}{2}\,R^{1}{}_{1} \,.
\label{eq:cG-Enn-3}
\end{align}

Now we turn off the structure constants associated with $\mathbb{Z}_2$-odd generators,
\begin{align}
 f_{\sfa}{}_2^{\sfb_1\sfb_2}=f_{\sfa}{}_{\bm{1}}^{\sfb_1\cdots\sfb_6}=f_{\sfa}{}_{\bm{1}}{}^{\bm{2}}=f_{\sfa}{}_{\bm{2}}{}^{\bm{1}}=0\,.
\label{eq:f-Z2-truncation}
\end{align}
Then the embedding tensors $X_{\cA}$ of the EDA associated with the $\mathbb{Z}_2$-even generators $T_{\hA}$ are
\begin{align}
\begin{split}
 X_{\sfa} &= f_{\sfa\sfb}{}^{\sfc}\,\tilde{K}^{\sfb}{}_{\sfc}
 + 2\,f_{\sfa \bm{1}}{}^{\bm{1}} \,R^{\bm{1}}{}_{\bm{1}}
 + \tfrac{1}{2!}\,f_{\sfa}{}_{1}^{\sfb_1\sfb_2}\,R^{\bm{1}}_{\sfb_1\sfb_2} 
 + f_{\sfa}{}_{2}^{1\cdots 6}\,R^{\bm{2}}_{1\cdots 6}
 -Z_{\sfa}\,(\tilde{K}^{\sfb}{}_{\sfb}+t_0)\,, 
\\
 X^{\sfa}_{\bm{1}} &= -f_{\sfb}{}_{\bm{1}}^{\sfc\sfa}\,\tilde{K}^{\sfb}{}_{\sfc}
 -\tfrac{1}{2!}\,\bigl[f_{\sfb_1\sfb_2}{}^{\sfa} + 2\,\delta^{\sfa}_{[\sfb_1}\,(f_{\sfb_2]\bm{1}}{}^{\bm{1}} +2\,Z_{\sfb_2]})\bigr]\,R_{\bm{1}}^{\sfb_1\sfb_2} \,,
\\
 X_{\bm{2}}^{\sfa_1\cdots \sfa_5} &= -f_{\sfb}{}_{\bm{2}}^{\sfc\sfa_1\cdots \sfa_5}\,\tilde{K}^{\sfb}{}_{\sfc}
 + \bigl(f_{\sfb\sfc}{}^{\sfc} + f_{\sfb\bm{1}}{}^{\bm{1}} - 6\,Z_{\sfb}\bigr)\, R_{\bm{2}}^{\sfa_1\cdots\sfa_5\sfb}\,, 
\\
 X^{\sfa_1\cdots \sfa_6,\sfa} &= -f_{\sfb}{}_{\bm{2}}^{\sfa_1\cdots \sfa_6}\,R_{\bm{1}}^{\sfa\sfb} 
 - f_{\sfb}{}_{\bm{1}}^{\sfa\sfb}\,R^{\sfa_1\cdots \sfa_6}_{\bm{2}} \,,
\end{split}
\label{eq:EDA-X-even}
\end{align}
while those associated with the $\mathbb{Z}_2$-odd generators are
\begin{align}
\begin{split}
 X^{\sfa}_{\bm{2}} &= 
 -\tfrac{1}{2!}\,\bigl[f_{\sfb_1\sfb_2}{}^{\sfa} - 2\,\delta^{\sfa}_{[\sfb_1}\,(f_{\sfb_2]\bm{1}}{}^{\bm{1}}-2\,Z_{\sfb_2]}) \bigr]\,R_{\bm{2}}^{\sfb_1\sfb_2} \,,
\\
 X^{\sfa_1\sfa_2\sfa_3} &= -3\, f_{\sfb}{}_{\bm{1}}^{[\sfa_1\sfa_2}\, R_{\bm{2}}^{\sfa_3]\sfb}
 -\tfrac{3}{2}\,f_{\sfb_1\sfb_2}{}^{[\sfa_1}\, R^{\sfa_2\sfa_3]\sfb_1\sfb_2}
 -4\,Z_{\sfb}\,R^{\sfa_1\sfa_2\sfa_3\sfb}\,,
\\
 X_{\bm{1}}^{\sfa_1\cdots \sfa_5} &= 
 10\,f_{\sfb}{}_{\bm{1}}^{[\sfa_1\sfa_2}\, R^{\sfa_3\sfa_4\sfa_5]\sfb} 
 + (f_{\sfb\sfc}{}^{\sfc}- f_{\sfb \bm{1}}{}^{\bm{1}}-6\,Z_{\sfb})\,R_{\bm{1}}^{\sfa_1\cdots\sfa_5\sfb} \,.
\end{split}
\label{eq:EDA-X-odd}
\end{align}
Here we have defined $\tilde{K}^{\sfa}{}_{\sfb}\equiv K^{\sfa}{}_{\sfb}+\beta_d\,t_0$\,. 
By comparing Eq.~\eqref{eq:EDA-X-even} with Eqs.~\eqref{eq:ExDA-X-1}--\eqref{eq:ExDA-X-6} under the identification \eqref{eq:cG-Enn-1}--\eqref{eq:cG-Enn-3} and the identification of generators
\begin{align}
\begin{cases}
 T_{a} = T_{\sfa}\,,\quad
 T^{a} = T_{\bm{1}}^{\sfa} & (d\geq 6)
\\
 T_{a} = T_{\sfa}\,,\quad
 T^{a} = T_{\bm{1}}^{\sfa}\,,\quad
 T_* = T_{\bm{2}}^{1\cdots 5} & (d=5)
\\
 T_{+a} = T_{\sfa}\,,\quad
 T_{+}{}^{a} = T_{\bm{1}}^{\sfa}\,,\quad
 T_{-}{}_{a} = \tfrac{1}{5!}\, \epsilon_{\sfa\sfb_1\cdots \sfb_5}\,T_{\bm{2}}^{\sfb_1\cdots \sfb_5}\,,\quad
 T_{-}{}^{a} = T^{1\cdots 6,\sfa} \quad & (d=4)
\end{cases},
\label{eq:half-to-maximal}
\end{align}
we can identify the structure constants of the half-maximal ExDA (LHS) with those of the EDA (RHS) as
\begin{align}
\begin{split}
 &f_{ab}{}^c = f_{\mathsf{a}\mathsf{b}}{}^{\mathsf{c}}\,, \quad
 f_a{}^{bc} = -f_{\mathsf{a}}{}_{\bm{1}}^{\mathsf{b}\mathsf{c}}\,,\quad
 Z_a = Z_{\mathsf{a}} + \tfrac{1}{2}\, f_{\mathsf{a}\bm{1}}{}^{\bm{1}} \quad (d\geq 4)\,,
\\
 &f_a = f_{\mathsf{a}\mathsf{b}}{}^{\mathsf{b}} + \tfrac{8-D}{2}\,f_{\mathsf{a}\bm{1}}{}^{\bm{1}} - D\,Z_{\mathsf{a}}\quad (d=4,5)\,, \qquad
 f_{a-}{}^+ = f_{\mathsf{a}}{}_{\bm{2}}^{1\cdots 6} \quad (d=4)\,.
\end{split}
\end{align}

In this way, the half-maximal ExDA with $n=0$ can be obtained from the $E_{D+1(D+1)}$ EDA through the $\mathbb{Z}_2$ truncation. 
However, it is noted that, due to the truncation, the closure conditions (i.e., the Leibniz identities) can become weaker. 
In other words, not all of the half-maximal ExDA (with $n=0$) can be embedded into the $E_{D+1(D+1)}$ EDA through \eqref{eq:half-to-maximal}. 
Further details on this point are discussed in section \ref{sec:uplift-EDA}. 

\section{Generalized frame fields}
\label{sec:frame}

In this section, we construct the generalized frame fields $E_{\hA}{}^{\hM}$ which satisfy the algebra $[E_{\hA},\,E_{\hB}]_{\text{D}}=-X_{\hA\hB}{}^{\hC}\,E_{\hC}$\,. 
Before we get into the details, let us introduce several setups that are common to all dimensions.

We construct a group element $g=\Exp{x^a\,T_a}\in G$ and define the left-/right-invariant 1-form/vector fields as
\begin{align}
 \ell = \ell_m^a\,\rmd x^m\,T_a = g^{-1}\,\rmd g\,,\qquad
 r = r_m^a\,\rmd x^m\,T_a = \rmd g\,g^{-1}\,, \qquad
 v_a^m\,\ell_m^b=\delta_a^b=e_a^m\,r_m^b\,.
\end{align}
We then define a matrix $M_{\hA}{}^{\hB}(g)$ through
\begin{align}
 g^{-1}\,\triangleright T_{\hA} \equiv M_{\hA}{}^{\hB}(g)\,T_{\hB}\,,
\label{eq:M-def}
\end{align}
where the product $\triangleright$ is defined as
\begin{align}
 g\,\triangleright T_{\hA} \equiv \Exp{x^b\,T_b\circ} T_{\hA} 
 = T_{\hA} + x^b\,T_b\circ T_{\hA} + \tfrac{1}{2!}\, x^b\,T_b\circ \bigl(x^c\,T_c\circ T_{\hA}\bigr) + \cdots\,.
\end{align}
By its construction, the matrix $M_{\hA}{}^{\hB}(g)$ enjoys the following three properties \cite{2009.04454}:
\begin{alignat}{2}
 &M_{\hA}{}^{\hB}(g=1) =\delta_{\hA}^{\hB}\,,&&
\label{eq:property-1}
\\
 &X_{\hA\hB}{}^{\hC} = M_{\hA}{}^{\hD}\,M_{\hB}{}^{\hE}\,(M^{-1})_{\hF}{}^{\hC}\,X_{\hD\hE}{}^{\hF}&\quad &\text{(algebraic identity)}\,,
\label{eq:property-2}
\\
 &M_{\hA}{}^{\hB}(gh) =M_{\hA}{}^{\hC}(g)\,M_{\hC}{}^{\hB}(h)&\quad &\text{(multiplicativity)}\,,
\label{eq:property-3}
\end{alignat}
where the first one follows from the last one by choosing $g=h=1$\,. 
In Eq.~\eqref{eq:property-3}, by considering an infinitesimal left translation $g=1+\epsilon^a\,T_a$\,, we obtain the differential identity
\begin{align}
 M_{\hA}{}^{\hD}\,D_c(M^{-1})_{\hD}{}^{\hB} = X_{c\hA}{}^{\hB}\,,
\label{eq:diff-id}
\end{align}
where $D_a\equiv e_a^m\,\partial_m$\,.
Combining the algebraic and the differential identities, we also find
\begin{align}
 v_c^m\,\partial_m(M^{-1})_{\hA}{}^{\hD}\,M_{\hD}{}^{\hB} = X_{c\hA}{}^{\hB} \,.
\label{eq:right-translation}
\end{align}
In the following, we elucidate these relations in each dimension. 
Then, we construct the generalized frame fields in section \ref{sec:frame-fields}. 
In section \ref{sec:G-para}, we show that the generalized frame fields satisfy the relation $[E_{\hA},\,E_{\hB}]_{\text{D}}=-X_{\hA\hB}{}^{\hC}\,E_{\hC}$\,. 

\subsection{Generalized Poisson--Lie structures}
\label{sec:gPL}

\subsubsection{$d\geq 5$}

In $d\geq 5$\,, we can parameterize the matrix $M_{\hA}{}^{\hB}$ as
\begin{align}
 M_{\hA}{}^{\hB} &= \bm{\Pi}_{\hA}{}^{\hC}\,\bm{A}_{\hC}{}^{\hB}\,,\qquad 
 \bm{\Pi} \equiv \Exp{-\pi^a_I\,R^I_a}\Exp{-\frac{1}{2!}\,\pi^{ab}\,R_{ab}},
\\
 \bm{A} &\equiv \Exp{\Delta\,(K+t_0)} \Exp{-\lambda\,(R_* + \beta_d\,t_0)} \Exp{-\frac{1}{2!}\,\beta^{IJ}\,R_{IJ}} \Exp{-\alpha_a{}^b\,K^a{}_b}.
\end{align}
When $d\geq 6$\,, we have $R_* + \beta_d\,t_0=0$ and $\lambda$ does not appear. 
In $d=5$\,, their explicit matrix forms are given by
\begin{align}
 \bm{A}_{\hA}{}^{\hB} &\equiv \begin{pmatrix} a_a{}^b & 0 & 0 & 0 \\
 0 & \Exp{-\Delta} \omega_{I}{}^{J} & 0 & 0 \\
 0 & 0 & \Exp{-2\Delta} (a^{-1})_b{}^a & 0 \\
 0 & 0 & 0 & \Exp{\lambda - \Delta} 
\end{pmatrix},
\\
 \bm{\Pi}_{\hA}{}^{\hB} &\equiv \begin{pmatrix} \delta_a^b & 0 & 0 & 0 \\
 \pi_I^b & \delta_{I}^{J} & 0 & 0 \\
 - \bigl(\pi^{ab}+\tfrac{1}{2}\,\delta^{KL}\,\pi^a_K\,\pi^b_L\bigr) & - \pi^{a}_K\,\delta^{KJ} & \delta^a_b & 0 \\
 0 & 0 & 0 & 1
\end{pmatrix},
\end{align}
where $\omega_I{}^J\in \OO(n)$\,, i.e., $\delta_{KL}\,\omega_I{}^K\,\omega_J{}^L=\delta_{IJ}$\,. 
In $d\geq 6$\,, they can be obtained by truncating the last row/column. 

The property \eqref{eq:property-1} shows that, at the unit element $g=1$ (whose position is called $x_0$), we have
\begin{align}
 \Delta(x_0)= 
 \pi^a_I(x_0)= 
 \pi^{ab}(x_0)=0\overset{(d=5)}{=}\lambda(x_0) \,,\qquad
 a_a{}^b(x_0)=\delta_a^b\,,\qquad
 \omega_I{}^J(x_0)=\delta_I^J\,,
\label{eq:id-6D}
\end{align}
In $d\geq 6$\,, the algebraic identity \eqref{eq:property-2} is equivalent to
\begin{align}
 &a_a{}^d\,a_b{}^e\,f_{de}{}^c = f_{ab}{}^d\,a_d{}^c\,,\qquad
 a_a{}^b\,\omega_I{}^K\,\omega_J{}^L\,f_{bKL} = f_{aIJ}\,,\qquad
  a_a{}^b\,Z_b = Z_a\,,
\label{eq:Alg6D-1}
\\
 &\Exp{-\Delta} a_a{}^c\,(a^{-1})_d{}^b\,f_c{}^d{}_J\, \omega_I{}^J
 = f_{a}{}^b{}_I
  - f_{aIJ}\,\pi^{bJ}
  + f_{ac}{}^b\,\pi^c_I
  - Z_a\,\pi^b_I\,,
\label{eq:Alg6D-2}
\\
 &\Exp{-2\Delta} a_a{}^e\, (a^{-1})_f{}^b\,(a^{-1})_g{}^c\, f_e{}^{fg} 
\nn\\
 &\quad = f_a{}^{bc} 
  - 2\,f_{ad}{}^{[b}\, \pi^{c] d} 
  - 2\,Z_a\,\pi^{bc} 
  + 2\,f_a{}^{[b}{}_I \, \pi^{c]I}
  + f_{ad}{}^{[b}\,\pi^{c]I}\, \pi^d_I 
  + f_a{}^{IJ}\, \pi^b_I\,\pi^c_J \,,
\label{eq:Alg6D-3}
\\
 &f_{a[IJ}\, \pi^a_{K]} =0\,,\qquad
  Z_a\,\pi^a_I = 0\,, \qquad
  Z_b\,\pi^{ba} = 0\,,
\label{eq:Alg6D-4}
\\
 &f_{bc}{}^a\, \pi^b_I\, \pi^c_J
 = 2\,f_b{}^a{}_{[I}\, \pi^b_{J]} 
  + f_{bIJ}\,\pi^{ba}
  - \tfrac{1}{2}\, f_{bIJ}\,\pi^{aK}\, \pi^b_K\,,
\label{eq:Alg6D-5}
\\
 &f_c{}^{ab}\,\pi^c_I
  + 2\,f_{cd}{}^{[a}\, \pi^{b]c}\, \pi^d_I
  + 2\,f_c{}^{[a}{}_I\, \pi^{b]c} 
  + f_e{}^{[a}{}_J\, \pi^{b]J} \pi^e_I
  + f_{cIJ}\,\pi^{[a|J}\, \bigl(\pi^{c|b]} - \tfrac{1}{2}\, \pi^{b]K}\, \pi^c_K\bigr) = 0\,,
\label{eq:Alg6D-6}
\\
 &3\,f_d{}^{[ab}\,\bigl(\pi^{c]d} + \tfrac{1}{6}\,\pi^{c] I}\, \pi^d_I\bigr)
  + 3\, f_{de}{}^{[a}\, \pi^{b |d|}\, \bigl( \pi^{c]e} + \tfrac{1}{3}\, \pi^{c]}_I\, \pi^{eI}\bigr) 
  - 6\, \pi^{[ab}\, \pi^{c] d}\, Z_d 
\nn\\
 &\quad + 2\,f_d{}^{[a}{}_I\, \pi^{b|I|}\, \bigl(\pi^{c]d} + \tfrac{1}{4}\, \pi^{c]J}\, \pi^d_J \bigr)
  + \tfrac{1}{4}\, f_d{}^{IJ}\, \pi^{[a}_I\,\pi^b_J\, \bigl(\pi^{c]d} + \tfrac{1}{2}\, \pi^{c]K}\, \pi^d_K\bigr) = 0\,.
\label{eq:Alg6D-7}
\end{align}
In $d=5$\,, we additionally have
\begin{align}
 a_a{}^b\,f_b = f_a \,,\qquad
 f_a\,\pi^a_I = f_a{}^a{}_I - \Exp{-\Delta} \omega_I{}^J\,f_a{}^a{}_J \,.
\label{eq:Alg6D-8}
\end{align}
The relation \eqref{eq:property-3} is equivalent to
\begin{align}
 (a_{gh})_a{}^b &=(a_{g})_a{}^c\,(a_{h})_c{}^b\,,\quad
 (\omega_{gh})_I{}^J=(\omega_{g})_I{}^K\,(\omega_{h})_K{}^J\,, \quad
 \Delta_{gh} =\Delta_{g}+\Delta_{h}\,,
\\
 (\pi_{gh})^a_I &= (\pi_{g})^a_I + \Exp{-\Delta_{g}}\,(a_{g}^{-1})_b{}^a\,(\pi_{h})^b_J\,(\omega_{g})_I{}^J\,,
\\
 \pi^{ab}_{gh}
 &= \pi^{ab}_{g}
 + \Exp{-2 \Delta_{g}}
       (a_{g}^{-1})_c{}^a\,(a_{g}^{-1})_d{}^b\,\pi_{h}^{cd} 
 + \Exp{-\Delta_{g}} \omega_{g}^{IJ} \, (\pi_{g})^{[a}_I\,(a_{g}^{-1})_c{}^{b]} \,(\pi_{h})^c_J \,,
\\
 \lambda_{gh}&=\lambda_{g}+\lambda_{h}\quad (\text{only in }d=5)\,.
\end{align}
They may be interpreted as the multiplicativity (see, for example, \cite{1511.02491}). 
The differential identity \eqref{eq:diff-id} is equivalent to
\begin{align}
 D_a a_b{}^c &= - f_{ab}{}^d\,a_d{}^c\,, \quad
 D_a \omega_{I}{}^{J} = - f_{aI}{}^K\,\omega_{K}{}^{J}\,, \quad
 D_a\Delta = Z_a\,,
\\
 D_a\pi^b_I &= f_a{}^b{}_I + f_{ac}{}^b\,\pi^c_I - f_{aI}{}^J\,\pi^b_J - Z_a\,\pi^b_I\,,
\\
 D_a\pi^{bc} &= f_a{}^{bc} + 2\,f_{ad}{}^{[b}\,\pi^{|d|c]} + \delta^{IJ}\,f_a{}^{[b}{}_I\,\pi^{c]}_J - 2\,Z_a\,\pi^{bc}\,,
\\
 D_a\lambda &= f_a\quad (\text{only in }d=5)\,.
\end{align}
Moreover, if we define the (generalized) Poisson--Lie structures as
\begin{align}
 \pi^{mn} \equiv \Exp{2\Delta}\pi^{ab}\,e_a^m\,e_b^n\,,\qquad
 \pi^m_I \equiv \Exp{\Delta}\pi^a_J\,\omega^J{}_I\,e_a^m\,,
\label{eq:gPL-def}
\end{align}
the relation \eqref{eq:right-translation} shows nice properties under an infinitesimal right translation
\begin{align}
 \Lie_{v_a}a_b{}^c &= -a_b{}^d\,f_{ad}{}^b\,, \quad
 \Lie_{v_a} \omega_I{}^{J} = -\omega_{IK}\, f_a{}^{KJ} \,, \quad
 \Lie_{v_a}\Delta = Z_a\,,
\\
 \Lie_{v_a}\pi^m_I &= f_a{}^b{}_I\,v_b^m - f_a{}^J{}_I\,\pi^m_J + Z_a\,\pi^{m}_I\,,
\label{eq:pi-1-5}
\\
 \Lie_{v_a}\pi^{mn} &= f_a{}^{bc}\,v_b^m\,v_c^n + \delta^{IJ}\,f_a{}^b{}_I\,\pi_J^{[m}\,v_b^{n]} + 2\,Z_a\,\pi^{mn}\,,
\label{eq:pi-2-5}
\\
 \Lie_{v_a}\lambda &= f_a \quad (\text{only in }d=5)\,.
\end{align}

\subsubsection{$d=4$}

In $d=4$\,, we can parameterize the matrix $M_{\hA}{}^{\hB}$ as
\begin{align}
 M_{\hA}{}^{\hB} &= \bm{\Pi}_{\hA}{}^{\hC}\,\bm{A}_{\hC}{}^{\hB}\,,\qquad 
 \bm{\Pi} \equiv \Exp{-\pi^a_I\,R^I_a}\Exp{-\frac{1}{2!}\,\pi^{ab}\,R_{ab}},
\\
 \bm{A} &\equiv \Exp{\Delta\,(K+t_0)} \Exp{-\gamma\,R^-{}_+} \Exp{-\lambda\,(R^+{}_+ +\frac{1}{2}\,t_0)} \Exp{-\frac{1}{2!}\,\beta^{IJ}\,R_{IJ}} \Exp{-\alpha_a{}^b\,K^a{}_b},
\end{align}
namely,
\begin{align}
 \bm{A}_{\hA}{}^{\hB} &\equiv \Exp{\frac{\lambda}{2}}\begin{pmatrix} \lambda_{\bba}{}^{\bbb}\,a_a{}^b & 0 & 0 \\
 0 & \Exp{-\Delta}\lambda_{\bba}{}^{\bbb}\,\omega_{I}{}^{J} & 0 \\
 0 & 0 & \Exp{-2\Delta}\lambda_{\bba}{}^{\bbb}\,(a^{-1})_b{}^a 
\end{pmatrix},\quad \lambda_{\bba}{}^{\bbb} \equiv \begin{pmatrix} \Exp{-\frac{\lambda}{2}} & 0 \\ -\Exp{-\frac{\lambda}{2}}\gamma & \Exp{\frac{\lambda}{2}} \end{pmatrix},
\\
 \bm{\Pi}_{\hA}{}^{\hB} &\equiv \begin{pmatrix} \delta_{\bba}^{\bbb}\,\delta_a^b & 0 & 0 \\
 \delta_{\bba}^{\bbb}\,\pi_I^b & \delta_{\bba}^{\bbb}\,\delta_{I}^{J} & 0 \\
 -\delta_{\bba}^{\bbb}\,\bigl(\pi^{ab}+\tfrac{1}{2}\,\delta^{KL}\,\pi^a_K\,\pi^b_L\bigr) & -\delta_{\bba}^{\bbb}\,\pi^{a}_K\,\delta^{KJ} & \delta_{\bba}^{\bbb}\,\delta^a_b
\end{pmatrix},
\end{align}
where $\omega_I{}^J\in \OO(n)$\,.
At the unit element $g=1$, we have
\begin{align}
 \lambda(x_0)= 
 \gamma(x_0)= 
 \Delta(x_0)= 
 \pi^a_I(x_0)= 
 \pi^{ab}(x_0)=0\,,\quad
 a_a{}^b(x_0)=\delta_a^b\,,\quad
 \omega_I{}^J(x_0)=\delta_I^J\,.
\end{align}

The algebraic identity is complicated in $d=4$\,. 
If we define
\begin{align}
 A_{\hA,\hB}{}^{\hC} \equiv X_{\hA\hB}{}^{\hD}\,M_{\hD}{}^{\hC}
 - M_{\hA}{}^{\hD}\,M_{\hB}{}^{\hE}\,X_{\hD\hE}{}^{\hC} \,,
\end{align}
the algebraic identity $A_{+a,+b}{}^{+c}=0$ gives $a_a{}^d\,a_b{}^e\,f_{de}{}^c = f_{ab}{}^d\,a_d{}^c$ and $A_{+a,-b}{}^{+c}=0$ gives
\begin{align}
 a_a{}^b\,f_{b-}{}^+ = 
 \Exp{- \lambda} \bigl(f_{a-}{}^+ + \gamma\, f_a\bigr)\,.
\end{align}
Then $A_{-a,+b}{}^{+c}=0$ further gives
\begin{align}
 \gamma\,\bigl(f_{ab}{}^c - 2\,f_{[a}\,\delta_{b]}^c\bigr) = 0\,,
\end{align}
which shows that $\gamma$ can be non-zero only when $f_{ab}{}^c$ takes a special form $f_{ab}{}^c = 2\,f_{[a}\,\delta_{b]}^c$\,. 
This is consistent with the discussion in section \ref{sec:Leibniz-4d}. 
The field $\gamma$ is produced by the structure constant $f_{a-}{}^+$\,, and the non-vanishing $f_{a-}{}^+$ strongly restricts the other structure constants. 
If we choose the branch $f_{a-}{}^+ = 0$ (where $\gamma=0$), we can identify the whole set of the algebraic identities. 
We find that the algebraic identity \eqref{eq:property-2} is equivalent to Eqs.~\eqref{eq:Alg6D-1}--\eqref{eq:Alg6D-8} and
\begin{align}
 f_b\,\pi^{ba} 
 = 2\,f_{bc}{}^{[a}\,\pi^{c]b}
 + f_{b}{}^a{}_I\,\pi^{bI} 
 + \tfrac{1}{2}\, f_{bIJ}\, \pi^{aI}\, \pi^{bJ}\,.
\end{align}
If we instead choose the other branch $f_{a-}{}^+ \neq 0$\,, the algebraic identity is equivalent to
\begin{align}
 a_a{}^b\,Z_b = 0\,,\qquad 
 a_a{}^b\,f_{b-}{}^+ = \Exp{- \lambda} \bigl( f_{a-}{}^+ + \gamma\,f_a\bigr)\,.
\end{align}

The multiplicativity \eqref{eq:property-3} can be easily identified. 
In general, it is equivalent to
\begin{align}
 (a_{gh})_a{}^b &=(a_{g})_a{}^c\,(a_{h})_c{}^b\,,\quad
 (\omega_{gh})_I{}^J=(\omega_{g})_I{}^K\,(\omega_{h})_K{}^J\,, 
\\
 \lambda_{gh}&=\lambda_{g}+\lambda_{h}\,,\quad
 \Delta_{gh}=\Delta_{g}+\Delta_{h}\,,\quad
 \gamma_{gh}=\gamma_{g}+\Exp{\lambda_{g}}\gamma_{h}\,,
\\
 (\pi_{gh})^a_I &= (\pi_{g})^a_I + \Exp{-\Delta_{g}}\,(a_{g}^{-1})_b{}^a\,(\pi_{h})^b_J\,(\omega_{g})_I{}^J\,,
\\
 \pi^{ab}_{gh}
 &= \pi^{ab}_{g}
 + \Exp{-2 \Delta_{g}}
       (a_{g}^{-1})_c{}^a\,(a_{g}^{-1})_d{}^b\,\pi_{h}^{cd} 
 + \Exp{-\Delta_{g}} \omega_{g}^{IJ} \, (\pi_{g})^{[a}_I\,(a_{g}^{-1})_c{}^{b]} \,(\pi_{h})^c_J \,.
\end{align}
The differential identity \eqref{eq:diff-id} reads
\begin{align}
 D_a a_b{}^c &= - f_{ab}{}^d\,a_d{}^c\,, \quad
 D_a \omega_{I}{}^{J} = - f_{aI}{}^K\,\omega_{K}{}^{J}\,, 
\\
 D_a\gamma &=f_{a-}{}^+ +f_a\,\gamma\,,\quad
 D_a\lambda =f_a\,,\quad 
 D_a\Delta = Z_a\,,
\\
 D_a\pi^b_I &= f_a{}^b{}_I + f_{ac}{}^b\,\pi^c_I - f_{aI}{}^J\,\pi^b_J - Z_a\,\pi^b_I\,,
\\
 D_a\pi^{bc} &= f_a{}^{bc} + 2\,f_{ad}{}^{[b}\,\pi^{|d|c]} + \delta^{IJ}\,f_a{}^{[b}{}_I\,\pi^{c]}_J - 2\,Z_a\,\pi^{bc}\,.
\end{align}
If we define the (generalized) Poisson--Lie structures as Eq.~\eqref{eq:gPL-def}, Eq.~\eqref{eq:right-translation} gives
\begin{align}
 \Lie_{v_a}a_b{}^c &= -a_b{}^d\,f_{ad}{}^b\,, \quad
 \Lie_{v_a} \omega_I{}^{J} = -\omega_{IK}\, f_a{}^{KJ} \,, 
\\
 \Lie_{v_a}\gamma &= \Exp{\lambda} f_{a-}{}^+ \,,\quad
 \Lie_{v_a}\lambda =f_a\,,\quad 
 \Lie_{v_a}\Delta = Z_a\,,
\\
 \Lie_{v_a}\pi^m_I &= f_a{}^b{}_I\,v_b^m - f_a{}^J{}_I\,\pi^m_J + Z_a\,\pi^{m}_I\,,
\label{eq:pi-1}
\\
 \Lie_{v_a}\pi^{mn} &= f_a{}^{bc}\,v_b^m\,v_c^n + \delta^{IJ}\,f_a{}^b{}_I\,\pi_J^{[m}\,v_b^{n]} + 2\,Z_a\,\pi^{mn}\,.
\label{eq:pi-2}
\end{align}

\subsubsection{Poisson--Lie structure for coboundary ExDAs}

For a general half-maximal ExDA, in order to find the explicit form of $\pi^{m}_I$ and $\pi^{mn}$\,, we need to compute the adjoint-like action \eqref{eq:M-def}. 
However, when the ExDA is a coboundary ExDA, we can generally solve the differential equations \eqref{eq:pi-1-5} and \eqref{eq:pi-2-5} (or \eqref{eq:pi-1} and \eqref{eq:pi-2} in $d=4$).
The solutions are
\begin{align}
 \pi^{m}_I = \rr^a_I\, v_c^m-\Exp{\Delta}\rr^a_J\,\omega^J{}_I\,e_a^m \,,\qquad 
 \pi^{mn} = \rr^{ab}\,\bigl(v_a^m\,v_b^n - \Exp{2\Delta}e_a^m\,e_b^n\bigr)
 + \Exp{\Delta}\omega^{IJ}\,\rr^a_I\,\pi_J^{[m}\,e_a^{n]}\,.
\end{align}
By using $\Lie_{v_a}v_b^m=f_{ab}{}^c\,v_c^m$\,, $\Lie_{v_a}e_b^m=0$\,, and the differential identities, one can easily see that they indeed satisfy Eqs.~\eqref{eq:pi-1-5} and \eqref{eq:pi-2-5} (or \eqref{eq:pi-1} and \eqref{eq:pi-2} in $d=4$). 
They also satisfy $\pi^m_I(x_0)=0$ and $\pi^{mn}(x_0)=0$\,. 

\subsection{Generalized frame fields}
\label{sec:frame-fields}

The generalized frame fields $E_{\hA}{}^{\hM}$ are constructed as
\begin{align}
 E_{\hA}{}^{\hM} \equiv M_{\hA}{}^{\hB}\, V_{\hB}{}^{\hM}\,,
\label{eq:E-MV}
\end{align}
by using certain generalized frame fields $V_{\hA}{}^{\hM}$\,. 
Here, by choosing $V_{\hA}{}^{\hM}$ appropriately, we can show that $E_{\hA}{}^{\hM}$ satisfy the frame algebra
\begin{align}
 [E_{\hA},\,E_{\hB}]_{\text{D}}=-X_{\hA\hB}{}^{\hC}\,E_{\hC}\,.
\label{eq:E-ExDA}
\end{align}
The explicit form of the generalized frame fields $V_{\hA}{}^{\hM}$ and $E_{\hA}{}^{\hM}$ in each dimension is found in the following subsections. 

\subsubsection{$d\geq 5$}

In $d\geq 5$\,, we introduce a set of generalized vector fields as
\begin{align}
 V_{\hA}{}^{\hM} = \begin{pmatrix}
 v_a^m & 0 & 0 & 0 \\
 0 & \delta_I^{\cI} & 0 & 0 \\
 0 & 0 & \ell^a_m &0 \\
 0 & 0 & 0 & \Exp{-2\,d} \end{pmatrix},\qquad 
 \Exp{-2\,d}\equiv \abs{\det \ell_m^a}\,.
\end{align}
Here we note that the last row/column vanishes in $d\geq 6$\,. 
We also note that $\Exp{-2\,d}$ behaves as a scalar density, which is consistent with the comments below Eq.~\eqref{eq:V-5d}.
We can easily show that they satisfy the algebra
\begin{align}
 [V_{\hA},\,V_{\hB}]_{\text{D}} = \mathring{X}_{\hA\hB}{}^{\hC}\,V_{\hC}\,,
\label{eq:V-alg}
\end{align}
where $\mathring{X}_{\hA\hB}{}^{\hC}$ denotes the structure constants of the half-maximal ExDA $X_{\hA\hB}{}^{\hC}$ with only $f_{ab}{}^c$ and $f_a=f_{ab}{}^b$ (i.e., the other structure constants are truncated).
Then using Eq.~\eqref{eq:E-MV} and the parameterization of $M_{\hA}{}^{\hB}$\,, we find that the generalized frame fields are given by
\begin{align}
 E_{\hA}{}^{\hM} = \begin{pmatrix} e_a^m & 0 & 0 & 0 \\
 \pi_I^b\,e_b^m & \Exp{-\Delta} \omega_{I}{}^{\cI} & 0 & 0 \\
 - \bigl(\pi^{ab}+\tfrac{1}{2}\,\delta^{KL}\,\pi^a_K\,\pi^b_L\bigr)\,e_b^m & - \Exp{-\Delta} \pi^{a}_K\,\omega^{K\cI} & \Exp{-2\Delta} r^a_m & 0 \\
 0 & 0 & 0 & \Exp{\bar{\lambda} - \Delta} 
\end{pmatrix},
\end{align}
where we have defined $\bar{\lambda}\equiv \lambda-2\,d$ and used $e_a^m=a_a{}^b\,v_b^m$ and $r^a_m=(a^{-1})_b{}^a\,\ell^b_m$\,. 
If we consider the generalized vector fields $E_{\hA}{}^{\hM}\,\partial_{\hM}$\,, they can be decomposed as
\begin{align}
\begin{split}
 E_a &= e_a \,,
\\
 E_I &= \Exp{-\Delta} \omega_{I}{}^{\cI}\,\bigl(\pi^m_{\cI}\,\partial_m + \partial_{\cI}\bigr) \,,
\\
 E^a &= \Exp{-2\Delta} r^a_m\,\bigl[- \bigl(\pi^{mn}+\tfrac{1}{2}\,\delta^{\cK\cL}\,\pi^m_{\cK}\,\pi^n_{\cL}\bigr)\,\partial_n - \pi^{m\cI}\,\partial_{\cI} + \tilde{\partial}^m\bigr] \,,
\\
 E_* &= \Exp{\bar{\lambda} - \Delta} \,\partial_* \,,
\end{split}
\end{align}
where $\pi^m_{\cI}\equiv \delta_{\cI}^J\,\pi^m_J$\,. 

\subsubsection{$d=4$}

In $d=4$\,, we consider
\begin{align}
 V_{\hA}{}^{\hM} = \Exp{-d}\begin{pmatrix}
 s_{\bba}{}^{\bbb}\,v_a^m & 0 & 0 \\
 0 & s_{\bba}{}^{\bbb}\,\delta_I^\cI & 0 \\
 0 & 0 & s_{\bba}{}^{\bbb}\,\ell^a_m \end{pmatrix},\quad
 s_{\bba}{}^{\bbb}\equiv \begin{pmatrix} \Exp{d} & 0 \\ 0 & \Exp{-d} \end{pmatrix},\quad
 \Exp{-2\,d}\equiv \abs{\det \ell_m^a}\,,
\end{align}
which again satisfy the algebra \eqref{eq:V-alg} in $d=4$\,. 
We then obtain the generalized frame fields as
\begin{align}
 E_{\hA}{}^{\hM} &= 
 \Exp{\frac{\bar{\lambda}}{2}}\begin{pmatrix} \bar{\lambda}_{\bba}{}^{\bbb}\,e_a^m & 0 & 0 \\
 \bar{\lambda}_{\bba}{}^{\bbb}\,\pi_I^b\,e_a^m & \Exp{-\Delta}\bar{\lambda}_{\bba}{}^{\bbb}\,\omega_{I}{}^{\cJ} & 0 \\
 -\bar{\lambda}_{\bba}{}^{\bbb}\,\bigl(\pi^{ab}+\tfrac{1}{2}\,\delta^{KL}\,\pi^a_K\,\pi^b_L\bigr)\,e_b^m & -\Exp{-\Delta}\bar{\lambda}_{\bba}{}^{\bbb}\,\pi^{a}_K\,\omega^{K\cJ} & \Exp{-2\Delta}\bar{\lambda}_{\bba}{}^{\bbb}\,r^a_m 
\end{pmatrix},
\end{align}
where
\begin{align}
 \bar{s} \equiv \begin{pmatrix} \Exp{-\frac{\bar{\lambda}}{2}} & 0 \\ -\Exp{-\frac{\bar{\lambda}}{2}}\gamma & \Exp{\frac{\bar{\lambda}}{2}} \end{pmatrix},\qquad 
 \bar{\lambda} \equiv \lambda-2\,d\,.
\end{align}

\subsubsection{Generalized parallelizability}
\label{sec:G-para}

Now we show the relation \eqref{eq:E-ExDA}, i.e., the generalized parallelizability. 
This can be shown by using the above parameterizations of $E_{\hA}{}^{\hM}$\,, the definition of the generalized Lie derivative, the algebraic identities, and the differential identities. 
However, this requires a relatively long calculation.
Here we show the relation by following the general proof given in \cite{2009.04454}. 

Let us define the Weitzenb\"ock connection associated with $V_{\hA}{}^{\hM}$ as
\begin{align}
 \hat{W}_{\hA\hB}{}^{\hC} \equiv V_{\hB}{}^{\hM}\, V_{\hA}{}^{\hN}\,\partial_{\hN} V_{\hM}{}^{\hC}\,.
\end{align}
Then, the Weitzenb\"ock connection $\bm{\Omega}_{\hA\hB}{}^{\hC}$ associated with $E_{\hA}{}^{\hM}$ can be expressed as
\begin{align}
 \bm{\Omega}_{\hA\hB}{}^{\hC}
 = M_{\hA}{}^{\hD}\,\bigl[M_{\hB}{}^{\hE}\,(M^{-1})_{\hF}{}^{\hC}\,\hat{W}_{\hD\hE}{}^{\hF} 
 + V_{\hD}{}^{\hM}\,(M\,\partial_{\hM}M^{-1})_{\hB}{}^{\hC}\bigr]\,,
\end{align}
and the generalized fluxes $\bm{X}_{\hA\hB}{}^{\hC}$ associated with $E_{\hA}{}^{\hM}$ become
\begin{align}
 \bm{X}_{\hA\hB}{}^{\hC} 
 = \bm{\Omega}_{\hA\hB}{}^{\hC} - \bm{\Omega}_{\hC\hA}{}^{\hB} + Y^{\hC\hE}_{\hD\hB}\,\bm{\Omega}_{\hE\hA}{}^{\hD} \,.
\end{align}
Our task is to prove that $\bm{X}_{\hA\hB}{}^{\hC}=X_{\hA\hB}{}^{\hC}$\,. 

Since we are choosing the section $\partial_{m}\neq 0$\,, we have $\hat{W}_{\hA\hB}{}^{\hC} = \delta_{\hA}^{d}\,\hat{W}_{d\hB}{}^{\hC}$\,, where the matrix $(\hat{W}_{a})_{\hB}{}^{\hC}\equiv \hat{W}_{a\hB}{}^{\hC}$ is given by
\begin{align}
 \hat{W}_{a} = \bm{k}_{ba}{}^c\,K^b{}_c + \bm{k}_{da}{}^d\,(\tilde{R} +\beta_d\,t_0) \,,\qquad 
 \bm{k}_{ab}{}^c\equiv v_a^m\,v_b^n\,\partial_n \ell_m^c\,.
\end{align}
Here $\tilde{R}\equiv R_*$ in $d\geq 5$ while $\tilde{R}\equiv R^+{}_+$ in $d=4$\,, and $\tilde{R} +\beta_d\,t_0=0$ in $d\geq 6$\,. 
Using the differential identity,
\begin{align}
 V_{\hD}{}^{\hM}\,(M\,\partial_{\hM}M^{-1})_{\hB}{}^{\hC}
 = \delta_{\hD}^{+e}\,(a^{-1})_e{}^f\,X_{e\hB}{}^{\hC} \,,
\end{align}
and the algebraic identity
\begin{align}
 (a^{-1})_e{}^f\,X_{e\hB}{}^{\hC} = M_{\hB}{}^{\hD}\,(M^{-1})_{\hE}{}^{\hC}\,X_{e\hD}{}^{\hE}\,,
\end{align}
the Weitzenb\"ock connection $\bm{\Omega}_{\hA\hB}{}^{\hC}$ can be expressed as
\begin{align}
 \bm{\Omega}_{\hA\hB}{}^{\hC}
 &= M_{\hA}{}^{\hD}\,M_{\hB}{}^{\hE}\,(M^{-1})_{\hF}{}^{\hC}\,\bm{\tilde\Omega}_{\hD\hE}{}^{\hF}\,,\qquad
 \bm{\tilde\Omega}_{\hA\hB}{}^{\hC} \equiv \delta_{\hA}^{d}\,\bigl(\bm{\tilde\Omega}_{d}\bigr)_{\hB}{}^{\hC}\,,
\\
 \bm{\tilde\Omega}_a
 &\equiv \hat{W}_a + X_a = X_a + \bm{k}_{ba}{}^c\,K^b{}_c + \bm{k}_{da}{}^d\,(\tilde{R} +\beta_d\,t_0) \,.
\end{align}
Now, we find a key identity
\begin{align}
 \Omega_{a} = X_{a} + k_{ba}{}^c\,K^b{}_c + k_{ba}{}^b \,(\tilde{R} + \beta_d\,t_0)\,,
\end{align}
which can be easily seen in each dimension by comparing the explicit forms of $\Omega_{a}$ and $X_{a}$ (for example, Eq.~\eqref{eq:Omega-a-4d} and Eq.~\eqref{eq:ExDA-X-1}, in $d=4$). 
Using $k_{[ab]}{}^c= \bm{k}_{[ab]}{}^c=\frac{1}{2}\,f_{ab}{}^c$\,, this becomes
\begin{align}
 \bm{\tilde\Omega}_a = \Omega_a + \bm{k}_{(ab)}{}^c\,\bigl[K^b{}_c + \delta_c^b\, (\tilde{R} +\beta_d\,t_0)\bigr] \,.
\end{align}
Moreover, as we observed in section \ref{sec:ExDA}, the symmetric part $k_{(ab)}{}^c$ of $k_{ab}{}^c$ does not contribute to the generalized fluxes. 
Taking this into account, we can conclude that
\begin{align}
 \bm{\tilde\Omega}_{\hA\hB}{}^{\hC}=\Omega_{\hA\hB}{}^{\hC} + \bm{\cN}_{\hA\hB}{}^{\hC}\,,
\end{align}
where $\bm{\cN}_{\hA\hB}{}^{\hC}$ (that contains $\bm{k}_{(ab)}{}^c$) does not contribute to the generalized fluxes:
\begin{align}
 \bm{\cN}_{\hA\hB}{}^{\hC} - \bm{\cN}_{\hC\hA}{}^{\hB} + Y^{\hC\hE}_{\hD\hB}\, \bm{\cN}_{\hE\hA}{}^{\hD} = 0\,.
\end{align}
Then, using the duality invariance of the $Y$-tensor, we find
\begin{align}
 \bm{X}_{\hA\hB}{}^{\hC} 
 &= M_{\hA}{}^{\hD}\,M_{\hB}{}^{\hE}\,(M^{-1})_{\hF}{}^{\hC}\, \bigl(\Omega_{\hA\hB}{}^{\hC} - \Omega_{\hC\hA}{}^{\hB} + Y^{\hC\hE}_{\hD\hB}\, \Omega_{\hE\hA}{}^{\hD}\bigr) 
\nn\\
 &= M_{\hA}{}^{\hD}\,M_{\hB}{}^{\hE}\,(M^{-1})_{\hF}{}^{\hC}\, X_{\hD\hE}{}^{\hF} = X_{\hA\hB}{}^{\hC} \,.
\end{align}
This completes the proof that the generalized fluxes $\bm{X}_{\hA\hB}{}^{\hC}$ coincide with the structure constants $X_{\hA\hB}{}^{\hC}$ of the half-maximal ExDA. 

Before closing this section, let us comment on the non-geometric $R$-fluxes. 
By looking at the structure constants of the half-maximal ExDA, we find that $X^{abc}=0$ and $X_{IJ}{}^a=0$\,. 
Then the generalized parallelizability shows that $\bm{X}^{abc}=0$ and $\bm{X}_{IJ}{}^a=0$\,. 
Here, $\bm{X}^{abc}$ is known as the non-geometric $R$-flux and $\bm{X}_{IJ}{}^a$ will be also a similar quantity. 
By computing these fluxes without using the algebraic or differential identity, we find
\begin{align}
 -\Exp{2\Delta}e_a^m\,\omega_I{}^K\,\omega_J{}^L\,\bm{X}_{KL}{}^a
 &= 2\,\pi^{n}_{[I}\,\partial_n\pi^m_{J]}
 +\bigl(\pi^{mn} + \tfrac{1}{2}\,\pi^{mL}\,\pi^n_L\bigr)\,\partial_n\omega_{KI}\,\omega^K{}_J
\nn\\
 &\quad 
 +\delta_{IJ}\,\bigl(\pi^{mn} + \tfrac{1}{2}\,\pi^{mK}\,\pi^n_K\bigr)\,\partial_n\Delta\,,
\\
 -\Exp{4\Delta}e_a^m\,e_b^n\,e_c^p\, \bm{X}^{abc} 
 &= 3\,\bigl(\pi^{[m|q} + \tfrac{1}{2}\, \pi^{qI}\,\pi^{[m}_I \bigr)\,\bigl(\partial_q \pi^{np]} + \pi^{n}_J\,\partial_q \pi^{p]J}\bigr) \,.
\end{align}
For example, if we set $n=0$\,, the disappearance of these fluxes reduce to
\begin{align}
 3\,\pi^{[m|q} \, \partial_q \pi^{np]} = 0\,,
\end{align}
which shows that $\pi^{mn}$ is a Poisson tensor. 
We thus regard
\begin{align}
 &2\,\pi^{n}_{[I}\,\partial_n\pi^m_{J]}
 +\bigl(\pi^{mn} + \tfrac{1}{2}\,\pi^{mL}\,\pi^n_L\bigr)\,\partial_n\omega_{KI}\,\omega^K{}_J +\delta_{IJ}\,\bigl(\pi^{mn} + \tfrac{1}{2}\,\pi^{mK}\,\pi^n_K\bigr)\,\partial_n\Delta = 0\,,
\\
 &3\,\bigl(\pi^{[m|q} + \tfrac{1}{2}\, \pi^{qI}\,\pi^{[m}_I \bigr)\,\bigl(\partial_q \pi^{np]} + \pi^{n}_J\,\partial_q \pi^{p]J}\bigr) = 0\,,
\end{align}
as the definition of the generalized Poisson tensors. 

\section{Subalgebra and upliftability}
\label{sec:uplift}

In this section, we restrict ourselves to the simplest case $n=0$\,. 

\subsection{Subalgebra DD$^+$}
\label{sec:uplift-DD+}

As was mentioned in section \ref{sec:ExDA}, the half-maximal ExDA in $d\geq 6$ is exactly the DD$^+$\,, whose algebra is
\begin{align}
\begin{split}
 T_{a}\circ T_{b} &= f_{ab}{}^c\,T_{c}\,,\qquad 
 T^a\circ T^b = f_c{}^{ab}\,T^c \,,
\\
 T_{a}\circ T^b &= f_a{}^{bc}\,T_{c} - f_{ac}{}^b\,T^c +2\,Z_a\,T^b\,,
\\
 T^a\circ T_b &= - f_b{}^{ac}\,T_{c} + \bigl(f_{bc}{}^a +2\,\delta^a_b\,Z_c -2\,\delta^a_c\,Z_b\bigr)\,T^c\,.
\end{split}
\label{eq:DD+}
\end{align}
The Leibniz identities can be summarized as
\begin{align}
 &f_{[ab}{}^e\,f_{c]e}{}^d = 0\,,\qquad
  4\,f_{[a}{}^{e[c}\,f_{b]e}{}^{d]}
  -f_{ab}{}^e\,f_e{}^{cd}
  +4\,f_{[a}{}^{cd}\,Z_{b]}=0\,,
\\
 &f_e{}^{[ab}\,f_d{}^{c]e}=0\,,\qquad
 f_{ab}{}^c\,Z_c = 0\,, \qquad
 f_a{}^{bc}\,Z_c=0\,.
\label{eq:DD+Lid}
\end{align}
If we consider $d=5,4$\,, the DD$^+$ is realized as a subalgebra of the half-maximal ExDA. 
However, the Leibniz identities \eqref{eq:LI-5d} or \eqref{eq:LI-4d} of the half-maximal ExDA give additional conditions on the structure constants of the DD$^+$,
\begin{align}
 f_a{}^{cd}\,f_{cd}{}^b = f_a{}^{bc}\,\eta_c \,, \qquad
 f_{ab}{}^c\,\eta_c = 0\qquad \bigl(\eta_a\equiv f_a - f_{ab}{}^{b} \bigr) \,.
\label{eq:half-maximal-condition}
\end{align}
This shows that a DD$^+$ which does not satisfy \eqref{eq:half-maximal-condition} cannot be embedded into (or uplifted to) a half-maximal ExDA in $d=5,4$\,.
Since $f_a$ (or $\eta_a$) does not appear in the subalgebra \eqref{eq:DD+}, it can be chosen arbitrarily such that the uplift condition \eqref{eq:half-maximal-condition} is satisfied. 
Here we consider three sufficient conditions for the conditions \eqref{eq:half-maximal-condition} to be satisfied.
\begin{enumerate}
\item[(i)] 
If we consider a DD$^+$ satisfying
\begin{align}
 f_a{}^{bc}\,f_{bc}{}^d=0\,,
\label{eq:maximal-uplift}
\end{align}
we can always find the trivial solution of Eq.~\eqref{eq:half-maximal-condition}, namely $\eta_a=0$ (or $f_a = f_{ab}{}^b$). 
Therefore, an arbitrary DD$^+$ satisfying \eqref{eq:maximal-uplift} can be embedded into a half-maximal ExDA. 

\item[(ii)] 
Let us also consider a DD$^+$ where the dual algebra is unimodular,
\begin{align}
 f_b{}^{ba}=0\,.
\end{align}
In this case, the condition \eqref{eq:half-maximal-condition} reduces to
\begin{align}
 f_{ab}{}^c\,f_c = 0\,,\qquad
 f_a{}^{bc}\,f_c = 0\,,
\end{align}
where we have used Eq.~\eqref{eq:DD+Lid}. 
Again, due to the existence of the trivial solution $f_a=0$\,, a DD$^+$ with $f_b{}^{ba}=0$ always has an uplift into a half-maximal ExDA. 

\item[(iii)] 
If we define
\begin{align}
 \zeta_a\equiv f_a + f_{ab}{}^b \,,
\end{align}
we find that the condition \eqref{eq:half-maximal-condition} is equivalent to
\begin{align}
 f_{ab}{}^c\,\zeta_c = 0\,,\qquad
 3\,f_a{}^{[bc}\,f_{bc}{}^{d]}=f_{a}{}^{dc}\,\zeta_c\,.
\end{align}
This shows that a DD$^+$ satisfying
\begin{align}
 3\,f_a{}^{[bc}\,f_{bc}{}^{d]}= 0\,,
\end{align}
also has an uplift into a half-maximal ExDA (with $f_a=-f_{ab}{}^b$).
\end{enumerate}

\subsection{Upliftability to EDA}
\label{sec:uplift-EDA}

As we discussed in section \ref{sec:ExDA-EDA}, a $\mathbb{Z}_2$ truncation of an $E_{D+1(D+1)}$ EDA gives a half-maximal ExDA. 
However, similar to the discussion given in the previous subsection, not all of the half-maximal ExDAs can be uplifted to an $E_{D+1(D+1)}$ EDA. 
Here we identify the condition that the half-maximal ExDA can be uplifted to an $E_{D+1(D+1)}$ EDA by restricting ourselves to the cases $D\leq 3$ (i.e., $d\geq 7$). 

For $D\leq 3$\,, the EDA is given by Eq.~\eqref{eq:IIB-EDA} with $f_{\sfa}{}^{\sfb_1\cdots \sfb_4}=0$\,.
As we explained in section \ref{sec:ExDA-EDA}, this EDA contains the half-maximal ExDA as a subalgebra when $f_{\sfa\bm{1}}{}^{\bm{2}}=0$ is satisfied. 
The whole set of Leibniz identities of the EDA can be found as
\begin{align}
 &f_{[\sfa\sfb}{}^{\sfe}\,f_{\sfc]\sfe}{}^{\sfd} =0\,,\qquad
 f_{\sfa\gamma}{}^\beta\,f_{\sfb\beta}{}^\delta - f_{\sfb\gamma}{}^\beta\,f_{\sfa\beta}{}^\delta
 + f_{\sfa\sfb}{}^{\sfc}\,f_{\sfc\gamma}{}^\delta = 0\,,\qquad f_{\sfa\sfb}{}^{\sfc}\,Z_{\sfc} =0\,,
\label{eq:EDA-LI1}
\\
 &4\,f_{[\sfa|\sfd}{}^{[\sfc_1|}\,f_{|\sfb]}{}^{\sfd|\sfc_2]}_\gamma - f_{\sfa\sfb}{}^{\sfd}\,f_{\sfd}{}^{\sfc_1\sfc_2}_\gamma -2\,f_{[\sfa|\gamma}{}^\delta\,f_{|\sfb]}{}^{\sfc_1\sfc_2}_\delta 
 -4\,Z_{[\sfa}\,f_{\sfb]}{}^{\sfc_1\sfc_2}_\gamma = 0\,,
\label{eq:EDA-LI2}
\\
 & f_{\sfa}{}_\gamma^{\sfb\sfc}\,f_{\sfc\alpha}{}^\gamma + 2\,f_{\sfa}{}_{\alpha}^{\sfb\sfc}\,Z_{\sfc}=0\,,
\qquad 
 f_{\sfc}{}^{\sfd\sfa}_\alpha\,f_{\sfd}{}^{\sfb_1\sfb_2}_\beta - 2\,f_{\sfd}{}^{\sfa[\sfb_1}_\alpha\,f_{\sfc}{}^{\sfb_2]\sfd}_\beta
 = 0 \,,
\label{eq:EDA-LI3}
\\
 &f_{\sfc_1\sfc_2}{}^{\sfa}\,f_{\sfb}{}_\alpha^{\sfc_1\sfc_2} = 4\,f_{\sfb}{}_\gamma^{\sfa\sfc}\,f_{\sfc\alpha}{}^\gamma \,,
\label{eq:EDA-LI4}
\end{align}
When $f_{\sfa\bm{1}}{}^{\bm{2}}=f_{\sfa\bm{2}}{}^{\bm{1}}=f_{\sfa}{}_{\bm{2}}^{\sfb_1\sfb_2}=0$ are satisfied, the identification
\begin{align}
 f_{ab}{}^c = f_{\mathsf{a}\mathsf{b}}{}^{\mathsf{c}}\,, \quad
 f_a{}^{bc} = -f_{\mathsf{a}}{}_{\bm{1}}^{\mathsf{b}\mathsf{c}}\,,\quad
 Z_a = Z_{\mathsf{a}} + \tfrac{1}{2}\, f_{\mathsf{a}\bm{1}}{}^{\bm{1}} \,,
\end{align}
shows that the conditions \eqref{eq:EDA-LI1}--\eqref{eq:EDA-LI3} are equivalent to the Leibniz identities of the half-maximal ExDA given in Eq.~\eqref{eq:LI-6d-ExDA}.
The condition \eqref{eq:EDA-LI4} additionally requires the following condition to the structure constants of the half-maximal ExDA:
\begin{align}
 f_{a}{}^{c_1c_2}\,f_{c_1c_2}{}^{b} = f_{a}{}^{bc}\,\eta_c \,,\qquad f_{ab}{}^c\,\eta_c = 0\,,
\label{eq:uplift}
\end{align}
where $\eta_a\equiv 4\,f_{a\bm{1}}{}^{\bm{1}}$\,. 
Namely, Eq.~\eqref{eq:uplift} is the uplift condition for $D\leq 3$\,, and interestingly, this has the same form as Eq.~\eqref{eq:half-maximal-condition}. 
If we can find a certain $\eta_{a}$ satisfying this condition, the half-maximal ExDA can be embedded into the EDA with $Z_{\mathsf{a}}$ and $f_{\sfa\bm{1}}{}^{\bm{1}}$ given by
\begin{align}
 Z_{\mathsf{a}} = Z_a - \tfrac{1}{8}\, \eta_a \,,\qquad
 f_{\sfa\bm{1}}{}^{\bm{1}} = \tfrac{1}{4}\,\eta_a \,.
\label{eq:uplift-Z-f}
\end{align}
If there is no solution for $\eta_a$\,, the half-maximal EDA does not have an uplift to EDA. 
For example, if $f_{c_1c_2}{}^{a}\,f_{b}{}^{c_1c_2} = 0$ is satisfied, there is a trivial solution $\eta_a=0$ and the half-maximal ExDA (or a DD$^+$) has an uplift to EDA. 

We note that the uplift condition \eqref{eq:uplift} is not duality covariant. 
It is not even symmetric under $f_{ab}{}^c \leftrightarrow f_c{}^{ab}$\,. 
Accordingly, the uplift condition \eqref{eq:uplift} (or \eqref{eq:half-maximal-condition}) depends on the choice of the Manin triple. 
Even if a Manin triple $(\mathfrak{g}|\tilde{\mathfrak{g}})$ is upliftable, the dual $(\tilde{\mathfrak{g}}|\mathfrak{g})$ may not be upliftable. 
The same Drinfel'd double may have another inequivalent Manin triple $(\mathfrak{g}'|\tilde{\mathfrak{g}}')$, but to check its upliftability, we again need to look at the condition \eqref{eq:uplift}. 

A similar uplift condition which is duality covariant has been discussed in \cite{1101.5954,1104.3587,1109.0290,1203.6562}. 
By assuming the absence of the trombone gauging, the condition for an embedding tensor of half-maximal supergravity to be upliftable to that of maximal supergravity can be written as
\begin{align}
 F_{ABC}\,F^{ABC} = 0\,.
\label{eq:geometric-general}
\end{align}
For the half-maximal ExDA, this can be also expressed as
\begin{align}
 f_{ab}{}^{c}\,f_{c}{}^{ab}=0\,.
\label{eq:geometric}
\end{align}
The similarity between \eqref{eq:uplift} and \eqref{eq:geometric} was pointed out in \cite{2006.12452}. 
As discussed in \cite{1109.0290,1203.6562}, the condition \eqref{eq:geometric-general} is a consequence of the section condition in DFT (if we assume the absence of the dilaton flux $F_A$), and a violation of this condition is a sign of non-geometry. 
In the context of the PL $T$-duality, for any Drinfel'd double we can construct the generalized frame fields $E_A{}^M$ satisfying the algebra $[E_A,\,E_B]_{\text{D}} = - X_{AB}{}^C\,E_C$ in such a way that $E_A{}^M$ depend only on the physical coordinates. 
Then, a natural question is why the section condition can be broken. 
The answer is related to the DFT dilation. 
Assuming the absence of the dilaton flux $F_A$\,, the DFT dilaton has a general form $\Exp{-2d}=\Exp{-\Delta} \abs{\det\ell^a_m}$\,. 
This depends only on the physical coordinates, but when we have $f_b{}^{ba}\neq 0$ we need to shift the derivative of the dilaton as \cite{1810.11446}
\begin{align}
 \partial_M d\to \partial_M d + (0,\,I^m)\,,\qquad
 I^m \equiv \tfrac{1}{2}\,f_b{}^{ba}\,v_a^m\,.
\label{eq:GSE-I}
\end{align}
This vector field $I^m$ should be identified as the Killing vector field in the generalized supergravity equations of motion and the Killing equations are equivalent to the section condition of DFT \cite{1611.05856,1703.09213}. 
The Killing equations for the metric, the $B$-field, and $\Delta$ are ensured by the Leibniz identities, but the Killing equation for the dilaton, in particular,
\begin{align}
 \Lie_I \ln \abs{\det\ell^a_m} = \tfrac{1}{2}\,f_b{}^{ba}\,\ell_c^m\,\Lie_{v_a} \ell^c_m = -\tfrac{1}{2}\,f_b{}^{ba}\, f_{ac}{}^c = 0\,,
\label{eq:dilaton-isometry}
\end{align}
is not ensured. 
Indeed, under the Leibniz identities, Eq.~\eqref{eq:dilaton-isometry} is exactly the condition \eqref{eq:geometric}. 

In the next section, we consider concrete examples where the condition \eqref{eq:geometric} is violated.
There, assuming $F_A=0$\,, the DFT dilaton indeed breaks the section condition. 
Since the condition \eqref{eq:geometric} is broken, the algebra does not have an uplift to any EDA with vanishing trombone gauging. 
However, if the uplift condition \eqref{eq:uplift} is satisfied, the half-maximal ExDA can be embedded into an EDA. 
This is possible because the EDA has a non-vanishing trombone gauging, and the condition \eqref{eq:geometric-general} does not apply. 
Thus, the two conditions \eqref{eq:uplift} and \eqref{eq:geometric-general} are different conditions. 
If one repeats the analysis of \cite{1101.5954,1104.3587,1109.0290,1203.6562} by allowing for the trombone gauging, one may find the upliftablity condition that modifies the condition \eqref{eq:geometric-general}. 
Then the condition will be weaker than \eqref{eq:uplift}. 

\section{Examples}
\label{sec:examples}

In this section, we show several examples of half-maximal ExDAs. 
We begin with four low-dimensional examples with $n=0$\,, where the half-maximal ExDA are DD$^+$\,. 
After that, we consider more complicated examples with $n>0$\,. 

\subsection{Examples with $n=0$}

\subsubsection*{Example 1 $(5.iii|6_0|b)$}

Let us consider a Manin triple $(5.iii|6_0|b)$ ($b\neq 0$) \cite{math:0202210} whose structure constants are
\begin{align}
 f_{23}{}^2 = -b \,,\quad 
 f_{13}{}^1 = -b \,,\quad 
 f_1{}^{23} = 1 \,,\quad 
 f_2{}^{13} = 1 \,.
\end{align}
We also introduce Abelian generators $\{T_4,\,T^4\}$ and consider an eight-dimensional Lie algebra. 
By introducing the coordinates $x^m=(x,y,z,w)$ and a parameterization $g=\Exp{x\,T_1}\Exp{y\,T_2}\Exp{z\,T_3}\Exp{w\,T_4}$, the right-invariant vector fields and the Poisson--Lie structure become
\begin{align}
 e_a{}^m = {\footnotesize\begin{pmatrix}
 1 & 0 & 0 & 0 \\
 0 & 1 & 0 & 0 \\
 b\,x & b\,y & 1 & 0 \\
 0 & 0 & 0 & 1\end{pmatrix}},\qquad
 \pi^{mn} = {\footnotesize\begin{pmatrix}
 0 & -\frac{b\,(x^2-y^2)}{2} & y & 0 \\
 \frac{b\,(x^2-y^2)}{2} & 0 & x & 0 \\
 -y & -x & 0 & 0 \\
 0 & 0 & 0 & 0\end{pmatrix}},\qquad 
 \Delta=0\,.
\end{align}
They construct the generalized frame fields $E_A{}^M$ that satisfies $[E_A,\,E_B]_{\text{D}}=-X_{AB}{}^C\,E_C$\,. 

This ExDA satisfies $f_1{}^{ab}\,f_{ab}{}^2 =-2\,b$ and $f_2{}^{ab}\,f_{ab}{}^1 =-2\,b$\,, and by introducing
\begin{align}
 \eta_1=\eta_2=0\,,\qquad \eta_3 =-2\,b\,,\qquad \eta_4:\text{arbitrary}
\end{align}
the uplift condition \eqref{eq:uplift} is satisfied.
Thus, this is upliftable to an $\SO(5,5)$ EDA. 

\subsubsection*{Example 2 $(6_0|5.iii|b)$}

Let us consider the $T$-dual of the previous example
\begin{align}
 f_{23}{}^1 = 1 \,,\quad 
 f_{13}{}^2 = 1 \,,\quad 
 f_2{}^{23} = -b \,,\quad 
 f_1{}^{13} = -b \,.
\label{eq:605iiib}
\end{align}
Using the same parameterization as in the previous example, we obtain
\begin{align}
 e_a{}^m = {\footnotesize\begin{pmatrix}
 1 & 0 & 0 & 0 \\
 0 & 1 & 0 & 0 \\
 -y & -x & 1 & 0 \\
 0 & 0 & 0 & 1 \end{pmatrix}},\qquad
 \pi^{mn} = {\footnotesize\begin{pmatrix}
 0 & \frac{b\,(x^2-y^2)}{2} & -b\,x & 0 \\
 -\frac{b\,(x^2-y^2)}{2} & 0 & -b\,y & 0 \\
 b\,x & b\,y & 0 & 0 \\
 0 & 0 & 0 & 0\end{pmatrix}},\qquad 
 \Delta=0\,.
\end{align}
In this case, the uplift condition cannot be satisfied for any $f_a$ and this does not have an uplift to the $\SO(5,5)$ EDA. 
However, since the condition \eqref{eq:geometric} is satisfied, this can be uplifted to some flux configuration in maximal supergravity. 
The explicit form of the algebra can be found by acting $T$-dualities in all directions to the $\SO(5,5)$ EDA obtained in the previous example (we choose $\eta_4=0$ for simplicity). 
The algebra is generated by $\{T_\sfa,\,T^{\sfa}_{\alpha},\,T^{\sfa_1\sfa_2\sfa_3}\}$ and the non-vanishing products can be found as follows:
{\footnotesize
\begin{align}
\begin{split}
T_1 \circ T_3 &=T_2\,,\quad
T_1 \circ T^1_{\bm{1}} =-b\,T_3\,,\quad
T_1 \circ T^2_{\bm{1}} =-T^3_{\bm{1}}\,,\quad
T_1 \circ T^3_{\bm{1}} =b\,T_1\,,\quad
\\
T_1 \circ T^2_{\bm{2}}&=-T^3_{\bm{2}}\,,\quad
T_1 \circ T^{123} =-b\,T^2_{\bm{2}}\,,\quad
T_1 \circ T^{124} =-T^{134}\,,\quad
T_1 \circ T^{134} =b\,T^4_{\bm{2}}\,,
\\
T_2 \circ T_3&=T_1\,,\quad
T_2 \circ T^1_{\bm{1}} =-T^3_{\bm{1}}\,,\quad
T_2 \circ T^2_{\bm{1}} =-b\,T_3\,,\quad
T_2 \circ T^3_{\bm{1}} =b\,T_2\,,
\\
T_2 \circ T^1_{\bm{2}}&=-T^3_{\bm{2}}\,,\quad
T_2 \circ T^{123} =b\,T^1_{\bm{2}}\,,\quad
T_2 \circ T^{124} =T^{234}\,,\quad
T_2 \circ T^{234} =b\,T^4_{\bm{2}}\,,
\\
T_3 \circ T_1&=-T_2\,,\quad
T_3 \circ T_2 =-T_1\,,\quad
T_3 \circ T^1_{\bm{1}} =T^2_{\bm{1}}\,,\quad
T_3 \circ T^2_{\bm{1}} =T^1_{\bm{1}}\,,
\\
T_3 \circ T^1_{\bm{2}}&=T^2_{\bm{2}}\,,\quad
T_3 \circ T^2_{\bm{2}} =T^1_{\bm{2}}\,,\quad
T_3 \circ T^{134} =T^{234}\,,\quad
T_3 \circ T^{234} =T^{134}\,,
\\
T^1_{\bm{1}} \circ T_1&=b\,T_3\,,\quad
T^1_{\bm{1}} \circ T_2 =T^3_{\bm{1}}\,,\quad
T^1_{\bm{1}} \circ T_3 =-T^2_{\bm{1}}\,,\quad
T^1_{\bm{1}} \circ T^3_{\bm{1}} =-b\,T^1_{\bm{1}}\,,
\\
T^1_{\bm{1}} \circ T^1_{\bm{2}}&=T^{123}\,,\quad
T^1_{\bm{1}} \circ T^3_{\bm{2}} =-b\,T^1_{\bm{2}}\,,\quad
T^1_{\bm{1}} \circ T^4_{\bm{2}} =T^{234}\,,\quad
T^1_{\bm{1}} \circ T^{234} =b\,T^{124}\,,
\\
T^2_{\bm{1}} \circ T_1&=T^3_{\bm{1}}\,,\quad
T^2_{\bm{1}} \circ T_2 =b\,T_3\,,\quad
T^2_{\bm{1}} \circ T_3 =-T^1_{\bm{1}}\,,\quad
T^2_{\bm{1}} \circ T^3_{\bm{1}} =-b\,T^2_{\bm{1}}\,,
\\
T^2_{\bm{1}} \circ T^2_{\bm{2}}&=-T^{123}\,,\quad
T^2_{\bm{1}} \circ T^3_{\bm{2}} =-b\,T^2_{\bm{2}}\,,\quad
T^2_{\bm{1}} \circ T^4_{\bm{2}} =T^{134}\,,\quad
T^2_{\bm{1}} \circ T^{134} =-b\,T^{124}\,,
\\
T^3_{\bm{1}} \circ T_1&=-b\,T_1\,,\quad
T^3_{\bm{1}} \circ T_2 =-b\,T_2\,,\quad
T^3_{\bm{1}} \circ T^1_{\bm{1}} =b\,T^1_{\bm{1}}\,,\quad
T^3_{\bm{1}} \circ T^2_{\bm{1}} =b\,T^2_{\bm{1}}\,,
\\
T^3_{\bm{1}} \circ T^3_{\bm{2}}&=-b\,T^3_{\bm{2}}\,,\quad
T^3_{\bm{1}} \circ T^4_{\bm{2}} =-b\,T^4_{\bm{2}}\,,\quad
T^3_{\bm{1}} \circ T^{123} =b\,T^{123}\,,\quad
T^3_{\bm{1}} \circ T^{124} =b\,T^{124}\,,
\\
T^1_{\bm{2}} \circ T_2&=T^3_{\bm{2}}\,,\quad
T^1_{\bm{2}} \circ T_3 =-T^2_{\bm{2}}\,,\quad
T^1_{\bm{2}} \circ T^1_{\bm{1}} =-T^{123}\,,\quad
T^1_{\bm{2}} \circ T^4_{\bm{1}} =-T^{234}\,,
\\
T^2_{\bm{2}} \circ T_1&=T^3_{\bm{2}}\,,\quad
T^2_{\bm{2}} \circ T_3 =-T^1_{\bm{2}}\,,\quad
T^2_{\bm{2}} \circ T^2_{\bm{1}} =T^{123}\,,\quad
T^2_{\bm{2}} \circ T^4_{\bm{1}} =-T^{134}\,,
\\
T^3_{\bm{2}} \circ T^1_{\bm{1}}&=b\,T^1_{\bm{2}}\,,\quad
T^3_{\bm{2}} \circ T^2_{\bm{1}} =b\,T^2_{\bm{2}}\,,\quad
T^3_{\bm{2}} \circ T^3_{\bm{1}} =b\,T^3_{\bm{2}}\,,\quad
T^3_{\bm{2}} \circ T^4_{\bm{1}} =b\,T^4_{\bm{2}}\,,
\\
T^{123} \circ T_1&=b\,T^2_{\bm{2}}\,,\quad
T^{123} \circ T_2 =-b\,T^1_{\bm{2}}\,,\quad
T^{123} \circ T^3_{\bm{1}} =-b\,T^{123}\,,\quad
T^{123} \circ T^4_{\bm{1}} =-b\,T^{124}\,.
\end{split}
\label{eq:non-EDA}
\end{align}}
This algebra satisfies the Leibniz identities. 
The subalgebra generated by $\{T_a\equiv T_{\sfa},\,T^a\equiv T^{\sfa}_{\bm{1}}\}$ is the Lie algebra of $(6_0|5.iii|b)$, and this is an uplift of the Manin triple $(6_0|5.iii|b)$. 

According to \cite{2009.04454}, the structure constants of the $\SO(5,5)$ EDA are given by
\begin{align}
 X_{\cA\cB}{}^\cC = \Theta_\cA{}^{\bm{\alpha}}\,(t_{\bm{\alpha}})_\cB{}^\cC
 -\bigl[\tfrac{1}{1+\beta_d}\,(t^{\bm{\alpha}})_\cA{}^\cD\,(t_{\bm{\alpha}})_\cB{}^\cC + \delta_\cA^\cD\,\delta_\cB^\cC\bigr]\,\vartheta_\cD\,,
\end{align}
where $\Theta_\cA{}^{\bm{\alpha}}$ and $\vartheta_\cA$ are defined as
\begin{align}
 \Theta_\cA{}^{\bm{\alpha}} \equiv \mathbb{P}_\cA{}^{\bm{\alpha}}{}^\cB{}_{\bm{\beta}}\,\Omega_\cB{}^{\bm{\beta}} \,,\qquad
 \vartheta_\cA = (1+\beta_d)\,\Omega_\cA{}^0 - \beta_d \, \Omega_\cD{}^{\bm{\alpha}}\,(t_{\bm{\alpha}})_\cA{}^\cD \,,
\end{align}
by using the structure constants $\Omega_\cA{}^{\bm{\alpha}}$ and $\Omega_\cA{}^0$ and a certain projector $\mathbb{P}_\cA{}^{\bm{\alpha}}{}^\cB{}_{\bm{\beta}}$\,. 
Here, by considering the section condition, $\Omega_\cA{}^{\bm{\alpha}}$ and $\Omega_\cA{}^0$ are supposed to have only the physical components $\Omega_{\sfa}{}^{\bm{\alpha}}$ and $\Omega_{\sfa}{}^0$\,. 
However, the algebra given in Eq.~\eqref{eq:non-EDA} is based on another section. 
If we construct the physical component $\Omega_{\sfa}{}^{\bm{\beta}}$ and $\Omega_{\sfa}{}^0$ by using
\begin{align}
 f_{23}{}^1 = 1\,,\quad
 f_{13}{}^2 = 1\,,\quad
 f_2{}_{\bm{1}}^{23} = \tfrac{2\,b}{3}\,,\quad
 f_1{}_{\bm{1}}^{13} = \tfrac{2\,b}{3}\,,\quad
 f_4{}_{\bm{1}}^{34} = \tfrac{b}{3}\,,
\end{align}
and also introduce the dual components of $\Omega_\cA{}^{\bm{\alpha}}$ and $\Omega_\cA{}^0$ as
\begin{align}
 \Omega_{\bm{2}}^{3}{}^{\bm{\alpha}}\,t_{\bm{\alpha}} = \tfrac{2\,b}{3}\,R^{\bm{1}}{}_{\bm{2}}\,,\qquad
 \Omega_{\bm{1}}^{3}{}^0 = \tfrac{b}{3}\,,
\end{align}
the resulting $X_{\cA\cB}{}^\cC$ reproduce the algebra \eqref{eq:non-EDA}. 

In summary, the Manin triple \eqref{eq:605iiib} does not satisfy the condition \eqref{eq:uplift} and it is not uplifted to any EDA. 
However, since the section condition \eqref{eq:geometric} is satisfied, it is uplifted to a flux configuration \eqref{eq:non-EDA} in maximal supergravity. 

\subsubsection*{Example 3 $(\{3.v, \frac{1}{2}\,(X_2-X_3)\}|\{3,0\})$}

Let us consider a coboundary-type DD$^+$
\begin{align}
\begin{split}
 f_{13}{}^1 &= 1\,,\quad
 f_{23}{}^1 = 1\,,\quad
 f_2{}^{12} = -1\,,\quad
 f_3{}^{12} = -1\,,
\\
 f_2{}^{13} &= -1\,,\quad
 f_3{}^{13} = -1\,,\quad
 Z_2 = \tfrac{1}{4}\,,\quad
 Z_3 = -\tfrac{1}{4}\,.
\end{split}
\end{align}
This satisfies the uplift condition \eqref{eq:uplift} if we introduce
\begin{align}
 \eta_a = (0, 2\,\xi, 2 - 2\,\xi)\,.
\end{align}
Indeed, this half-maximal ExDA is uplifted to the $\SL(5)$ EDA with the structure constants
\begin{align}
\begin{split}
 f_{13}{}^1 &= 1\,,\quad
 f_{23}{}^1 = 1\,,\quad
 f_2{}_{\bm{1}}^{12} = 1\,,\quad
 f_3{}_{\bm{1}}^{12} = 1\,,\quad
 f_2{}_{\bm{1}}^{13} = 1\,,\quad
 f_3{}_{\bm{1}}^{13} = 1\,,
\\
 Z_2 &= \tfrac{1-\xi}{4}\,,\quad
 Z_3 = -\tfrac{2-\xi}{4}\,,\quad
 f_{2\bm{1}}{}^{\bm{1}} = \tfrac{\xi}{2}\,,\quad
 f_{3\bm{1}}{}^{\bm{1}} = \tfrac{1-\xi}{2}\,.
\end{split}
\end{align}

\subsubsection*{Example 4 $(3|3.i|b)$}

According to the classification of the six-dimensional Drinfel'd doubles \cite{math:0202210}, there are 22 Drinfel'd doubles and among these, three Drinfel'd doubles, DD3, DD4, and DD8 break the condition \eqref{eq:geometric}. 
The corresponding Manin triples are $(7_a|7_{1/a}|b)$, $(6_a|6_{1/a}.i|b)$, and $(3|3.i|b)$\,. 
Here we consider $(3|3.i|b)$ as an example. 

The structure constants are given by
\begin{align}
\begin{split}
 f_{12}{}^2 &= -1\,,\quad
 f_{12}{}^3 = -1\,,\quad
 f_{13}{}^2 = -1\,,\quad
 f_{13}{}^3 = -1\,,
\\
 f_2{}^{12} &= -b\,,\quad
 f_3{}^{12} = -b\,,\quad
 f_2{}^{13} = -b\,,\quad
 f_3{}^{13} = -b\quad (b\neq0),
\end{split}
\end{align}
and one can check that the condition \eqref{eq:geometric} is broken: $f_a{}^{bc}\,f_{bc}{}^a=8\,b$\,. 

In order to show that the section condition is violated, let us construct the generalized frame fields by using the parameterizations $x^m=(x,y,z)$ and $g=\Exp{z\,T_3}\Exp{y\,T_2}\Exp{x\,T_1}$. 
We find
\begin{align}
 E_A{}^M = {\footnotesize\begin{pmatrix}
 1 & -y-z & -y-z & 0 & 0 & 0 \\
 0 & 1 & 0 & 0 & 0 & 0 \\
 0 & 0 & 1 & 0 & 0 & 0 \\
 0 & b\,(y+z) & b\,(y+z) & 1 & 0 & 0 \\
 -b\,(y+z) & b\,(y+z)^2 & b\,(y+z)^2 & y+z & 1 & 0 \\
 -b\,(y+z) & b\,(y+z)^2 & b\,(y+z)^2 & y+z & 0 & 1 
\end{pmatrix}},
\label{eq:Ex4-EAM}
\end{align}
and this, of course, does not break the section condition. 
Requiring the absence of the dilaton flux $F_A$\,, the DFT dilaton takes the form
\begin{align}
 \Exp{-2\,d} = \abs{\det \ell^a_m} = \Exp{2\,x},
\end{align}
and the vector field $I$ given in Eq.~\eqref{eq:GSE-I} becomes
\begin{align}
 I=\tfrac{1}{2}\,f_b{}^{ba}\,v_a^m = b\,\partial_x\,.
\end{align}
Then we can clearly see that the dilaton $d(x)$ is not isometric along the $I$-direction. 
In other words, if we include the $I$ into the DFT dilaton, we find $d = -x + b\,\tilde{x}$\,, and this clearly breaks the section condition: $\partial_M\,\partial^M d\neq 0$\,. 

Although the condition \eqref{eq:geometric} is broken, the condition \eqref{eq:uplift} is satisfied for $\eta_a=(4,\,\xi,\,-\xi)$ with an arbitrary $\xi$\,. 
Choosing $\xi=0$ and using \eqref{eq:uplift-Z-f}, we find that this ExDA can be uplifted to the $\SL(5)$ EDA with
\begin{align}
\begin{split}
 f_{12}{}^2 &= -1\,,\quad
 f_{12}{}^3 = -1\,,\quad
 f_{13}{}^2 = -1\,,\quad
 f_{13}{}^3 = -1\,,
\\
 f_2{}_{\bm{1}}^{12} &= b\,,\quad
 f_3{}_{\bm{1}}^{12} = b\,,\quad
 f_2{}_{\bm{1}}^{13} = b\,,\quad
 f_3{}_{\bm{1}}^{13} = b\,,\quad
 Z_1 =-\tfrac{1}{2}\,,\quad 
 f_{1\bm{1}}{}^{\bm{1}} = 1\,,
\end{split}
\end{align}
which has non-vanishing trombone gauging $X_{\cA\cB}{}^{\cB}\neq 0$\,.
Using $x^m=(x,y,z)$ and $g=\Exp{z\,T_3}\Exp{y\,T_2}\Exp{x\,T_1}$, we obtain the generalized frame fields as
\begin{align}
 E_{\cA}{}^{\cM} &= \Exp{-\frac{8}{5}\Delta} \abs{\det e_a^m}^{\frac{1}{5}}\begin{pmatrix}
 e_a^m & 0 & 0 \\
 \pi^{ab}_\alpha\,e_b^m & \lambda_{\alpha}{}^{\beta}\,r^a_m & 0 \\
 0 & -\frac{3\,\epsilon^{\gamma\delta}\,r_{m}^{[\sfa_1}\pi^{\sfa_2\sfa_3]}_\gamma\,\lambda_{\delta}{}^{\beta}}{\sqrt{3!}} & r^{a_1}_{[m_1}\,r^{a_2}_{m_2}\,r^{a_3}_{m_3]}
\end{pmatrix}
\\
 &={\footnotesize \left(\!\!\begin{array}{ccc|ccc:ccc|c}
 1 & -y-z & -y-z & 0 & 0 & 0 & 0 & 0 & 0 & 0 \\
 0 & 1 & 0 & 0 & 0 & 0 & 0 & 0 & 0 & 0 \\
 0 & 0 & 1 & 0 & 0 & 0 & 0 & 0 & 0 & 0 \\ \hline
 0 & b\,(y+z) & b\,(y+z) & 1 & 0 & 0 & 0 & 0 & 0 & 0 \\
 -b\,(y+z) & b\,(y+z)^2 & b\,(y+z)^2 & y+z & 1 & 0 & 0 & 0 & 0 & 0 \\
 -b\,(y+z) & b\,(y+z)^2 & b\,(y+z)^2 & y+z & 0 & 1 & 0 & 0 & 0 & 0 \\ \hdashline
 0 & 0 & 0 & 0 & 0 & 0 & \Exp{2 x} & 0 & 0 & 0 \\
 0 & 0 & 0 & 0 & 0 & 0 & \Exp{2 x} (y+z) & \Exp{2 x} & 0 & 0 \\
 0 & 0 & 0 & 0 & 0 & 0 & \Exp{2 x} (y+z) & 0 & \Exp{2 x} & 0 \\ \hline
 0 & 0 & 0 & 0 & 0 & 0 & 0 & b \Exp{2 x} (y+z) & -b \Exp{2 x} (y+z) & \Exp{2 x}\end{array}\!\!\right)},\nn
\end{align}
which is an uplift of Eq.~\eqref{eq:Ex4-EAM}. 

If we consider $(7_a|7_{1/a}|b)$ and $(6_a|6_{1/a}.i|b)$, both conditions \eqref{eq:uplift} and \eqref{eq:geometric} are broken, and we do not find any uplift to the maximal theory.

\subsection{Examples with $n\neq 0$}

\subsubsection*{Example 5}

Let us consider a half-maximal EDA in $d=4$ satisfying $f_{a}{}^b{}_I\neq 0$ and $\vartheta_{\hA}=0$,
\begin{align}
\begin{split}
 f_{12}{}^2 &= 1\,,\quad
 f_{13}{}^3 = -1\,,\quad
 f_{45}{}^6 = 1\,,\quad
 f_{56}{}^4 = 1\,,\quad
 f_{46}{}^5 = -1\,, 
\\
 f_1{}^{23} &= 1\,,\quad
 f_1{}^{26} = 2\,,\quad
 f_4{}^{25} = -1\,,\quad
 f_4{}^{35} = 1\,,\quad
 f_5{}^{24} = 1\,,\quad
 f_5{}^{34} = -1\,,
\\
 f_1{}^2{}_I &= p_I\,,\quad
 f_1{}^3{}_I = q_I\,,\quad
 Z_1 = -\tfrac{1}{2}\,,\quad
 f_1 = -1\,.
\end{split}
\end{align}
This satisfies the Leibniz identities. 
The structure constants $f_a{}^b{}_I$ can be expressed as in Eq.~\eqref{eq:f-r-1} by using
\begin{align}
 \rr^2_I = \tfrac{2}{3}\,p_I\,,\qquad 
 \rr^3_I = -2\,q_I\,,
\end{align}
but $f_a{}^{bc}$ cannot be expressed as in Eq.~\eqref{eq:f-r-2}, and this algebra is not of coboundary type. 

If we provide the parameterization,
\begin{align}
 x^m=(x,y,z,\theta,\phi,\psi)\,,\qquad
 g= \Exp{x\,T_1}\Exp{y\,T_2}\Exp{z\,T_3}\Exp{\phi\,T_6}\Exp{\theta\,T_5}\Exp{\phi\,T_6},
\end{align}
we find
\begin{align}
 \omega_I{}^J &= \delta_I^J\,,\quad \Exp{\lambda}=\Exp{-x},\quad \Exp{-2\Delta} = \Exp{x},\quad \gamma=0\,,
\\
 e_a{}^m &= {\footnotesize\begin{pmatrix}
 1 & 0 & 0 & 0 & 0 & 0 \\
 0 & \Exp{-x} & 0 & 0 & 0 & 0 \\
 0 & 0 & \Exp{x} & 0 & 0 & 0 \\
 0 & 0 & 0 & -\sin \phi & -\cot \theta \cos \phi & \csc \theta \cos \phi \\
 0 & 0 & 0 & \cos \phi & -\cot \theta \sin \phi & \csc \theta \sin \phi \\
 0 & 0 & 0 & 0 & 1 & 0\end{pmatrix}},\quad
 \pi_I^m = {\footnotesize\begin{pmatrix} 0 \\ \tfrac{2}{3}\, (1 - \Exp{-\frac{3 x}{2}})\,p_I \\ 2\,(\Exp{\frac{x}{2}}-1) q_I \\ 0 \\ 0 \\ 0 \end{pmatrix}} ,
\\
 \pi^{mn} &= {\footnotesize\begin{pmatrix}
 0 & 0 & 0 & 0 & 0 & 0 \\
 & 0 & \frac{2\Exp{-\frac{x}{2}}}{3} \bigl[\sinh (x/2)\,(4\,p_I\,q^I +3) - 2\,p_I\,q^I\sinh x \bigr] & 0 & -\Exp{-2 x} & 1 \\
 & & 0 & 0 & 1 & -1 \\
 & & & 0 & 0 & 0 \\
 & & & & 0 & 0 \\
 & & & & & 0 \end{pmatrix}} .
\end{align}
Then the resulting generalized frame fields $E_{\hA}{}^{\hM}$ satisfy the algebra $[E_{\hA},\,E_{\hB}]_{\text{D}}=-X_{\hA\hB}{}^{\hC}\,E_{\hC}$\,.

\subsubsection*{Example 6}

Let us consider the branch $f_{a-}{}^+ = p_a \neq 0$ in $d=4$ with $f_{ab}{}^c\neq 0$\,. 
Performing a redefinition of generators $T_a$\,, we can always realize $Z_a=f_a=\delta_a^1$ and $f_{ab}{}^c=2\,\delta^1_{[a}\,\delta_{b]}^c$\,. 
Using a parameterization
\begin{align}
 x^m = (x,\,y^i)\quad (i=2,\dotsc,6)\,,\qquad 
 g =\Exp{x\,T_1} \Exp{y^2\,T_2}\cdots\Exp{y^6\,T_6}, 
\end{align}
we find the general expression for various tensors:
\begin{align}
\begin{split}
 e_a{}^m &= \diag(1,\Exp{-x},\dotsc,\Exp{-x})\,,\qquad
 \omega_I{}^J=\delta_I^J\,,\qquad 
 \Exp{\lambda}=\Exp{x},\qquad 
 \Exp{-2\,\Delta}=\Exp{-2\,x},
\\
 \gamma &=(\Exp{x}-1)\,p_1 + \Exp{x}\bigl(p_2\,y^2 + \cdots + p_6\,y^6\bigr)\,,\qquad 
 \pi^m_I=\pi^{mn}=0\,.
\end{split}
\end{align}
They construct the generalized frame fields satisfying $[E_{\hA},\,E_{\hB}]_{\text{D}}=-X_{\hA\hB}{}^{\hC}\,E_{\hC}$\,.

\subsubsection*{Example 7}

Here we consider an example with $f_{aIJ}\neq 0$\,. 
For example, if we consider $d=7$ and $n=3$\,, we find that the ExDA with
\begin{align}
 f_{12}{}^3 = 1\,,\quad f_{23}{}^1 = 1\,,\quad f_{13}{}^2 = -1\,,\quad
 f_{1\dot{1}\dot{3}} = 1\,,\quad 
 f_{2\dot{2}\dot{3}} = 1\,,\quad 
 f_{3\dot{1}\dot{2}} = 1\,,
\end{align}
satisfies the Leibniz identities. 
Using $x^m=(x,y,z)$ and $g=\Exp{x\,T_1}\Exp{y\,T_2}\Exp{z\,T_3}$, we obtain
\begin{align}
\begin{split}
 e_a^m &= {\footnotesize\begin{pmatrix} 1 & 0 & 0 \\ \sin x \tan y & \cos x & - \frac{\sin x}{\cos y} \\
 -\cos x \tan y & \sin x & \frac{\cos x}{\cos y} \end{pmatrix}},\qquad
 \Delta=\pi^m_I=\pi^{mn}=0\,,
\\
 \omega_I{}^J &= {\footnotesize\begin{pmatrix}
 \cos x \cos z-\sin x \sin y \sin z & -\sin x \sin y \cos z -\cos x \sin z & - \sin x \cos y \\
 \cos y \sin z & \cos y \cos z & -\sin y \\
 \cos x \sin y \sin z+\sin x \cos z & \cos x \sin y \cos z-\sin x \sin z & \cos x \cos y\end{pmatrix}}.
\end{split}
\end{align}

This ExDA has vanishing $f_a{}^{bc}$ and $f_a{}^b{}_I$\,, and we can consider the Yang--Baxter deformation, i.e., the $\OO(3,6)$ transformation given by Eq.~\eqref{eq:cR-def}. 
However, in this case, there is no solution to the homogeneous CYBE, i.e., Eqs.~\eqref{eq:YB-1}--\eqref{eq:YB-5}. 

\subsubsection*{Example 8}

Here we consider the case where $T_a\circ T_b=f_{ab}{}^c\,T_c$ is a non-semisimple Lie algebra,
\begin{align}
 f_{12}{}^2 = 1\,,\quad
 f_{13}{}^3 = -1\,,\quad
 f_1{}^{12} = f_1{}^{13} = \tfrac{1}{\sqrt{2}}\,.
\end{align}
In this case, we find a solution of the homogeneous CYBE
\begin{align}
 \rr^a{}_I={\small\begin{pmatrix}
 0 & 0 & 0 \\
 \frac{\eta_1}{\sqrt{2}} & -\frac{\eta_1}{2} & \frac{\eta_1}{2} \\
 \frac{\eta_2}{\sqrt{2}} & -\frac{\eta_2}{2} & \frac{\eta_2}{2}
\end{pmatrix}},\qquad
 \rr^{ab}={\small\begin{pmatrix}
 0 & 0 & 0 \\
 0 & 0 & \eta_3 \\
 0 & -\eta_3 & 0
\end{pmatrix}}.
\end{align}
They produce the structure constants $f_a{}^b{}_I$ and $f_a{}^{bc}$ through Eqs.~\eqref{eq:f-r-1} and \eqref{eq:f-r-2} as
\begin{align}
 f_1{}^2{}_1 = \frac{\eta_1}{\sqrt{2}}\,,\quad
 f_1{}^2{}_3 = \eta_1\,,\quad
 f_1{}^3{}_1 = -\frac{\eta_2}{\sqrt{2}}\,,\quad
 f_1{}^3{}_2 = \eta_2\,,\quad
 f_1{}^{23} = \eta_1 \eta_2\,,
\end{align}
where $\eta_3$ does not appear in the structure constants (similar to the case of Abelian Yang--Baxter deformation). 

Using the deformed half-maximal ExDA and the parameterization,
\begin{align}
 x^m=(x,y,z)\,,\qquad g=\Exp{x\,T_1}\Exp{y\,T_2}\Exp{z\,T_3},
\end{align}
we can compute various tensors as
\begin{align}
\begin{split}
 e_a{}^m &= {\footnotesize\begin{pmatrix} 1 & 0 & 0 \\ 0 & \Exp{-x} & 0 \\ 0 & 0 & \Exp{x}\end{pmatrix}},\qquad
 \omega_I{}^J = {\footnotesize\begin{pmatrix} \cos x & -\frac{\sin x}{\sqrt{2}} & -\frac{\sin x}{\sqrt{2}} \\
 \frac{\sin x}{\sqrt{2}} & \cos^2(\frac{x}{2}) & \frac{\cos x-1}{2} \\
 \frac{\sin x}{\sqrt{2}} & \frac{\cos x-1}{2} & \cos^2(\frac{x}{2}) \end{pmatrix}},\qquad \Delta =0\,,
\\
 \pi_I{}^m &= {\footnotesize\begin{pmatrix}
 0 & \frac{\Exp{-x} \eta_1 (\Exp{x}-\cos x)}{\sqrt{2}} & \frac{\eta_2(1-\Exp{x} \cos x)}{\sqrt{2}} \\
 0 & \frac{\eta_1}{2} \Exp{-x} (\sin x-\Exp{x}+1) & \frac{\eta_2}{2} (\Exp{x} \sin x+\Exp{x}-1) \\
 0 & \frac{\eta_1}{2} \Exp{-x} (\sin x+\Exp{x}-1) & \frac{\eta_2}{2} (\Exp{x} \sin x-\Exp{x}+1)\end{pmatrix}},
\\
 \pi^{mn} &= \eta_1 \eta_2 \sinh x \cos^2\bigl(\tfrac{x}{2}\bigr) {\footnotesize\begin{pmatrix} 0 & 0 & 0 \\ 0 & 0 & 1 \\ 0 & -1 & 0\end{pmatrix}}.
\end{split}
\end{align}
They construct the generalized frame fields satisfying $[E_{\hA},\,E_{\hB}]_{\text{D}}=-X_{\hA\hB}{}^{\hC}\,E_{\hC}$\,.

\section{Conclusions}
\label{sec:conclusion}

We have constructed the ExDA for half-maximal supergravities in $d\geq 4$\,. 
Then, following the general discussion \cite{2009.04454}, we have proven that the half-maximal ExDA systematically provides a set of generalized frame fields $E_{\hA}{}^{\hM}$ satisfying $[E_{\hA},\,E_{\hB}]_{\text{D}} = - X_{\hA\hB}{}^{\hC}\,E_{\hC}$\,. 
We have also computed the generalized CYBE associated with the half-maximal ExDA, and provided the general form of the generalized Poisson--Lie structures for coboundary-type ExDAs. 

A possible future direction is to extend the half-maximal ExDA to $d=3$\,. 
In $d=3$\,, the duality group is $\cG=\OO(d+1,d+1+n)$ and the corresponding ExFT has been studied in \cite{1707.06693}. 
In $d\geq 4$\,, the half-maximal ExDA with $n=0$ was obtained by truncating $2^{9-d}$ generators from the generators of the $E_{d+1(d+1)}$ EDA via the $\mathbb{Z}_2$ truncation.
In $d=3$\,, the number of $\mathbb{Z}_2$-odd generators becomes $2^{10-d}$ and the dimension of the half-maximal ExDA with $n=0$ should be $(248-2^{10-3})=120$\,. 
Then the generators can be parameterized as $T_{\hA}=T_{[A_1A_2]}$ \cite{1707.06693} where $A=1,\dotsc,2(d+1)$ denotes the vector index of $\OO(d+1,d+1)$\,. 
In \cite{2009.04454}, the $E_{8(8)}$ EDA in the type IIB picture has been already determined, and it will be not difficult to determine the explicit form of the half-maximal ExDA (with $n=0$) through the $\mathbb{Z}_2$ truncation. 
Its further extension to $n>0$ will be also straightforward. 

Another interesting direction is to study the $\mathbb{Z}_2$ truncation of the $E_{d+1(d+1)}$ EDA in the M-theory picture. 
The $\mathbb{Z}_2$ projection \eqref{eq:Z2-action} corresponds to putting the $S$-dual of O9-planes (and the $S$-dual of D9-branes) in type IIB theory. 
Under $U$-duality transformations, O9-planes can be mapped to certain orientifold planes in M-theory. 
They introduce a different $\mathbb{Z}_2$ projection which reduces the $E_{d+1(d+1)}$ EDA in M-theory picture to a certain half-maximal ExDA. 
It is interesting to find the explicit form of such a half-maximal ExDA. 

Here we have concentrated on algebra and the generalized frame fields $E_{\hA}{}^{\hM}$\,. 
Using $E_{\hA}{}^{\hM}$ and a constant matrix $\hat{\cH}_{\hA\hB}$\,, we can construct the generalized metric $\cH_{\hM\hN}$ of the ExFT, and by using some parameterization of the generalized metric, we can identify the corresponding supergravity fields. 
Then, we can study the extension of the PL $T$-duality, which rotates the generators of the Drinfel'd double. 
A redefinition $T_{\hA}\to T'_{\hA} = C_{\hA}{}^{\hB}\,T_{\hB}$ ($C_{\hA}{}^{\hB}\in \cG$) can map a half-maximal ExDA to another half-maximal ExDA, and the new generators $T'_{\hA}$ construct new generalized frame fields $E'_{\hA}{}^{\hM}$\,. 
Then we obtain a new generalized metric $\cH'_{\hM\hN}$ which describes the dual background. 
It is an important future work to prove that the non-Abelian duality, $\cH_{\hM\hN}\to \cH'_{\hM\hN}$\,, is a symmetry of ExFT. 
To this end, it is useful to study the flux formulation of ExFTs in detail. 
In addition, $\cH_{\hM\hN}$ can be parameterized in terms of several theories, such as heterotic/$T^D$, type I/$T^D$, or type II/K3$\times T^{D-4}$\,. 
To study non-Abelian dualities among these theories, it is important to study the parameterizations in detail. 
Moreover, to find various examples of non-Abelian duality, it is also important to study the classification of inequivalent redefinitions of generators, similar to \cite{math:0202210}. 

\subsection*{Acknowledgments}

We thank Jose J.\ Fernandez-Melgarejo for useful discussion in the early stage of the work. 
This work is supported by JSPS Grant-in-Aids for Scientific Research (C) 18K13540 and (B) 18H01214. 

\appendix

\section{Conventions}
\label{app:conv}

\subsection{Summary of indices}

Here we summarize the convention for various indices used in this paper. 
The generalized coordinates in the half-maximal ExFTs are parameterized as
\begin{align}
 x^{\hM} =
\begin{cases}
 x^{M} =\bigl(x^{m},\,x^{\cI},\,x_m\bigr) & (d\geq 6)
\\
 \bigl(x^{M},\,x^*\bigr) =\bigl(x^{m},\,x^{\cI},\,x_m,\,x^*\bigr)\quad & (d=5) 
\\
 x^{\dbba M} =\bigl(x^{\dbba m},\,x^{\dbba \cI},\,x^{\dbba}{}_m\bigr) & (d=4)
\end{cases},
\end{align}
where $M=1,\dotsc,2D+n$, $\cI=\dot{1},\dotsc,\dot{n}$, $m=1,\dotsc,D$, and $\dbba=+,-$, where $D\equiv 10-d$\,. 
The index $\cI$ may be raised/lowered by using the Kronecker delta $\delta_{\cI\cJ}$\,. 
In $E_{D+1(D+1)}$ EFT in the type IIB picture, the generalized coordinates are denoted as $x^{\cM}$\,. 

The generators of the half-maximal ExDA are parameterized as
\begin{align}
 T_{\hA} =
\begin{cases}
 T_A =\bigl(T_a,\,T_I,\,T^a\bigr) & (d\geq 6)
\\
 \bigl(T_A,\,T_*\bigr) =\bigl(T_a,\,T_I,\,T^a,\,T_*\bigr)\quad & (d=5) 
\\
 T_{\bba A} =\bigl(T_{\bba a},\,T_{\bba I},\,T_{\bba}{}^a\bigr) & (d=4)
\end{cases},
\end{align}
where $A=1,\dotsc,2D+n$, $I=\dot{1},\dotsc,\dot{n}$, $a=1,\dotsc,D$, and $\bba=+,-$. 
The index $I$ is raised/lowered by using the Kronecker delta $\delta_{IJ}$\,. 
The generators of the $E_{D+1(D+1)}$ EDA in the type IIB picture is parameterized as
\begin{align}
 T_{\cA} = \{T_{\sfa},\,T^{\sfa}_{\bm{\alpha}},\,\tfrac{T^{\sfa_1\sfa_2\sfa_3}}{\sqrt{3!}},\,\tfrac{T_{\bm{\alpha}}^{\sfa_1\cdots \sfa_5}}{\sqrt{5!}},\,T^{\sfa_1\cdots \sfa_6,\sfa}\}\,,
\end{align}
where $\sfa=1,\dotsc,D$ and $\bm{\alpha}=\bm{1},\bm{2}$\,. 
Here the multiple indices are totally antisymmetric. 
When we make the matrix representation, the indices are further decomposed as
\begin{align}
 T_{\cA} = \{T_{\sfa},\,T^{\sfa}_{\bm{1}},\,T^{\sfa}_{\bm{2}},\,\tfrac{T^{\sfa_1\sfa_2\sfa_3}}{\sqrt{3!}},\,\tfrac{T_{\bm{1}}^{\sfa_1\cdots \sfa_5}}{\sqrt{5!}},\,\tfrac{T_{\bm{2}}^{\sfa_1\cdots \sfa_5}}{\sqrt{5!}},\,T^{\sfa_1\cdots \sfa_6,\sfa}\}\,.
\end{align}

The generalized Poisson--Lie structures $\pi^{mn}$ and $\pi^m_I$ are related to $\pi^{ab}$ and $\pi^a_I$ as in Eq.~\eqref{eq:gPL-def}. 
Sometimes we also use the notation, such as $\pi^m_\cI=\delta_\cI^I\,\pi^m_I$ and $\omega_{I}{}^{\cI}= \omega_I{}^J\,\delta_J^\cI$\,. 

\subsection{Duality algebra in $d\geq 5$}

In $d\geq 5$\,, the duality group is $\mathbb{R}^+\times\OO(D,D+n)$ and the generators are decomposed as
\begin{align}
 \{t_{\bm{a}}\} = \bigl\{R_*,\,\tfrac{R_{a_1a_2}}{\sqrt{2!}},\,R_a^I,\,K^{a_1}{}_{a_2},\,\tfrac{R_{IJ}}{\sqrt{2}},\,R^a_I,\,\tfrac{R^{a_1a_2}}{\sqrt{2!}}\bigr\}\,,
\end{align}
where $a=1,\dotsc,D$\,. 
The $R_*$ is the generator of $\mathbb{R}^+_d$ and it commutes with other generators. 
The other $\OO(D,D+n)$ generators satisfy the following algebra:
{\small
\begin{align}
\begin{split}
 [K^a{}_b,\,K^{c}{}_d] &= \delta^c_b\,K^a{}_d-\delta^a_d\,K^c{}_b\,,\quad
 [K^a{}_b,\,R_{KL}]=0\,,\quad
 [K^a{}_b,\,R^{c}_K] = \delta^c_b\,R^{a}_K\,,\quad
\\
 [K^a{}_b,\,R_c^K] &= -\delta_c^a\,R_b^K\,,\quad
 [K^a{}_b,\,R^{cd}] = 2\,\delta^{cd}_{be}\,R^{ae}\,,\quad
 [K^a{}_b,\,R_{cd}] = -2\,\delta_{cd}^{ae}\,R_{be}\,,
\\
 [R_{IJ},\,R_{KL}] &= -2\,\bigl(\delta_{K[I}\,R_{J]L}-\delta_{L[I}\,R_{J]K}\bigr) \,,\quad
 [R_{IJ},\,R^{cd}] = 0\,,\quad
 [R_{IJ},\,R_{cd}] = 0 \,,
\\
 [R_{IJ},\,R^c_K] &= - 2\,\delta_{K[I}\,\delta_{J]}^L\,R^c_L \,,\quad
 [R_{IJ},\,R_c^K]  = - 2\,\delta^K_{[I}\,\delta_{J]L} \,R_c^L \,, 
\\
 [R^{ab},\,R^{cd}] &= 0\,,\quad
 [R^{ab},\,R_{cd}] = -4\,\delta^{[a}_{[c}\,K^{b]}{}_{d]} \,,\quad
 [R^{ab},\,R^{c}_K] = 0 \,,\quad
 [R^{ab},\,R_{c}^K] = -2\,\delta^{KL}\,\delta^{[a}_c\,R^{b]}_L\,,
\\
 [R_{ab},\,R_{cd}] &= 0\,,\quad
 [R_{ab},\,R_{c}^K] = 0 \,,\qquad
 [R_{ab},\,R^{c}_K] = -2\,\delta_{KL}\,\delta_{[a}^c\,R_{b]}^L\,, 
\\
 [R^a_I,\,R^b_J] &= - \delta_{IJ}\,R^{ab}\,,\quad
 [R^a_I,\,R_b^J] = -\delta_I^J\,K^a{}_b - \delta^a_b\,\delta^{JK}\,R_{IK}\,, \quad
 [R_a^I,\,R_b^J] = - \delta^{IJ}\,R_{ab}\,.
\end{split}
\label{eq:cG-algebra}
\end{align}}

In $d=5$\,, we can construct the matrix representations of these generators in the vector representation as follows. 
\begin{align}
 R_* &= \beta_d {\small\begin{pmatrix} \delta_a^b & 0 & 0 & 0 \\
 0 & \delta_I^J & 0 & 0
\\
 0 & 0 & \delta^a_b & 0\\
 0 & 0 & 0 & -2
\end{pmatrix}},
\\
 K^c{}_d &\equiv 
 {\small\begin{pmatrix} \delta_a^c\,\delta_d^b & 0 & 0 & 0 \\ 0 &0& 0& 0 \\ 0 & 0 & - \delta_d^a\,\delta_b^c & 0 \\ 0&0&0&0 \end{pmatrix}},\quad
 R_{KL} \equiv {\small\begin{pmatrix} 0 & 0 & 0 & 0 \\
 0 & \delta_{KI}\,\delta_L^J - \delta_{LI}\,\delta_K^J & 0 & 0 \\
 0 & 0 & 0 & 0 \\
 0 & 0 & 0 & 0\end{pmatrix}},
\\
 R^{c_1c_2} &\equiv {\small\begin{pmatrix} 0 & 0 & 2\, \delta^{c_1c_2}_{ab} & 0 \\
 0 & 0 & 0 & 0 \\
 0 & 0 & 0 & 0 \\
 0 & 0 & 0 & 0 \end{pmatrix}},\quad
 R_{c_1c_2} \equiv {\small\begin{pmatrix} 0 & 0 & 0 & 0 \\
 0 & 0 & 0& 0\\
 2\,\delta_{c_1c_2}^{ab} & 0 & 0& 0\\
 0 & 0 & 0& 0\end{pmatrix}},
\\
 R^c_K &\equiv {\small\begin{pmatrix} 0 & \delta^c_a\,\delta_K^J & 0 & 0\\
 0 & 0 & -\delta_{KI}\,\delta_b^c & 0\\
 0 & 0 & 0 & 0\\
 0 & 0 & 0 & 0 \end{pmatrix}},\quad
 R_c^K \equiv {\small\begin{pmatrix} 0 & 0 & 0 & 0 \\
 - \delta_c^b\,\delta^K_I & 0 & 0 & 0 \\
 0 & \delta^{KJ}\,\delta^a_c & 0 & 0 \\
 0 & 0 & 0& 0\end{pmatrix}}.
\end{align}
In $d\geq 6$, we can obtain the matrix representations by truncating the last row/column of the above matrices. 
Consequently, the generator $R_*$ is proportional to the identity matrix. 

We also define the dual generators as
\begin{align}
 \{t^{\bm{a}}\} = \bigl\{R^*,\,\tfrac{R^{a_1a_2}}{\sqrt{2!}},\,R^a_I,\,K_{a_1}{}^{a_2},\,\tfrac{R^{IJ}}{\sqrt{2}},\,R_a^I,\,\tfrac{R_{a_1a_2}}{\sqrt{2!}}\bigr\}\,,
\end{align}
where $R^* \equiv - (d-2)\,R_*$ and $K_a{}^b \equiv - K^b{}_a$\,.
Then the $Y$-tensor computed from Eq.~\eqref{eq:Y-tensor} coincides with the $\mathbb{Z}_2$-truncation of the $Y$-tensor in $E_{D+1(D+1)}$ EFT. 

\subsection{Duality algebra in $d=4$}

We denote the generators of the duality group $\cG=\SL(2)\times\OO(6,6+n)$ collectively as
\begin{align}
 \{t_{\bm{a}}\} = \bigl\{R^{\bba_1}{}_{\bba_2},\,\tfrac{R_{a_1a_2}}{\sqrt{2!}},\,R_a^I,\,K^{a_1}{}_{a_2},\,\tfrac{R_{IJ}}{\sqrt{2}},\,R^a_I,\,\tfrac{R^{a_1a_2}}{\sqrt{2!}}\bigr\}\,,
\end{align}
where $a,b=1,\dotsc,6$\,, $I,J=1,\dotsc,n$ and $\bba,\bbb=+,-$\,.
The $\SL(2)$ generators $R^{\bba}{}_{\bbb}$ ($R^{\bba}{}_{\bba}=0$) satisfy the commutation relations
\begin{align}
 [R^{\bba}{}_{\bbb},\,R^{\bbc}{}_{\bbd}] = \delta_{\bbb}^{\bbc}\,R^{\bba}{}_{\bbd} - \delta_{\bbd}^{\bba}\,R^{\bbc}{}_{\bbb}\,,\qquad
 [R^{\bba}{}_{\bbb},\,(\text{others})] = 0\,,
\end{align}
and the other $\OO(6,6+n)$ generators satisfy the same algebra as \eqref{eq:cG-algebra}. 
Their matrix representations are as follows:
\begin{align}
 R^\gamma{}_\delta &= {\small\begin{pmatrix} \bigl(\delta_{\bba}^{\bbc}\,\delta^{\bbb}_{\bbd}-\tfrac{1}{2}\delta_{\bba}^{\bbb}\delta^{\bbc}_{\bbd}\bigr)\,\delta_a^b & 0 & 0 
\\
 0 & \bigl(\delta_{\bba}^{\bbc}\,\delta^{\bbb}_{\bbd}-\tfrac{1}{2}\delta_{\bba}^{\bbb}\delta^{\bbc}_{\bbd}\bigr)\,\delta_I^J & 0 
\\
 0 & 0 & \bigl(\delta_{\bba}^{\bbc}\,\delta^{\bbb}_{\bbd}-\tfrac{1}{2}\delta_{\bba}^{\bbb}\delta^{\bbc}_{\bbd}\bigr)\,\delta^a_b 
\end{pmatrix}},
\\
 K^c{}_d &\equiv 
 {\small\begin{pmatrix} \delta_{\bba}^{\bbb}\,\delta_a^c\,\delta_d^b & 0 & 0 \\ 0 &0& 0\\ 0 & 0 & -\delta_{\bba}^{\bbb}\,\delta_d^a\,\delta_b^c \end{pmatrix}},\quad
 R_{KL} \equiv {\small\begin{pmatrix} 0 & 0 & 0 \\
 0 & \delta_{\bba}^{\bbb}\,\bigl(\delta_{KI}\,\delta_L^J - \delta_{LI}\,\delta_K^J\bigr) & 0 \\
 0 & 0 & 0 \end{pmatrix}},
\\
 R^{c_1c_2} &\equiv {\small\begin{pmatrix} 0 & 0 & 2\,\delta_{\bba}^{\bbb}\,\delta^{c_1c_2}_{ab} \\
 0 & 0 & 0 \\
 0 & 0 & 0\end{pmatrix}},\quad
 R_{c_1c_2} \equiv {\small\begin{pmatrix} 0 & 0 & 0 \\
 0 & 0 & 0\\
 2\,\delta_{\bba}^{\bbb}\,\delta_{c_1c_2}^{ab} & 0 & 0\end{pmatrix}},
\\
 R^c_K &\equiv {\small\begin{pmatrix} 0 & \delta_{\bba}^{\bbb}\,\delta^c_a\,\delta_K^J & 0 \\
 0 & 0 & -\delta_{KI}\,\delta_{\bba}^{\bbb}\,\delta_b^c \\
 0 & 0 & 0 \end{pmatrix}},\quad
 R_c^K \equiv {\small\begin{pmatrix} 0 & 0 & 0 \\
 -\delta_{\bba}^{\bbb}\,\delta_c^b\,\delta^K_I & 0 & 0 \\
 0 & \delta^{KJ}\,\delta_{\bba}^{\bbb}\,\delta^a_c & 0 \end{pmatrix}}.
\end{align}
We also define the dual generators as
\begin{align}
 \{t^{\bm{a}}\} = \bigl\{R_{\bba_1}{}^{\bba_2},\,\tfrac{R^{a_1a_2}}{\sqrt{2!}},\,R^a_I,\,K_{a_1}{}^{a_2},\,\tfrac{R^{IJ}}{\sqrt{2}},\,R_a^I,\,\tfrac{R_{a_1a_2}}{\sqrt{2!}}\bigr\}\,,
\end{align}
where
\begin{align}
 R_{\bba}{}^{\bbb} \equiv - R^{\bbb}{}_{\bba} \,,\qquad K_a{}^b \equiv - K^b{}_a\,.
\end{align}
They satisfy, for example,
\begin{align}
 (t^{\bm{a}})_{\hA}{}^{\hC}\,(t_{\bm{a}})_{\hC}{}^{\hB} &= -\bigl(\tfrac{25}{2}+n\bigr)\,\delta_{\hA}^{\hB}\,,
\\
 (t^{\bm{a}})_{\hA}{}^{\hC}\,(t_{\bm{b}})_{\hC}{}^{\hB} &= \begin{pmatrix}
 -(12+n)\,(\delta_{\bba_1}^{\bbb_1}\,\delta^{\bba_2}_{\bbb_2}-\tfrac{1}{2}\,\delta_{\bba_1}^{\bba_2}\,\delta^{\bbb_1}_{\bbb_2})\,\delta_{\hA}^{\hB} & 0 \\ 0 & -4\,\delta^{\bar{\bm{a}}}_{\bar{\bm{b}}}\,\delta_{\hA}^{\hB}\end{pmatrix},
\end{align}
where $\delta^{\bar{\bm{a}}}_{\bar{\bm{b}}}$ is a restriction of $\delta^{\bm{a}}_{\bm{b}}$ to $\OO(6,6+n)$ generators. 
We also find
\begin{align}
 (t^{\bm{a}})_{\hA}{}^{\hB}\,(t_{\bm{a}})_{\hC}{}^{\hD} = -28\,(\mathbb{P}_{(\bm{3},\bm{1})})_{\hA}{}^{\hB}{}_{\hC}{}^{\hD} -4\,(\mathbb{P}_{(\bm{1},\bm{ad})})_{\hA}{}^{\hB}{}_{\hC}{}^{\hD}\,,
\end{align}
where we have defined the projectors to the adjoint representations as
\begin{align}
\begin{split}
 (\mathbb{P}_{(\bm{3},\bm{1})})_{\hA}{}^{\hB}{}_{\hC}{}^{\hD} &= \tfrac{1}{2(12+n)}\,\bigl(\delta_{\bba}^{\bbd}\,\delta^{\bbb}_{\bbc}-\epsilon_{\bba\bbc}\,\epsilon^{\bbb\bbd}\bigr)\,\delta_A^B\,\delta_C^D \qquad (\epsilon_{+-}=1=\epsilon^{+-})\,,
\\
 (\mathbb{P}_{(\bm{1},\bm{ad})})_{\hA}{}^{\hB}{}_{\hC}{}^{\hD} &= \tfrac{1}{4}\,\delta_{\bba}^{\bbb}\,\delta_{\bbc}^{\bbd}\,\bigl(\delta_A^D\,\delta_C^B-\eta_{AC}\,\eta^{BD}\bigr) \,.
\end{split}
\end{align}

\section{Explicit form of half-maximal ExDA}
\label{app:4D-ExDA}

In this appendix, we summarize the explicit form of half-maximal ExDA in each dimension. 

In $d\geq 6$\,, the half-maximal ExDA is given by
\begin{align}
\begin{split}
 T_{a}\circ T_{b} &= f_{ab}{}^c\,T_{c}\,,
\\
 T_{a}\circ T_{J} &= -f_{a}{}^c{}_J \,T_{c} + f_{aJ}{}^{K}\,T_{K} + Z_a\,T_{J} \,,
\\
 T_{a}\circ T^b &= f_a{}^{bc}\,T_{c} + f_a{}^{bK}\,T_{K} - f_{ac}{}^b\,T^c + 2\,Z_a\,T^b\,,
\\
 T_{I} \circ T_{b} &= f_b{}^{c}{}_I\,T_{c} - f_b{}_I{}^K\,T_{K} -Z_b\,T_{I} \,,
\\
 T_{I}\circ T_{J} &= f_{cIJ}\,T^c + \delta_{IJ}\,Z_c\,T^c\,,
\\
 T_{I}\circ T^b &= -f_c{}^b{}_I\,T^c\,,
\\
 T^a\circ T_{b} &= -f_b{}^{ac}\,T_{c} - f_b{}^{aK}\,T_K + \bigl(f_{bc}{}^a+2\,\delta^a_b\,Z_c-2\,\delta^a_c\,Z_b\bigr)\,T^c \,,
\\
 T^a\circ T_{J} &= f_c{}^{a}{}_J\,T^c \,,
\\
 T^a\circ T^b &= f_c{}^{ab}\,T^c \,,
\end{split}
\end{align}
where $a,b=1,\cdots,D\equiv 10-d$ and $I,J=1,\dotsc,n$\,. 

In $d=5$\,, there is an additional generator $T_*$ and the ExDA has the form,
\begin{align}
\begin{split}
 T_{a}\circ T_{b} &= f_{ab}{}^c\,T_{c}\,,
\\
 T_{a}\circ T_{J} &= -f_{a}{}^c{}_J \,T_{c} + f_{aJ}{}^{K}\,T_{K} + Z_a\,T_{J} \,,
\\
 T_{a}\circ T^b &= f_a{}^{bc}\,T_{c} + f_a{}^{bK}\,T_{K} - f_{ac}{}^b\,T^c + 2\,Z_a\,T^b\,,
\\
 T_a \circ T_* &= (Z_a - f_a)\,T_*\,,
\\
 T_{I} \circ T_{b} &= f_b{}^{c}{}_I\,T_{c} - f_b{}_I{}^K\,T_{K} -Z_b\,T_{I} \,,
\\
 T_{I}\circ T_{J} &= f_{cIJ}\,T^c + \delta_{IJ}\,Z_c\,T^c\,,
\\
 T_{I}\circ T^b &= -f_c{}^b{}_I\,T^c\,,
\\
 T_I \circ T_* &= - f_c{}^c{}_I\,T_*\,, 
\\
 T^a\circ T_{b} &= -f_b{}^{ac}\,T_{c} - f_b{}^{aK}\,T_K + \bigl(f_{bc}{}^a+2\,\delta^a_b\,Z_c-2\,\delta^a_c\,Z_b\bigr)\,T^c \,,
\\
 T^a\circ T_{J} &= f_c{}^{a}{}_J\,T^c \,,
\\
 T^a\circ T^b &= f_c{}^{ab}\,T^c \,,
\\
 T^a \circ T_* &= - f_c{}^{ca}\,T_*\,,
\\
 T_*\circ T_b &=0\,,
\\
 T_*\circ T_J &=0\,,
\\
 T_*\circ T^b &=0\,,
\\
 T_*\circ T_* &=0\,.
\end{split}
\end{align}

In $d=4$\,, the half-maximal ExDA has the following form:
{\footnotesize\begin{align}
\begin{split}
 T_{+a}\circ T_{+b} &= f_{ab}{}^c\,T_{+c}\,,
\\
 T_{+a}\circ T_{-b} &= f_{a-}{}^+\,T_{+b} + \bigl(f_{ab}{}^c-f_a\,\delta_b^c \bigr)\,T_{-c} \,,
\\
 T_{+a}\circ T_{+J} &= -f_{a}{}^c{}_J \,T_{+c} + f_{aJ}{}^{K}\,T_{+K} + Z_a\,T_{+J} \,,
\\
 T_{+a}\circ T_{-J} &= -f_{a}{}^c{}_J \,T_{-c} + f_{a-}{}^+\,T_{+J} + f_{aJ}{}^{K}\,T_{-K} +(Z_a-f_a)\,T_{-J} \,,
\\
 T_{+a}\circ T_{+}{}^b &= f_a{}^{bc}\,T_{+c} + f_a{}^{bK}\,T_{+K} - f_{ac}{}^b\,T_{+}{}^c + 2\,Z_a\,T_+{}^b\,,
\\
 T_{+a}\circ T_{-}{}^b &= f_a{}^{bc}\,T_{-c} + f_a{}^{bK}\,T_{-K} + f_{a-}{}^+\,T_{+}{}^b - f_{ac}{}^b\,T_{-}{}^c + (2\,Z_a-f_a)\,T_-{}^b\,,
\\
 T_{-a}\circ T_{+b} &= -f_{b-}{}^+\,T_{+a} + f_a\,T_{-b} \,,
\\
 T_{-a}\circ T_{-b} &= -f_{b-}{}^+\,T_{-a} + f_{a-}{}^+\,T_{-b} \,,
\\
 T_{-a}\circ T_{+J} &= f_a\,T_{-J} \,,
\\
 T_{-a}\circ T_{-J} &= f_{a-}{}^+\,T_{-J} \,,
\\
 T_{-a}\circ T_{+}{}^b &= \delta_a^b\,f_{c-}{}^+\,T_+{}^c + f_a\,T_-{}^b \,,
\\
 T_{-a}\circ T_{-}{}^b &= f_{a-}{}^+\,T_-{}^b + \delta_a^b\,f_{c-}{}^+\,T_-{}^c \,,
\\
 T_{+I} \circ T_{+b} &= f_b{}^{c}{}_I\,T_{+c} - f_b{}_I{}^K\,T_{+K} -Z_b\,T_{+I} \,,
\\
 T_{+I} \circ T_{-b} &= f_b{}^{c}{}_I\,T_{-c} -f_c{}^{c}{}_I\,T_{-b} - f_b{}_I{}^K\,T_{-K} -Z_b\,T_{-I} \,,
\\
 T_{+I}\circ T_{+J} &= f_{cIJ}\,T_+{}^c + \delta_{IJ}\,Z_c\,T_+{}^c\,,
\\
 T_{+I}\circ T_{-J} &= -f_c{}^c{}_I\,T_{-J} + f_{cIJ}\,T_-{}^c + \delta_{IJ}\,Z_c\,T_-{}^c\,,
\\
 T_{+I}\circ T_{+}{}^b &= -f_c{}^b{}_I\,T_+{}^c\,,
\\
 T_{+I}\circ T_{-}{}^b &= -f_c{}^b{}_I\,T_-{}^c - f_c{}^{c}{}_I\,T_-{}^b\,,
\\
 T_{-I}\circ T_{+b} &= f_c{}^c{}_I\,T_{-b} -f_{b-}{}^+\,T_{+I} \,,
\\
 T_{-I}\circ T_{-b} &= -f_{b-}{}^+\,T_{-I} \,,
\\
 T_{-I}\circ T_{+J} &= f_c{}^c{}_I\,T_{-J} + \delta_{IJ}\,f_{c-}{}^+\,T_{+}{}^c\,,
\\
 T_{-I}\circ T_{-J} &= \delta_{IJ}\,f_{c-}{}^+\,T_{-}{}^c \,,
\\
 T_{-I}\circ T_{+}{}^b &= f_c{}^c{}_I\,T_{-}{}^b\,,
\\
 T_{-I}\circ T_{-}{}^b &= 0\,,
\\
 T_{+}{}^a\circ T_{+b} &= -f_b{}^{ac}\,T_{+c} - f_b{}^{aK}\,T_{+K} + \bigl(f_{bc}{}^a+2\,\delta^a_b\,Z_c-2\,\delta^a_c\,Z_b\bigr)\,T_{+}{}^c \,,
\\
 T_{+}{}^a\circ T_{-b} &= -f_b{}^{ac}\,T_{-c} - f_c{}^{ca}\,T_{-b} - f_b{}^{aK}\,T_{-K} + \bigl(f_{bc}{}^a+2\,\delta^a_b\,Z_c-2\,\delta^a_c\,Z_b\bigr)\,T_{-}{}^c \,,
\\
 T_{+}{}^a\circ T_{+J} &= f_c{}^{a}{}_J\,T_{+}{}^c \,,
\\
 T_{+}{}^a\circ T_{-J} &= -f_c{}^{ca}\,T_{-J} + f_c{}^{a}{}_J\,T_{-}{}^c\,,
\\
 T_{+}{}^a\circ T_{+}{}^b &= f_c{}^{ab}\,T_{+}{}^c \,,
\\
 T_{+}{}^a\circ T_{-}{}^b &= f_c{}^{ab}\,T_{-}{}^c - f_c{}^{ca}\,T_{-}{}^b \,,
\\
 T_{-}{}^a\circ T_{+b} &= f_c{}^{ca}\,T_{-b} + \bigl(\delta^a_b\,f_{c-}{}^+ - \delta_c^a\,f_{b-}{}^+\bigr)\,T_{+}{}^c \,,
\\
 T_{-}{}^a\circ T_{-b} &= \bigl(\delta^a_b\,f_{c-}{}^+ - \delta_c^a\,f_{b-}{}^+\bigr)\,T_{-}{}^c \,,
\\
 T_{-}{}^a\circ T_{+J} &= f_c{}^{ca}\,T_{-J} \,,
\\
 T_{-}{}^a\circ T_{-J} &= 0 \,,
\\
 T_{-}{}^a\circ T_{+}{}^b &= f_c{}^{ca}\,T_{-}{}^b \,,
\\
 T_{-}{}^a\circ T_{-}{}^b &= 0 \,.
\end{split}
\end{align}}


\begin{thebibliography}{99}
\bibitem{hep-th:9502122} 
  C.~Klimcik and P.~Severa,
  ``Dual nonAbelian duality and the Drinfeld double,''
  Phys.\ Lett.\ B {\bf 351}, 455 (1995)
  [hep-th/9502122].



\bibitem{hep-th:9302036} 
  W.~Siegel,
  ``Two vierbein formalism for string inspired axionic gravity,''
  Phys.\ Rev.\ D {\bf 47}, 5453 (1993)
  [hep-th/9302036].



\bibitem{hep-th:9305073} 
  W.~Siegel,
  ``Superspace duality in low-energy superstrings,''
  Phys.\ Rev.\ D {\bf 48}, 2826 (1993)
  [hep-th/9305073].



\bibitem{hep-th:9308133} 
  W.~Siegel,
  ``Manifest duality in low-energy superstrings,''
  hep-th/9308133.



\bibitem{0904.4664} 
  C.~Hull and B.~Zwiebach,
  ``Double Field Theory,''
  JHEP {\bf 0909}, 099 (2009)
  [arXiv:0904.4664 [hep-th]].



\bibitem{1006.4823} 
  O.~Hohm, C.~Hull and B.~Zwiebach,
  ``Generalized metric formulation of double field theory,''
  JHEP {\bf 1008}, 008 (2010)
  [arXiv:1006.4823 [hep-th]].



\bibitem{1008.1763} 
  D.~S.~Berman and M.~J.~Perry,
  ``Generalized Geometry and M theory,''
  JHEP {\bf 1106}, 074 (2011)
  [arXiv:1008.1763 [hep-th]].



\bibitem{1111.0459} 
  D.~S.~Berman, H.~Godazgar, M.~J.~Perry and P.~West,
  ``Duality Invariant Actions and Generalised Geometry,''
  JHEP {\bf 1202}, 108 (2012)
  [arXiv:1111.0459 [hep-th]].



\bibitem{1206.7045} 
  P.~West,
  ``E11, generalised space-time and equations of motion in four dimensions,''
  JHEP {\bf 1212}, 068 (2012)
  [arXiv:1206.7045 [hep-th]].



\bibitem{1208.5884} 
  D.~S.~Berman, M.~Cederwall, A.~Kleinschmidt and D.~C.~Thompson,
  ``The gauge structure of generalised diffeomorphisms,''
  JHEP {\bf 1301}, 064 (2013)
  [arXiv:1208.5884 [hep-th]].



\bibitem{1308.1673} 
  O.~Hohm and H.~Samtleben,
  ``Exceptional Form of D=11 Supergravity,''
  Phys.\ Rev.\ Lett.\  {\bf 111}, 231601 (2013)
  [arXiv:1308.1673 [hep-th]].



\bibitem{1312.0614} 
  O.~Hohm and H.~Samtleben,
  ``Exceptional Field Theory I: $E_{6(6)}$ covariant Form of M-Theory and Type IIB,''
  Phys.\ Rev.\ D {\bf 89}, no. 6, 066016 (2014)
  [arXiv:1312.0614 [hep-th]].



\bibitem{1312.4542} 
  O.~Hohm and H.~Samtleben,
  ``Exceptional field theory. II. E$_{7(7)}$,''
  Phys.\ Rev.\ D {\bf 89}, 066017 (2014)
  [arXiv:1312.4542 [hep-th]].



\bibitem{1406.3348} 
  O.~Hohm and H.~Samtleben,
  ``Exceptional field theory. III. E$_{8(8)}$,''
  Phys.\ Rev.\ D {\bf 90}, 066002 (2014)
  [arXiv:1406.3348 [hep-th]].



\bibitem{1707.08624} 
  F.~Hassler,
  ``Poisson-Lie T-Duality in Double Field Theory,''
  Phys.\ Lett.\ B {\bf 807}, 135455 (2020)
  [arXiv:1707.08624 [hep-th]].



\bibitem{1810.11446} 
  S.~Demulder, F.~Hassler and D.~C.~Thompson,
  ``Doubled aspects of generalised dualities and integrable deformations,''
  JHEP {\bf 1902}, 189 (2019)
  [arXiv:1810.11446 [hep-th]].



\bibitem{1903.12175} 
  Y.~Sakatani,
  ``Type II DFT solutions from Poisson-Lie T-duality/plurality,''
  PTEP, 073B04 (2019)
  [arXiv:1903.12175 [hep-th]].



\bibitem{2007.07897} 
  F.~Hassler and T.~Rochais,
  ``$\alpha'$‐Corrected Poisson‐Lie T‐Duality,''
  Fortsch.\ Phys.\  {\bf 68}, no. 9, 2000063 (2020)
  [arXiv:2007.07897 [hep-th]].



\bibitem{2007.07902} 
  R.~Borsato and L.~Wulff,
  ``Quantum Correction to Generalized $T$ Dualities,''
  Phys.\ Rev.\ Lett.\  {\bf 125}, no. 20, 201603 (2020)
  [arXiv:2007.07902 [hep-th]].



\bibitem{2007.09494} 
  T.~Codina and D.~Marques,
  ``Generalized Dualities and Higher Derivatives,''
  JHEP {\bf 2010}, 002 (2020)
  [arXiv:2007.09494 [hep-th]].



\bibitem{1911.06320} 
  Y.~Sakatani,
  ``$U$-duality extension of Drinfel’d double,''
  PTEP {\bf 2020}, no. 2, 023B08 (2020)
  [arXiv:1911.06320 [hep-th]].



\bibitem{1911.07833} 
  E.~Malek and D.~C.~Thompson,
  ``Poisson-Lie U-duality in Exceptional Field Theory,''
  JHEP {\bf 2004}, 058 (2020)
  [arXiv:1911.07833 [hep-th]].



\bibitem{2007.08510} 
  E.~Malek, Y.~Sakatani and D.~C.~Thompson,
  ``E$_{6(6)}$ exceptional Drinfel’d algebras,''
  JHEP {\bf 2101}, 020 (2021)
  [arXiv:2007.08510 [hep-th]].



\bibitem{2009.04454} 
  Y.~Sakatani,
  ``Extended Drinfel'd algebras and non-Abelian duality,''
  arXiv:2009.04454 [hep-th].



\bibitem{2012.13263} 
  E.~T.~Musaev and Y.~Sakatani,
  ``Non-abelian U-duality at work,''
  arXiv:2012.13263 [hep-th].



\bibitem{2001.09983} 
  Y.~Sakatani and S.~Uehara,
  ``Non-Abelian $U$-duality for membranes,''
  PTEP {\bf 2020}, no. 7, 073B01 (2020)
  [arXiv:2001.09983 [hep-th]].



\bibitem{1103.2136} 
  O.~Hohm and S.~K.~Kwak,
  ``Double Field Theory Formulation of Heterotic Strings,''
  JHEP {\bf 1106}, 096 (2011)
  [arXiv:1103.2136 [hep-th]].



\bibitem{1109.4280} 
  D.~Geissbuhler,
  ``Double Field Theory and N=4 Gauged Supergravity,''
  JHEP {\bf 1111}, 116 (2011)
  [arXiv:1109.4280 [hep-th]].



\bibitem{1201.2924} 
  M.~Grana and D.~Marques,
  ``Gauged Double Field Theory,''
  JHEP {\bf 1204}, 020 (2012)
  [arXiv:1201.2924 [hep-th]].



\bibitem{1612.05230} 
  F.~Ciceri, G.~Dibitetto, J.~J.~Fernandez-Melgarejo, A.~Guarino and G.~Inverso,
  ``Double Field Theory at SL(2) angles,''
  JHEP {\bf 1705}, 028 (2017)
  [arXiv:1612.05230 [hep-th]].



\bibitem{1104.3587} 
  G.~Dibitetto, A.~Guarino and D.~Roest,
  ``How to halve maximal supergravity,''
  JHEP {\bf 1106}, 030 (2011)
  [arXiv:1104.3587 [hep-th]].



\bibitem{1707.00714} 
  E.~Malek,
  ``Half‐Maximal Supersymmetry from Exceptional Field Theory,''
  Fortsch.\ Phys.\  {\bf 65}, no. 10-11, 1700061 (2017)
  [arXiv:1707.00714 [hep-th]].



\bibitem{1805.04524} 
  C.~D.~A.~Blair, E.~Malek and D.~C.~Thompson,
  ``O-folds: Orientifolds and Orbifolds in Exceptional Field Theory,''
  JHEP {\bf 1809}, 157 (2018)
  [arXiv:1805.04524 [hep-th]].



\bibitem{1109.0290} 
  G.~Aldazabal, W.~Baron, D.~Marques and C.~Nunez,
  ``The effective action of Double Field Theory,''
  JHEP {\bf 1111}, 052 (2011)
  Erratum: [JHEP {\bf 1111}, 109 (2011)]
  [arXiv:1109.0290 [hep-th]].



\bibitem{1203.6562} 
  G.~Dibitetto, J.~J.~Fernandez-Melgarejo, D.~Marques and D.~Roest,
  ``Duality orbits of non-geometric fluxes,''
  Fortsch.\ Phys.\  {\bf 60}, 1123 (2012)
  [arXiv:1203.6562 [hep-th]].



\bibitem{2104.00007} 
  J.~J.~Fernández-Melgarejo and Y.~Sakatani,
  ``Jacobi-Lie T-plurality,''
  arXiv:2104.00007 [hep-th].



\bibitem{EDA.nb}
  Y.~Sakatani,
  ``A Mathematica notebook for half-maximal ExDA in $d\geq 4$'',
  Ancillary files of the arXiv submission.



\bibitem{0807.4527} 
  M.~Grana, R.~Minasian, M.~Petrini and D.~Waldram,
  ``T-duality, Generalized Geometry and Non-Geometric Backgrounds,''
  JHEP {\bf 0904}, 075 (2009)
  [arXiv:0807.4527 [hep-th]].



\bibitem{1401.3360} 
  K.~Lee, C.~Strickland‐Constable and D.~Waldram,
  ``Spheres, generalised parallelisability and consistent truncations,''
  Fortsch.\ Phys.\  {\bf 65}, no. 10-11, 1700048 (2017)
  [arXiv:1401.3360 [hep-th]].



\bibitem{hep-th:0602024} 
  J.~Schon and M.~Weidner,
  ``Gauged N=4 supergravities,''
  JHEP {\bf 0605}, 034 (2006)
  [hep-th/0602024].



\bibitem{0809.5180} 
  A.~Le Diffon and H.~Samtleben,
  ``Supergravities without an Action: Gauging the Trombone,''
  Nucl.\ Phys.\ B {\bf 811}, 1 (2009)
  [arXiv:0809.5180 [hep-th]].



\bibitem{1407.4236}
  A.~Rezaei-Aghdam and M.~Sephid,
  ``Classification of real low-dimensional Jacobi (generalized)--Lie bialgebras,''
  Int.\ J.\ Geom.\ Methods Mod.\ Phys.\ 14 (2016) 1750007
  [arXiv:1407.4236 [math-ph]].



\bibitem{1511.02491}
  Y.~Kosmann-Schwarzbach,
  ``Multiplicativity, from Lie groups to generalized geometry,''
  arXiv:1511.02491 [math.SG].



\bibitem{1101.5954} 
  G.~Aldazabal, D.~Marques, C.~Nunez and J.~A.~Rosabal,
  ``On Type IIB moduli stabilization and N = 4, 8 supergravities,''
  Nucl.\ Phys.\ B {\bf 849}, 80 (2011)
  [arXiv:1101.5954 [hep-th]].



\bibitem{2006.12452} 
  C.~D.~A.~Blair, D.~C.~Thompson and S.~Zhidkova,
  ``Exploring Exceptional Drinfeld Geometries,''
  JHEP {\bf 2009}, 151 (2020)
  [arXiv:2006.12452 [hep-th]].



\bibitem{1611.05856} 
  Y.~Sakatani, S.~Uehara and K.~Yoshida,
  ``Generalized gravity from modified DFT,''
  JHEP {\bf 1704}, 123 (2017)
  [arXiv:1611.05856 [hep-th]].



\bibitem{1703.09213} 
  J.~Sakamoto, Y.~Sakatani and K.~Yoshida,
  ``Weyl invariance for generalized supergravity backgrounds from the doubled formalism,''
  PTEP {\bf 2017}, no. 5, 053B07 (2017)
  [arXiv:1703.09213 [hep-th]].



\bibitem{math:0202210} 
  L.~Snobl and L.~Hlavaty,
  ``Classification of six-dimensional real Drinfeld doubles,''
  Int.\ J.\ Mod.\ Phys.\ A {\bf 17}, 4043 (2002)
  [math/0202210 [math-qa]].



\bibitem{1707.06693} 
  O.~Hohm, E.~T.~Musaev and H.~Samtleben,
  ``O($d+1, d+1$) enhanced double field theory,''
  JHEP {\bf 1710}, 086 (2017)
  [arXiv:1707.06693 [hep-th]].
\end{thebibliography}
\end{document}